\documentclass[iop]{emulateapj}

\usepackage[utf8]{inputenc}

\usepackage{natbib}
\bibpunct{(}{)}{;}{a}{}{,}

\usepackage{amsfonts}
\usepackage{amsmath}
\usepackage{amssymb}
\usepackage{booktabs}
\usepackage{color}
\usepackage{graphicx}
\usepackage{multirow}
\usepackage{textcomp}
\usepackage{wasysym}

\usepackage{xcolor}

\definecolor{forestgreen}{rgb}{0.133, 0.545, 0.133}

\newcommand*{\revision}[1]{{\color{black}{#1}}}
\newcommand*{\change}[1]{{\color{black}{}}}

\newcommand{\be}{\begin{equation}}
\newcommand{\ee}{\end{equation}}
\newcommand{\bea}{\begin{eqnarray}}
\newcommand{\eea}{\end{eqnarray}}

\newcommand{\mm}{\,\hbox{mm}}

\newcommand{\mum}{\,\hbox{\textmu{}m}}
\newcommand{\cm}{\,\hbox{cm}}

\newcommand{\AU}{\,\hbox{AU}}
\newcommand{\g}{\,\hbox{g}}

\newcommand{\pc}{\,\hbox{pc}}

\newcommand{\Jy}{\,\hbox{Jy}}

\newcommand{\K}{\,\hbox{K}}

\newcommand*{\smin}{s_\mathrm{min}}
\newcommand*{\smax}{s_\mathrm{max}}
\newcommand*{\sblow}{s_\mathrm{blow}}

\newcommand*{\rbb}{R_\mathrm{BB}}
\newcommand*{\rdisc}{R_\mathrm{cold}}
\newcommand*{\rwarm}{R_\mathrm{warm}}
\newcommand*{\tbb}{T_\mathrm{BB}}
\newcommand*{\td}{T_\mathrm{dust}}
\newcommand*{\teff}{T_\mathrm{eff}}
\newcommand*{\Qpr}{Q_\mathrm{pr}}
\newcommand*{\Qabs}{Q_\mathrm{abs}}

\shorttitle{Properties of Herschel-resolved debris disks}
\shortauthors{Pawellek et al.}

\begin{document}

\title{Disk Radii and Grain Sizes in {\it Herschel}-Resolved Debris Disks}

\author{Nicole Pawellek and Alexander V. Krivov}
\affil{Astrophysikalisches Institut und Universit\"atssternwarte, Friedrich-Schiller-Universit\"at Jena, Schillerg\"a{\ss}chen~2--3, 07745 Jena, Germany}

\author{Jonathan P. Marshall}
\affil{School of Physics, University of New South Wales, Sydney NSW 2052, Australia}
\affil{Australian Centre for Astrobiology, University of New South Wales, Sydney, NSW 2052, Australia}
\affil{Departamento de F\'isica Te\'orica, Facultad de Ciencias, Universidad Aut\'onoma de Madrid, Cantoblanco, 28049 Madrid, Spain}

\author{Benjamin Montesinos}
\affil{Departmento de Astrof\'isica, Centro de Astrobiolog\'ia (CAB, CSIC-INTA), ESAC Campus, PO Box 78, 28691 Villanueva de la Ca\~nada, Madrid, Spain}

\author{P\'eter \'Abrah\'am and Attila Mo\'or}
\affil{Konkoly Observatory, Research Centre for Astronomy and Earth Sciences, Hungarian Academy of Sciences, PO Box 67, H-1525 Budapest, Hungary}

\author{Geoffrey Bryden}
\affil{Jet Propulsion Laboratory, California Institute of Technology, Pasadena, CA 91109, USA}

\and

\author{Carlos Eiroa}
\affil{Departamento de F\'isica Te\'orica, Facultad de Ciencias, Universidad Aut\'onoma de Madrid, Cantoblanco, 28049 Madrid, Spain}


\begin{abstract}
The radii of debris disks and the sizes of their dust grains 
are important tracers of the \revision{planetesimal} formation mechanisms and physical 
processes operating in these systems.
Here we use a representative sample of 34 debris disks resolved in various 
{\it Herschel Space Observatory}\footnote{Herschel
is an ESA space observatory with science instruments provided by 
European-led Principal Investigator consortia and with important participation from NASA.}
programs to constrain the disk radii and the size distribution of their dust.
\revision{While we modeled disks with both warm and cold components, and identified warm
inner disks around about two-thirds of the stars,
we focus our analysis only on the cold outer disks, i.e. Kuiper-belt analogs.}
We derive the disk radii from the resolved images and find
a large dispersion for host stars of any spectral class, but no significant
trend with the stellar luminosity.
This argues against ice lines as a dominant player in setting the
debris disk sizes, since the ice line location varies with the luminosity of the 
central star.
Fixing the disk radii to those inferred from the resolved images,
we model the spectral energy distribution to determine the dust temperature 
and the \revision{grain} size distribution for each target.
While the dust temperature systematically increases towards earlier spectral types,
the ratio of the dust temperature to the blackbody temperature at the disk radius
decreases with the stellar luminosity.
This is explained by a clear trend of typical sizes increasing towards more luminous 
stars.
The typical grain sizes are compared to the radiation pressure blowout limit $\sblow$
that is proportional to the stellar luminosity-to-mass ratio and thus also increases 
towards earlier spectral classes.
The grain sizes in the disks of \revision{G}- to A-stars are inferred to be
several times $\sblow$ at all stellar luminosities, in agreement with collisional 
models of debris disks.
The sizes, measured in the units of $\sblow$, appear to decrease with the luminosity,
which may be suggestive of the disk's stirring level increasing towards earlier-type stars.
The dust opacity \revision{index} $\beta$ ranges between zero and two, and the size 
distribution index $q$ varies between three and five for all the disks in the sample.
\change{Sentence deleted}
\end{abstract}



\section{Introduction}

Protoplanetary disks are known to exhibit broad, smooth radial profiles
\citep{williams-cieza-2011}.
However, many of their successors, debris disks, have most of the material confined to
one or more distinct belts \citep{matthews-et-al-2013}.
The evidence comes from the resolved images of prominent debris disks
such as that of Fomalhaut \citep{acke-et-al-2012,boley-et-al-2012},
\revision{HD~181327 \citep{lebreton-et-al-2012}
or 49~Cet \citep{roberge-et-al-2013},
and from the spectral energy distributions (SEDs) of unresolved disks that are fitted
reasonably well by assuming one or two narrow dust rings at distinct locations
\citep[e.g.][]{morales-et-al-2009,morales-et-al-2011,%
donaldson-et-al-2013,ballering-et-al-2013,chen-et-al-2014}.
}

\change{Indent}\revision{Observation of debris rings} at specific locations around 
the star implies that planetesimals were able to form there by one or another mechanism
\citep{johansen-et-al-2014}, \revision{or that} they formed elsewhere and were moved to 
their current locations by migration processes or planetary scattering 
\citep{raymond-et-al-2011}.
For the rings to be seen, it is also necessary that those regions
are dynamically stable against planetary perturbations for a significant fraction
of the lifetime of the central star, and that planetesimal belts
retained much of their mass despite collisional depletion.
Furthermore, the planetesimals at these 
locations have to be sufficiently excited dynamically \citep{wyatt-2008},
either by neighboring planets \citep{mustill-wyatt-2009}
or embedded perturbers \citep{kenyon-bromley-2008},
to make planetesimal collisions destructive and to enable production 
of visible debris dust.
It is likely that the radii of debris rings are set by planets that clear up 
their inner regions \citep{kennedy-wyatt-2010} and may also be solely or partly responsible
for triggering the collisional cascade in the debris zone.
Alternatively, the debris zones may mark the locations at which planetesimals can most
efficiently form.
Either way, it is indisputable that the radii of debris disks bear valuable 
information on planetesimals and planets and their formation processes.

Another important diagnostic of debris disks
is the size distribution of their dust. It reflects an interplay
between grain-grain collisions, radiation pressure, various transport processes, as well
as mechanisms that lead to modification of dust grains
\citep{wyatt-et-al-2011}.

\change{Indent}An essential theoretical expectation is
that in the disks around solar- or earlier-type stars, the dominant
grain size is $\smin \approx b \sblow$, where 
$\sblow$ is the blowout size, below which the grains are expelled by
direct radiation pressure, and
$b$ is a numerical factor.
The latter depends, in particular, on the dynamical excitation of dust-producing 
planetesimals in the disk.
For collisionally active disks, collisional simulations suggest
$b \sim 2$ \citep[e.g.][]{krivov-et-al-2006,thebault-augereau-2007}.
However, for dynamically
cold disks with planetesimals in low-eccentricity, low-inclination orbits, $b$ can be 
much larger \citep{thebault-wu-2008,krivov-et-al-2013}.
For instance, \citet{loehne-et-al-2011}
infer $b \approx 8$ for the disk around HD~207129 \citep{marshall-et-al-2011}.
Around late-type stars, radiation pressure may be too weak to eliminate grains at all sizes,
so that $\smin$ should be set by size-dependent transport mechanisms
such as the Poynting-Robertson (P-R) force
\citep[see, e.g.,][]{kuchner-stark-2010,vitense-et-al-2010,wyatt-et-al-2011}
or stellar wind drag \citep{plavchan-et-al-2005,reidemeister-et-al-2011}.
Further effects may come from dust interactions with gas which is present
at detectable levels in some young disks
\citep[e.g.][]{roberge-et-al-2006,moor-et-al-2011,donaldson-et-al-2013,roberge-et-al-2013}.
Altogether, $\smin$ should sensitively depend on the mechanical and optical 
properties of 
dust, the dynamical excitation of the disk, and the 
radiation field of the central star and stellar winds.
Therefore, retrieving the dust sizes from observations and 
comparing them with model predictions enables constraints of all these parameters 
and the disk physics to be probed.

A difficulty of deducing the disk radii and grain sizes from observations is 
that there is  a degeneracy between the particle size, the disk radius, and the optical 
properties of dust. Grains are not perfect absorbers and emitters, so that their temperature
differs from the blackbody temperature. Smaller grains are usually warmer
than larger ones, and their temperature at a given size and distance depends on their
optical properties. As a consequence, the same SED
can be reproduced equally well with smaller grains placed farther out from the star,
larger grains closer in, and also with the same-sized grains of different composition
at another distance from the star.

The degeneracy with the optical properties is hard to break, but the one between sizes and
distance can be removed if resolved images are available and thus the location of the
emitting dust is known.
The case studies of individual resolved disks reveal no statistically significant 
trend with stellar parameters \citep{matthews-et-al-2013}.
The radii of the cold debris rings range from a few tens of AU to more than $200\AU$,
without any obvious correlation with stellar age, spectral type, or metallicity.
However, \revision{a few small studies} done so far convincingly show
that measured sizes of debris disks from resolved images are 
systematically larger than those inferred from SED blackbody analyses.
The ratio of the radius measured from the image to that inferred from
the disk temperature assuming blackbody grains, $\Gamma$, 
is greater than unity. 
\citet{rodriguez-zuckerman-2012} report $\Gamma \sim 1$--$5$ for a handful of thermally resolved 
disks. 
\citet{booth-et-al-2013} show compiled $\Gamma$-ratios for nine
resolved A-type stars in the {\it Herschel}/DEBRIS survey to range between 1 and 2.5.
\citet{marshall-et-al-2011} and \citet{wyatt-et-al-2012}
find $\Gamma \approx 4$ for G-stars HD 207129 and 61~Vir, respectively.
\citet{lestrade-et-al-2012} determine $\Gamma \approx 10$ for an M-star 
GJ~581.
These results are consistent with the
typical grain size being on the order of microns, roughly comparable
with several times the radiation pressure blowout limit for commonly assumed
dust material compositions and compact or moderately porous grains.

The number of resolved debris disks has dramatically increased to about 100
over the last few years \citep{matthews-et-al-2013}\footnote{At the time of this 
writing, about a half of these have been published.},
thanks particularly to the large-scale {\it Herschel} \citep{pilbratt-et-al-2010} surveys
\citep[see][for a review]{matthews-et-al-2013}, carried out with the PACS \citep{poglitsch-et-al-2010}
and SPIRE \citep{griffin-et-al-2010} instruments.
These surveys also provided accurate photometry
across the far-infrared (far-IR) and sub-millimeter spanning 
60--$500\mum$ and covering the region of the
source SED where dust emission peaks.
The resolved images along with the densely populated SEDs can be advantageously 
used to measure the disk radii and to better constrain the grain sizes.
As said above, the results could shed more light onto the disks' physics.
Besides, they can then be applied to 
hundreds of disks that as yet remain unresolved or marginally resolved, serving as a 
``calibrator'' for their SED modeling.

This paper tries to constrain both disk radii and typical grain sizes from 
observational data for
a selection of well-resolved disks with well-sampled SEDs, for which {\it Herschel} data are 
available.
We extend the study by \citet{booth-et-al-2013} from nine disks around A-type stars
to 34 disks around stars of spectral types from A to M.
Section~2 describes and characterizes our sample.
Section~3 explains the modeling procedure.
The results are presented in Section~4.
Section~5 contains a discussion.
Conclusions are drawn in Section~6.


\section{Sample selection and characterization}

\subsection{Selection criteria and resulting sample}

We focus on the main, cold component of the systems, i.e. Kuiper-belt analogs.
For selection, we require that
\begin{itemize}
\item
  the target has \revision{high signal-to-noise} (S/N$>10$)
 {\it Herschel} fluxes \revision{in at least two} out of three PACS bands
 (70 $\mum$, 100 $\mum$, 160 $\mum$);
\item
  the disk is spatially resolved with PACS in at least one of these wavebands; 
\item
  the system does not reveal peculiarities,
  such as a known nearby galaxy contaminating the image
  \citep[e.g. 61~Vir,][]{wyatt-et-al-2012},
  substantial asymmetries in the resolved image
  \citep[e.g. $\beta$~Pic,][]{golimowski-et-al-2006};
  or centrally peaked disk
  \citep[e.g. $\epsilon$~Eri,][]{backman-et-al-2009}.
\end{itemize}

The resulting selection of stars is listed in Table~\ref{tab:list}.
This list ensures a broad coverage of luminosities (except for the late-type stars),
distances, and ages.
We are aware that our sample is incomplete and may be biased in one or another way.
Yet it is the largest sample of resolved disks ever analyzed, which we
\revision{deem sufficient}
for the goals of our paper.

\begin{table}
\begin{center}
\caption{
\label{tab:list}
Our Sample of Resolved Debris Disks}
{
\begin{tabular}{rrcccl}
\hline
HD       &HIP      &Name     		&Program   	&Ref	&\\
\hline
 10647   &  7978   & q$^1$ Eri  	&  a   		&1	&\\  
 23484   & 17439   & -      	 	&  a   		&2	&\\
 48682   & 32480   & 56 Aur  		&  a   		&3	&\\    
207129   &107649   & -       		&  a   		&4	&\\   
         &         &     		&          	&	&\\
 95418   & 53910   & $\beta$ Uma 	&  b  		&5	&\\    
102647   & 57632   & $\beta$ Leo 	&  b  		&5	&\\
109085   & 61174   & $\eta$ Crv         &  b            &5      &\\
         &         &         		&          	&       &\\
 13161   & 10064   & $\beta$ Tri 	&  b  		&6      &\\    
 14055   & 10670   & $\gamma$ Tri 	&  b  		&6      &\\    
 20320   & 15197   & $\zeta$ Eri 	&  b  		&6      &\\    
 71155   & 41307   & 30 Mon  		&  b  		&6      &\\    
110411   & 61960   & $\varrho$ Vir 	&  b  		&6      &\\    
125162   & 69732   & $\lambda$ Boo 	&  b  		&6      &\\    
139006   & 76267   & $\alpha$ CrB 	&  b  		&6      &\\    
188228   & 98495   & $\epsilon$ Pav 	&  b  		&6      &\\    
         &         &        		&          	&       &\\
 -	 & 74995   & GJ 581   		&  b  		&7      &\\ 
 27290   & 19893   & $\gamma$ Dor 	&  b  		&8	&\\
         &         &         		&          	&       &\\
218396   &114189   & HR 8799 		&  c		&9      &\\
         &         &         		&          	&       &\\
172167   & 91262   & Vega    		&  d  		&10     &\\   
197481   &102409   & AU Mic             &  d            &11     &\\
216956   &113368   & Fomalhaut		&  d  		&12     &\\
         &         &         		&          	&       &\\
  9672   &  7345   & 49 Cet  		&  e   		&13     &\\               
         &         &         		&          	&       &\\
 10939   &  8241   & q$^2$ Eri  	&  f 		&14     &\\       
 17848   & 13141   & $\nu$ Hor  	&  f 		&14     &\\       
 21997   & 16449   & HR 1082 		&  f 		&15     &\\       
 50571   & 32775   & HR 2562 		&  f 		&14     &\\       
 95086   & 53524   & -       		&  f 		&16     &\\       
161868   & 87108   & $\gamma$ Oph	&  f 		&14     &\\       
170773   & 90936   & HR 6948 		&  f 		&14     &\\       
182681   & 95619   & HR 7380 		&  f 		&14     &\\       
195627   &101612   & $\varphi^1$ Pav	&  f 		&14     &\\       
         &         &         		&          	&       &\\
104860   & 58876   & -       		&  g  		&17     &\\          
         &         &         		&          	&       &\\
142091   & 77655   & $\kappa$ CrB 	&  h  		&18     &\\
         &         &                    &               &       &\\
 71722   & 41373   & HR 3341            &  i            &17     &\\  
\hline
\end{tabular}
}
\end{center}
\begin{flushleft}
\noindent
{\em Programs:}\\[0mm]
[a]~DUNES;
[b]~DEBRIS;
[c]~OT1\_bmatthew\_4;
[d]~KPGT\_golofs01\_1;
[e]~GASPS;
[f]~OT1\_pabraham\_2;
[g]~OT1\_gbryden\_1;
[h]~OT1\_abonsor\_1;
[i]~OT2\_fmorales\_3.
\end{flushleft}

\noindent
{\em References:}\\[0mm]
[1]~\citet{liseau-et-al-2010};
[2]~\citet{ertel-et-al-2013};
[3]~\citet{eiroa-et-al-2013};
[4]~\citet{marshall-et-al-2011};
[5]~\citet{matthews-et-al-2010};
[6]~\citet{booth-et-al-2013};
[7]~\citet{lestrade-et-al-2012};
[8]~\citet{broekhovenfiene-et-al-2013};
[9]~\citet{matthews-et-al-2013b};
[10]~\citet{sibthorpe-et-al-2010};
[11]~Matthews et al. (in prep.);
[12]~\citet{acke-et-al-2012};
[13]~\citet{roberge-et-al-2013};
[14]~Mo\'or et al (in prep.);
[15]~\citet{moor-et-al-2013b};
[16]~\citet{moor-et-al-2013};
[17]~\citet{morales-et-al-2013};
[18]~\citet{bonsor-et-al-2013b}.

\end{table}

\subsection{Stellar parameters and photospheres}

For all the stars in our sample, we have collected stellar data from
the literature (Table~\ref{tab:stars}).  These stellar parameters
were used to calculate the synthetic stellar spectra by interpolation
in the PHOENIX/GAIA model grid \citep{brott-hauschildt-2005}.  For two
stars with $\teff \ge 10000\K$, ATLAS9 models
\citep{castelli-kurucz-2004} were used instead, because the PHOENIX
grid does not go beyond 10000 K.

\begin{table*}
\caption{Stellar Parameters Sorted by Stellar Luminosity
\label{tab:stars}
}
\tabcolsep 3pt
  \begin{center}
    \begin{tabular}{rrcrcrrccrc}
    \toprule
    HD    	& HIP   	& Name  	& d [pc] & SpT   & $L/L_\odot$ 	& $T_\text{eff}$ [K]& $M/M_\odot$	& log(g) & [Fe/H] & Ref \\
    \midrule
    -	 	& 74995 	& GJ 581 	& 6.2   & M5V  	 	& 0.012 	& 3498  	& 0.31  	& 4.90  & 0.00  & 1 \\
    197481      & 102409        & AU Mic        & 9.9   & M1Ve          & 0.062         & 3600	        & 0.48          & 4.60  & 0.00  & 2\\
    23484 	& 17439 	& -     	& 16.0  & K2V   	& 0.41  	& 5166  	& 0.79  	& 4.44  & 0.05  & 3 \\
    104860 	& 58876 	& -     	& 45.5  & F8    	& 1.16  	& 5930  	& 1.04  	& 4.39  & -0.26 & 4, 5, 6 \\
    207129 	& 107649 	& -     	& 16.0  & G2V   	& 1.25  	& 5912  	& 1.06  	& 4.44  & -0.01 & 3 \\
    10647 	& 7978  	& q1 Eri 	& 17.4  & F9V   	& 1.52  	& 6155  	& 1.12  	& 4.48  & -0.04 & 3 \\
    48682 	& 32480 	& 56 Aur 	& 16.7  & G0V   	& 1.83  	& 6086  	& 1.17  	& 4.35  & 0.09  & 3 \\
    50571 	& 32775 	& HR 2562 	& 33.6  & F5VFe+0.4 	& 3.17  	& 6490  	& 1.35  	& 4.23  & -0.01 & 4, 5, 6, 7 \\
    170773 	& 90936 	& HR 6948 	& 37.0  & F5V   	& 3.44  	& 6590  	& 1.38  	& 4.29  & -0.05 & 4, 5, 6, 8 \\
    218396 	& 114189 	& HR 8799 	& 39.4  & A5V   	& 4.81  	& 7380  	& 1.51  	& 4.29  & -0.50 & 4, 6, 8 \\
    109085	& 61174	        & $\eta$ Crv	& 18.2  & F2V           & 4.87          & 6950          & 1.52          & 4.14  & -0.08 & 9\\
    27290 	& 19893 	& $\gamma$ Dor 	& 20.5  & F1V   	& 6.27  	& 7070  	& 1.62  	& 4.10  & -0.13 & 4, 6, 7 \\
    95086 	& 53524 	& -     	& 90.4  & A8III 	& 7.04  	& 7530  	& 1.67  	& 4.29  & 0.00  & 4, 6 \\
    195627 	& 101612 	& $\phi$1 Pav 	& 27.8  & F0V   	& 7.36  	& 7200  	& 1.69  	& 4.05  & -0.12 & 4, 6, 7 \\
    20320 	& 15197 	& $\zeta$ Eri 	& 33.6  & kA4hA9mA9V$^a$& 10.3  	& 7575  	& 1.85  	& 4.05  & 0.04  & 4, 6, 7 \\
    21997 	& 16449 	& HR 1082 	& 71.9  & A2IV/V 	& 11.2  	& 8325  	& 1.89  	& 4.30  & 0.00  & 4, 6, 10 \\
    110411 	& 61960 	& $\varrho$ Vir	& 36.3  & A0V   	& 11.7  	& 8710  	& 1.91  	& 4.18  & -1.10 & 4, 6, 8 \\
    142091 	& 77655 	& $\kappa$ CrB 	& 30.5  & K1IVa 	& 12.5  	& 4815  	& 1.94  	& 3.12  & -0.09 & 4, 6, 8 \\
    102647 	& 57632 	& $\beta$ Leo 	& 11.0  & A3Va  	& 13.2  	& 8490  	& 1.97  	& 4.26  & 0.00  & 4, 6, 8 \\
    125162 	& 69732 	& $\lambda$ Boo	& 30.4  & A0p   	& 15.4  	& 8550  	& 2.05  	& 4.11  & -1.86 & 4, 6, 8 \\
    216956 	& 113368 	& Fomalhaut 	& 7.7   & A4V   	& 15.5  	& 8195  	& 2.06  	& 4.17  & 0.10  & 4, 6, 7 \\
    17848 	& 13141 	& $\nu$ Hor 	& 50.5  & A2V   	& 15.7  	& 8400  	& 2.07  	& 4.20  & 0.00  & 4, 6, 10 \\
    9672  	& 7345  	& 49 Cet 	& 59.4  & A1V   	& 16.0  	& 9000  	& 2.07  	& 4.30  & 0.10  & 11 \\
    71722       & 41373         & HR 3341       & 71.7  & A0V           & 18.5          & 8925          & 2.16          & 4.29  & 0.00  & 12 \\
    182681 	& 95619 	& HR 7380 	& 69.9  & B9V   	& 24.9  	& 10000 	& 2.33  	& 4.30  & 0.00  & 4, 6 \\
    14055 	& 10670 	& $\gamma$ Tri 	& 34.4  & A1Vnn 	& 25.0  	& 9350  	& 2.33  	& 4.19  & 0.00  & 4, 6 \\
    161868 	& 87108 	& $\gamma$ Oph 	& 31.5  & A0V   	& 26.0  	& 9020  	& 2.36  	& 4.12  & -0.81 & 4, 6, 8 \\
    188228 	& 98495 	& $\epsilon$ Pav& 32.2  & A0Va  	& 26.6  	& 10190 	& 2.37  	& 4.23  & -0.04 & 4, 6, 7 \\
    10939 	& 8241  	& q2 Eri  	& 62.0  & A1V   	& 31.3  	& 9200  	& 2.47  	& 4.17  & 0.00  & 4, 6, 10 \\
    71155 	& 41307 	& 30 Mon 	& 37.5  & A0V   	& 35.7  	& 9770  	& 2.56  	& 4.06  & -0.44 & 4, 6, 6 \\
    172167 	& 91262 	& Vega  	& 7.7   & A0V   	& 51.8  	& 9530  	& 2.83  	& 3.93  & -0.43 & 4, 6, 8 \\
    139006 	& 76267 	& $\alpha$ CrB 	& 23.0  & A0V   	& 57.7  	& 9220  	& 2.91  	& 3.77  & 0.00  & 4, 6, 8 \\
    95418 	& 53910 	& $\beta$ Uma 	& 24.4  & A1IVps 	& 58.2  	& 9130  	& 2.91  	& 3.76  & 0.06  & 4, 6, 8 \\
    13161 	& 10064 	& $\beta$ Tri 	& 38.9  & A5III 	& 73.8  	& 8010  	& 3.10  	& 3.62  & 0.20  & 4, 6, 8 \\
    \bottomrule
    \end{tabular}
  \end{center}

\noindent
{\em Notes:}\\[0mm]
The effective temperatures, metallicities and gravities are averaged over the listed 
literature values.\\
The stellar masses were computed from the luminosities by means of  a standard relation 
$M\propto L^{1/3.8}$.\\
$^a$Gray-Corbally notation. See App.~A2 in \citet{trilling-et-al-2007}
for its explanation. 

\medskip
\noindent
{\em References:}\\[0mm]
[1]~\citet{braun-et-al-2011};
[2]~\citet{torres-2010};
[3]~\citet{eiroa-et-al-2013} and references therein;
[4]~\citet{gray-1992};
[5]~\citet{holmberg-et-al-2009};
[6]~\citet{allende-prieto-et-al-1999};
[7]~\citet{gray-et-al-2006};
[8]~\citet{gray-et-al-2003};
[9]~\citet{duchene-et-al-2014};
[10]~\citet{paunzen-et-al-2006};
[11]~\citet{roberge-et-al-2013};
[12]~\citet{morales-et-al-2013}.

\end{table*}

\looseness=-1
We also collected optical and near-IR photometry to build the SEDs.
Johnson $BV$ and Cousins $I_{\rm c}$ photometry was taken from the
Hipparcos catalogue (CDS catalogue {\tt I/239/hip\_main}), and 2MASS
$JHK_{\rm s}$ from \citet{cutri-et-al-2003}
(CDS catalogue {\tt II/246}). The
magnitudes were transformed into fluxes using the calibrations
by \citet{bessell-1979} ($BVI_{\rm c}$) and \citet{cohen-et-al-2003}
($JHK_{\rm s}$).  For each star, the synthetic model photosphere was normalized
to that photometry taking the flux at I$_c$ as a reference, since
2MASS photometry was found not to have the best quality for a number of
targets. The only exception was GJ~581 where the normalization was
done to the flux at 2MASS K$_s$.

\change{Indent}We did not correct the photospheres for extinction, since
this would have a small effect on mid- and far-IR fluxes.
Indeed, in the Local Bubble ($d<100\pc$), where all of our targets are located,
the extinction in the optical is $A_V \la 0.1$~mag
\citep[e.g.][]{frisch-et-al-2011,reis-et-al-2011},
which is comparable to the uncertainties in the measured magnitudes.
We take $A_V = 0.2$~mag as the worst case.
At I$_c$ to which our model photospheres were normalized,
assuming a ratio $A(\text{I}_c)/A_V \approx 0.5$ from \citet{rieke-lebofsky-1985},
the extinction for our targets should be $\la 0.1$~mag.
Thus, without dereddening, we may be underestimating the true photospheric flux
by $\la 10$\%.
In the mid-IR where the excess fluxes may be only slightly above the photospheric flux,
this would lead to overestimating the excess fluxes in the mid-IR by
the same \revision{10 percent}. However, this is the maximum possible uncertainty that is only
achieved for stars with the largest extinction {\em and} nearly photospheric mid-IR fluxes.
For most of the stars, the uncertainty is several percent at most,
which is comparable to the measurement uncertainties of the mid-IR fluxes.
Specific checks were done for nine stars observed by the OT1\_pabraham\_2 program
(see Table~\ref{tab:list}). These include
the two most distant objects in our sample,
HD~95086 (A8 III, d=90.4 pc) and
HD 21997 (A2IV/V, d=71.9 pc).
Of these nine stars, reddening was only found for HD 21997.
In this case, we derived $A_V \approx 0.16$~mag,
$A(\text{I}_c) = 0.08$~mag and, since the observed $24\mum$
flux is 2.2 times the photospheric one, estimated the $24\mum$ excess flux to be
uncertain at 4\%.
In the far-IR where the excess fluxes exceed
the photospheric fluxes by one to three orders of magnitude, the effect
would be proportionally smaller and thus completely negligible.

\subsection{Photometry}

\subsubsection{Mid-IR photometry}

\looseness=1
For SED modeling, fluxes at wavelengths longward of 10~$\mum$ were used.
The mid-IR data included {\it WISE}/12 and /22 
(WISE All-Sky Release Catalog, \citeauthor{wright-et-al-2010} \citeyear{wright-et-al-2010}),
{\it AKARI}/18
(AKARI All-Sky Catalogue, \citeauthor{ishihara-et-al-2010} \citeyear{ishihara-et-al-2010}), 
MIPS/24 measurements
\citep[e.g.][]{su-et-al-2006,chen-et-al-2012},
as well as IRS data 
(e.g. Spitzer Heritage Archive, \citeauthor{chen-et-al-2014} \citeyear{chen-et-al-2014},
\revision{and the CASSIS archive,
\citeauthor{lebouteiller-et-al-2011} \citeyear{lebouteiller-et-al-2011}).
}
For some objects also 
Gemini/MICHELLE \citep[e.g.][]{churcher-et-al-2011},
MMT/BLINC \citep{stock-et-al-2010}, 
and Keck/MIRLIN data \citep{wahhaj-et-al-2007} were used.
The resulting photometry table is given in Appendix~A (Table~\ref{tab:disks_MIR_phot}).
Where available, the mid-IR fluxes were taken from the papers listed in
Table~\ref{tab:disks_MIR_phot}.
Where no flux values were given, we took them from the catalogues mentioned above,
accessed through Vizier and NASA/IPAC Infrared Science Archive.

\subsubsection{Far-IR and sub-mm photometry}

{\it Herschel}/PACS and SPIRE fluxes for all targets in the sample have been derived in the original papers,
cited in Table~\ref{tab:list}.
Since, however, different groups employed different reductions, we cross-checked those fluxes
by our own analysis, as described below.

In our own analysis,
the data were reduced with the {\it Herschel} Interactive 
Processing Environment \citep[HIPE,][]{ott-2010}, user release 10.0.0 and PACS calibration 
version 45.
Reduction was started from the level 1 data, which were obtained 
from the Herschel Science Archive (HSA). The level 1 (basic calibrated) products were processed 
using the standard pipeline 
script, with a pixel fraction of 1.0, pixel sizes of 1\arcsec~ at 70 and 100~$\mum$ and 2\arcsec~ at 
160~$\mum$. High 
pass filter widths of 15, 20 and 25 frames were adopted.
A region 20\arcsec\ in radius centered on
the expected star location, and other sources with a 
signal-to-noise $\ge$~5 (as measured by sextractor in the observation's
level 2 scan from the HSA) were masked from the high pass filtering process to avoid skewing the
background measurement.
Deglitching was performed using the second level deglitching task, as appropriate for bright 
sources. In the cases of 
targets that were observed with both 70/160 and 100/160 channel combinations, the four 160~$\mum$ 
scans were combined to produce the final mosaic.

The fluxes were measured using an IDL script based on the APER photometry 
routine from the IDL astronomy library\footnote{idlastro.gsfc.nasa.gov}. In the PACS 
images, the radius of the flux aperture varied 
depending on the extent of the disk, typically being between 15\arcsec--20\arcsec\ in radius.
Aperture corrections, as provided in 
Table 2 of \citet{balog-et-al-2013}, were applied.
For 15\arcsec\ aperture radius these are 0.829 at $70\mum$, 
0.818 at  $100\mum$, and 0.729 at $160\mum$. For 20\arcsec\ aperture radius the 
correction factors are 0.863, 0.847 and 0.800, respectively.
The sky background 
and r.m.s. variation were estimated from the mean and standard deviation of five square
apertures with sizes matched to the same area as the flux aperture at
each wavelength and for each target. The sky apertures were randomly  
scattered at distances of 30\arcsec--60\arcsec\ from the source peak  
(larger for Vega \revision{and HR~8799 that have very extended disks) to avoid}
the central region where the target disk lay 
and \revision{to elude} any identified background sources in the image.
\revision{A correction factor for the correlated noise was not used as it is not required if 
the apertures are sufficiently large (i.e. much bigger than a single native pixel, which they 
are in this case~--- 35\arcsec\ sky boxes versus 3.2\arcsec\ and 6.4\arcsec\ pixels at 
$100\mum$ and $160\mum$, respectively) and there 
are sufficient numbers of them (at least five). See \citet{eiroa-et-al-2013} for 
details of  noise measurement using multiple apertures.}
A calibration uncertainty of 5\%
was assumed for all three PACS bands.
In nearly all cases the calibration  
uncertainty dominates the total measured uncertainty of the target.
It is only for the faintest sources with a PACS $100\mum$ flux less than $\sim 150$~mJy
that the sky noise sometimes makes the larger contribution to the total uncertainty.

SPIRE fluxes were measured using an aperture 
with radius of 22\arcsec~ at 250~$\mum$, 30\arcsec~ at 350~$\mum$ and 42\arcsec~at 500~$\mum$.
The sky background and r.m.s. were estimated using a sky annulus between 60\arcsec--90\arcsec 
centered on the source position.
A calibration uncertainty of 7.5\%\footnote{Taken from SPIRE observer's manual, Sect. 5.2.12}
was assumed for the three SPIRE bands.

The results were found to agree reasonably well (within $\sim 10\%$) with the fluxes reported
in the literature.
We have preferred to use the values from literature sources in the construction
of our SEDs, as they represent the result of detailed individual analysis rather than the
more general, uniform approach taken in our own reduction.
The resulting photometry table \revision{and references are} given in Appendix~A (Table~\ref{tab:disks_FIR_phot}).
This table also lists the non-{\it Herschel} far-IR photometry,
namely {\it Spitzer}/MIPS data at $70\mum$,
and sub-mm fluxes from JCMT/SCUBA and APEX/LABOCA.

\subsection{PACS images}

Besides the photometry points, the PACS images also provide direct estimates of the disk radii.
This, too, has been done in the original papers listed in Table~\ref{tab:list}.
However, the heterogeneity of the procedures used in the literature to deduce the disk radii
is higher than those of the flux derivation.
Even the definition of the ``disk radius'' differs from one paper to another.
This has motivated us to obtain our own disk radii estimates with a uniform procedure for all the sources
in the sample.
To this end, we used the same PACS images as for the photometry analysis (see Sect. 2.3.2).

Note that the PACS observations of different targets were not all obtained in 
the same mode.
This is potentially important, since the observation parameters affect the shape of the
Point Spread Function (PSF)\footnote{See 
herschel.esac.esa.int/twiki/bin/view/Public/\\PacsCalibrationWeb\#PACS\_calibration\_and\_performance}, 
which is a key measure for determination of the disk radii.
For instance, the Guaranteed Time program maps were uniformly deep across the image, whereas all the others
(see Table~\ref{tab:list}) had lower coverage towards the edges and were all done 
in mini scan map 
mode. However, the scan speed, which would have the biggest effect on the PSF,
was the same in all programs.

\subsection{Extraction of disk radii}

The disk radii were derived from the resolved PACS images as follows.
First, we considered a grid of fiducial disks
at a distance of $d=20\pc$
with 11 dust orbital radii $\rdisc$ from 10--210~AU
and a dust annulus width of 10~AU
at two extreme inclinations of $0^\circ$ (face-on) and $90^\circ$ (edge-on).
The latter was needed to check by how much the long-axis \revision{extent} of the
disk image can be affected by the disk orientation if all other parameters are kept the same.
\revision{Convolved images were thereafter combined with a Gaussian noise component to examine 
the influence of the source S/N on determination of the disk extent in our grid of models.
The  magnitude of the noise covered a range of amplitudes spanning S/N values between 3--300 
for the  peak of the source (i.e. the $\sigma$ of the noise component was varied between 1/3 
to 1/300 of the source peak brightness).}
We then produced synthetic images of these disks 
at $100\mum$, fixing the total ring flux to a typical value of $0.5\Jy$ for all disk sizes.
The $100\mum$ wavelength was considered the most suitable of all three PACS wavelengths,
since it is a reasonable compromise 
between having the best angular resolution (which would be at $70\mum$)
and getting a possibility to detect
the largest (coldest) 
grains and thus trace the parent bodies
(which would be at $160\mum$).
No $100\mum$ image of Fomalhaut was taken and we therefore used the $70\mum$ image in our analysis.
Then, we convolved the ring models with
the PSF for the PACS/100 band from the calibrator star $\alpha$~Boo
and added stellar contributions at different $F_{\star}/F_{\rm dust}$ 
levels from 0 to 10.0.

The resulting source profiles were fitted with 2D Gaussian profiles to find 
the FWHMs of the convolved synthetic images in the long-axis direction. 
This resulted in a grid of fiducial disks that gives the FWHM as a function of
$\rdisc/d$, orientation (face-on or edge-on), S/N, and $F_{\star}/F_{\rm dust}$.

Second, the $100\mum$-resolved {\em observed} image of each disk in our sample was fitted by a 
Gaussian profile.
The ratio of the long-axis to short-axis 
FWHM of that profile was used to roughly judge whether the disk is closer to edge-on or to face-on.
Then, the long-axis FWHM of the disk from the sample was compared with the long-axis FWHMs of those 
fiducial disks in the grid that have the selected orientation and
have the stellar flux to dust flux ratio closest to the one determined from the SED.
The radius of the grid disk that provides the best match
was then taken as the ``true'' (physical) disk radius.

Given the relatively low spatial resolution of Herschel, disks are often unresolved in 
the minor axis direction. Indeed, 13 out of 34 disks have minor-axis FWHM$<1.2$PSF.
The concern is that some disks that are actually closer to edge-on than face-on could 
be wrongly designated as face-on. 
However, we found the effect of the orientation to be only minor.
The true disk radii
retrieved from the measured disk extent under face-on and edge-on assumptions differ by
$\la 5\%$.

Also, S/N has little effect on the measured extent.
At S/N $> 9$, which is the case for the $100\mum$ flux of all our targets,
the measured extent for a given true disk radius is uncertain at $\la 2$\%.
Conversely, the uncertainty of the true disk radius determined from the measured FWHM 
was found to be below 1\%.

The $F_{\star}/F_{\rm dust}$ ratio does not influence the results either,
as long as it is below unity.
\revision{
For values of $F_{\star}/F_{\rm dust} \le 0.1$, which is the case for all targets in our 
sample, the addition of a point source scaled to the expected stellar photosphere flux to the 
center of the disk model changes the measured disk radius of the final convolved model 
by $\la 2$\% compared to a convolved model without a stellar component included.} 
Combining the $\la 4$\% uncertainty with which the long-axis FWHM at $100\mum$ is measured
\citep{kennedy-et-al-2012b} with the $\la 5$\% uncertainty induced by the orientation,
the $\la 1$\% S/N-related uncertainty, and the $\la 2$\% star-related uncertainty, 
we conservatively estimate that the true disk radii
are accurate to $\la 7$\%.

Some of the disks in our sample have been previously resolved in scattered light.
\revision{Although the scattered-light images trace much smaller grains than the thermal emission 
observations do, such small grains should be most abundant in the parent planetesimal belt. Thus
the} scattered-light images, in principle, could be used to retrieve the disk radii
with a better accuracy than from the PACS images.
However, for the sake of homogeneity, we decided to use radii from the PACS images
in all the cases.

\section{SED fitting procedure}

\subsection{Basics of the fitting}
~ 
We included in the fitting all available photometry points from the mid-IR through 
(sub-)mm where excess is seen, as described in Sect.~2.3.
We modeled the SEDs with the SEDUCE code \citep{mueller-et-al-2009}, complemented with a fitting
tool using the ``simulated annealing''algorithm \citep{numerical-recipes-2002}.

Although
the focus of this study is on the main, cold, Kuiper belt-type, component of 
the disks, an additional warm component is known or at least suspected to be present in some 
systems.
That warm component may contribute to the fluxes, affecting the parameters of the cold one.
Accordingly, we perform a two-component fitting, as was commonly done in previous studies
\citep[e.g.][among others]{morales-et-al-2009,morales-et-al-2011,%
lebreton-et-al-2012,donaldson-et-al-2013,roberge-et-al-2013,ballering-et-al-2013,chen-et-al-2014}.
We assume that all the dust in each of the two components is confined to a narrow annulus
at a radius $\rdisc$ or $\rwarm$ from the star.
For the warm component, this assumption is sufficient, given large 
uncertainties of the fluxes and the photospheric subtraction in the mid-IR.
For the cold one, it is justified by the fact that most of the emitting
dust is located in the region of the underlying planetesimal belt, which
is expected to be relatively narrow \citep[see, e.g.,][]{kennedy-wyatt-2010}.
Potential limitations of this assumption,
including corrections to the results 
arising from a possible finite extent of
the ``main'' (cold) dust disk, are discussed in
Sect.~5.3.

\change{The former subsection 3.2 split into two subsections}

\subsection{Modified blackbody method}

The warm component is fitted by a pure blackbody.
For the cold one, we employ
two methods of the SED fitting, both of which are commonly used.
A simpler one is to use the modified blackbody \citep[MBB,][]{backman-paresce-1993}.
In this approach, the disk material is assumed to emit the specific intensity 
\revision{proportional to}
\be
 B_\nu (\lambda, \td)
 \times
 \left[
 H(\lambda_0 - \lambda)+ H(\lambda - \lambda_0)(\lambda / \lambda_0)^{-\beta}
 \right] ,
 \label{mod_BB}
\ee
where $B_\nu (\lambda, \td)$ is the Planck function,
$H$ is the Heaviside step function equal to unity for non-negative arguments and to zero 
for negative ones,
$\lambda_0$ is the characteristic wavelength,
and $\beta$ is the opacity index.
The temperature $\td$ in  Eq.~(\ref{mod_BB}) plays the same role as effective
temperature in the pure blackbody description and thus
can be called {\em modified effective temperature}.

The MBB emission law (\ref{mod_BB}) reflects the fact that dust grains
are inefficient emitters at wavelengths much longer than the grain size and so,
$s = \lambda_0/(2\pi)$ can be considered as the characteristic grain radius in the disk.
Thus in the MBB method the disk can be thought of as composed of like-sized grains
of size $s$, which have the absorption efficiency
\be
  \Qabs(\lambda, s)  =
  \left\{
  \begin{array}{l l}
    1,                              & \qquad \lambda \le \lambda_0\\
    (\lambda / \lambda_0)^{-\beta}, & \qquad \lambda >   \lambda_0\\
  \end{array} \right. ,
\label{Qabs_MBB}
\ee
where $\lambda_0 \equiv 2 \pi s$.

The particular relation between $\lambda_0$ and $s$ needs to be explained.
\citet{backman-paresce-1993} note that
$\lambda_0 = 2 \pi s$ and $1/(2\pi) s$  for 
strongly and weakly absorbing materials, respectively, while $\lambda_0 = s$ for moderately absorbing 
dielectrics.
We adopt $\lambda_0 = 2\pi s$, because for micron-sized and larger grains it matches the best
$\Qabs$ of the astrosilicate \citep[see, e.g., Figure 2 in][]{krivov-et-al-2008}. 
Since astrosilicate is used in another SED fitting method,
described below, this ensures consistency of the results obtained with the two methods.

\revision{Denoting by $N$ the total number of grains of size $s$ in the disk, the
dust specific luminosity is given by
\bea
 L_\nu^\mathrm{MBB} &=&
 N
 \times 4 \pi s^2
 \nonumber\\
 &\times&
 \Qabs(\lambda, s)
 \times \pi 
 B_\nu (\lambda, \td (s)) .
 \label{L_MBB}
\eea
The flux density measured at Earth is simply
$F_\nu^\mathrm{MBB} = L_\nu^\mathrm{MBB}/(4\pi d^2)$.
}

The above description of the MBB method is complete,
as long as the disk under study is unresolved and only an SED is available for analysis.
However,
in the case where the disk radius $\rdisc$ is known from the resolved image,
one can also calculate
the {\em physical temperature} of grains with the radius
$s = \lambda_0/(2\pi)$ at a distance $\rdisc$ from the star.
This can be done by solving the thermal balance equation of grains with
absorption efficiency (\ref{Qabs_MBB}) in the stellar radiation field.
A question arises, whether 
the physical temperature of these grains is equal to the
modified effective temperature $\td$ found from fitting the SED by
Eq.~(\ref{mod_BB}).
It is generally not, unless this requirement is incorporated in the fitting routine!
The fitting algorithm implemented in SEDUCE
makes sure it is, thus providing a
self-consistent solution.
This implies that the MBB results obtained here may differ from those obtained
by other authors with the MBB method.

\subsection{Size distribution method}

Another method is to assume that grains in a disk have
a size distribution (SD). It is commonly approximated by a power law
\be
  N(s) \propto s^{-q} \qquad (s \ge \smin) ,
\label{n(s)}
\ee
where
\revision{$N(s)ds$ is the number of grains with radii from $s$ to $s+ds$},
$\smin$ is a certain cutoff size,
and $q$ is a size distribution index.
The latter is equal to $3.5$ for an ideal infinite collisional cascade with a 
size-independent critical fragmentation  energy of solids \citep{dohnanyi-1969}
and is expected to lie approximately in the range $3 \la q \la 4$
under more realistic assumptions 
\citep[e.g.][]{krivov-et-al-2006,thebault-augereau-2007,loehne-et-al-2007,wyatt-et-al-2011}.
\revision{The dust specific luminosity is now computed as
\bea
 L_\nu^\mathrm{SD} &=&
 \int_\mathrm{s_\mathrm{min}}^\mathrm{s_\mathrm{max}}
 N(s) ds
 \times 4 \pi s^2
 \nonumber\\
 &\times&
 \Qabs(\lambda, s)
 \times \pi 
 B_\nu (\lambda, \td (s)) ,
 \label{L_SD}
\eea
and the flux density measured at Earth is again
$F_\nu^\mathrm{SD} = L_\nu^\mathrm{SD}/(4\pi d^2)$.
}

\revision{The absorption efficiency $\Qabs(\lambda, s)$ needed
to compute the thermal emission of dust}
is usually calculated with the Mie theory, with
a set of optical constants as input.
These constants have to be generated with one or another method which
requires, in turn, assumptions about the material composition and possibly, also 
porosity of the dust grains.
We assume compact astrosilicate particles \citep{draine-2003}
with a bulk density $\rho = 3.3\g\cm^{-3}$.

For comparison with the MBB method and with the other studies, it is reasonable
to define  the {\em dust temperature} in the SD method, too.
This is not straightforward, since in this case the disk is comprised of dust grains
that all have different sizes and thus different equilibrium temperatures in the stellar 
radiation field.
For instance, one can define $\td$ as the average physical temperature of 
different-sized grains weighted by their contribution to the disk's total
cross section.
In this paper, we use another way, simply measuring the wavelength 
$\lambda_\mathrm{max}$ where the modeled SED peaks and applying the Wien
displacement law to define $\td \equiv 5100\K (\mum/\lambda_\mathrm{max})$.

\begin{table*}[htb!]
  \begin{center}
  \tabcolsep 2pt
  \caption{\change{Some values slightly changed}Identification of the Warm Component}
  \scriptsize    
    \begin{tabular}{rc|rrrrrrc|rrrrrrc}
    \toprule
    
    \multirow{3}[10]{*}{HD} & \multirow{3}[10]{*}{$R_\text{sub}$} & \multicolumn{7}{c}{MBB} & \multicolumn{7}{c}{SD} \\
     \toprule
           &       & \multicolumn{4}{c}{Excess}            & \multirow{2}[0]{*}{$\chi^2_\text{one}/\chi^2_\text{two}$} & \multirow{2}[0]{*}{$R_\text{warm}$} & \multirow{2}[0]{*}{Warm} 
           & \multicolumn{4}{c}{Excess}            & \multirow{2}[0]{*}{$\chi^2_\text{one}/\chi^2_\text{two}$} & \multirow{2}[0]{*}{$R_\text{warm}$} & \multirow{2}[0]{*}{Warm} \\
           &       & IRS22 & WISE22 & MIPS24 & IRS31 &       &       &component?       & IRS22 & WISE22 & MIPS24 & IRS31 &       &       &component?  \\
           & [AU]  & [$\sigma$] & [$\sigma$] & [$\sigma$] & [$\sigma$] &       & [AU]      &       & [$\sigma$] & [$\sigma$] &[$\sigma$] & [$\sigma$] &       & [AU]       &  \\
       \toprule    
    GJ 581 & 0.01  & -0.2  & 0.4   & -     & 0.1   & -     & -     & N     & 0.4   & 0.5   & -     & -0.3  & -     & -     & N \\
    197481 & 0.01  & -     & 1.0   & 6.0   & -     & 7.8   & 2.3   & Y     & -     & 0.9   & 2.5   & -     & -     & -     & N \\
    23484 & 0.03  & -0.1  & -0.5  & 0.2   & 3.2   & 3.3   & 4.8   & Y     & 0.1   & 0.3   & 1.2   & 2.8   & -     & -     & N \\
    104860 & 0.05  & 0.9   & 1.9   & 1.1   & 0.4   & -     & -     & N     & 0.7   & 0.8   & 1.0   & 0.5   & -     & -     & N \\
    207129 & 0.05  & 2.3   & 2.1   & -1.2  & 0.7   & -     & -     & N     & 3.0   & 2.7   & -1.3  & 2.9   & -     & -     & N \\
    10647 & 0.06  & -     & 9.2   & 6.7   & 3.3   & 4.0   & 4.8   & Y     & -     & 8.9   & 4.7   & -1.0  & 3.6   & 5.5   & Y \\
    48682 & 0.06  & 2.6   & 4.0   & -0.1  & 4.6   & 1.3   & 8.9   & N     & 3.3   & 4.7   & -0.5  & 4.5   & 1.7   & 9.0   & N \\
    50571 & 0.08  & -0.1  & 1.3   & -0.1  & -1.0  & -     & -     & N     & -0.6  & -0.1  & -0.2  & -0.2  & -     & -     & N \\
    170773 & 0.08  & 1.3   & 2.5   & 1.9   & 0.8   & -   & -      & N     & 1.5   & 3.0   & 1.7   & 2.4   & -     & -     & N \\
    218396 & 0.10  & 12.7  & 12.0  & 19.9  & 10.6  & 14.7  & 2.3   & Y     & 13.1  & 12.4  & 19.9  & 13.9  & 17.7  & 2.5   & Y \\
    109085 & 0.10  & -     & 7.6   & 13.6  & -     & 23.4  & 2.2   & Y     & -     & 7.7   & 13.2  & -     & 37.9  & 2.7   & Y \\
    27290 & 0.11  & 2.7   & 3.9   & 6.3   & 0.1   & 5.5   & 0.1   & N     & 2.8   & 4.2   & 6.2   & -0.7  & 4.5   & 0.1   & N \\
    95086 & 0.12  & -     & 9.3   & 10.8  & 2.8   & 10.1  & 3.3   & Y     & -     & 5.6   & 0.7   & -4.8  & 5.3   & 6.7   & Y \\
    195627 & 0.12  & 6.5   & 7.4   & 1.6   & 5.3   & 8.8   & 1.1   & Y     & 6.9   & 7.0   & 1.2   & 4.1   & 8.1   & 1.3   & Y \\
    20320 & 0.15  & 3.6   & -     & 4.4   & 3.3   & 10.4  & 2.3   & Y     & 2.7   & -     & 4.7   & 8.2   & 45.3  & 5.2   & Y \\
    21997 & 0.15  & -     & 7.1   & 12.1  & 4.9   & 19.2  & 9.6   & Y     & -     & 6.3   & 9.4   & 1.0   & 6.7   & 8.1   & Y \\
    110411 & 0.16  & 6.4   & -     & 8.2   & 0.0   & 5.0   & 2.0   & Y     & 6.3   & -     & 7.4   & 0.6   & 3.1   & 1.7   & Y \\
    142091 & 0.16  & -     & -1.8  & 2.4   & -     & -     & -     & N     & -     & -1.7  & -0.6  & -     & -     & -     & N \\
    102647 & 0.17  & 8.5   & -     & 11.3  & -     & 42.6  & 0.7   & Y     & 8.2   & -     & 12.3  & -     & 465.9 & 4.2   & Y \\
    125162 & 0.18  & 16.2  & -     & 35.1  & 18.7  & 68.4  & 8.3   & Y     & 15.6  & -     & 30.0  & 12.6  & 51.2  & 9.3   & Y \\
    216956* & 0.18  & 2.0   & -     & 3.7   & -     & 1.1   & 57.5  & N     & 3.2   & -     & 3.8   & -     & 1.6   & 52.8  & N \\
    17848 & 0.18  & 6.4   & 9.4   & 7.4   & 6.1   & 10.2  & 2.2   & Y     & 5.4   & 7.3   & 7.3   & 6.8   & 9.9   & 2.2   & Y \\
    9672  & 0.18  & -     & 6.3   & 14.3  & -     & 40.2  & 11.9  & Y     & -     & 3.1   & 3.1   & -     & 3.5   & 13.1  & Y \\
    71722 & 0.20  & 15.3  & 12.7  & 16.9  & 21.5  & 29.6  & 2.5   & Y     & 15.1  & 12.0  & 17.4  & 22.8  & 30.1  & 9.5   & Y \\
    182681 & 0.23  & -     & 21.9  & -     & -     & 30.2  & 0.5   & Y     & -     & 25.9  & -     & -     & 113.0 & 2.3   & Y \\
    14055 & 0.23  & 17.1  & -     & 15.6  & 26.7  & 47.9  & 24.3  & Y     & 17.7  & -     & 15.6  & 27.9  & 50.0  & 23.1  & Y \\
    161868 & 0.23  & 8.7   & 13.0  & 6.5   & -     & 9.9   & 1.6   & Y     & 12.7  & 17.9  & 14.7  & -     & 22.1  & 1.1   & Y \\
    188228 & 0.24  & 1.8   & -     & 1.0   & 0.7   & -     & -     & N     & -0.6  & -     & -59.9 & -1.2  & -     & -     & N \\
    10939 & 0.26  & 5.9   & 12.2  & 13.0  & 9.1   & 19.1  & 20.1  & Y     & 6.4   & 13.0  & 13.5  & 11.0  & 16.8  & 26.0  & Y \\
    71155 & 0.27  & 1.2   & -     & -4.4  & -6.4  & 22.1  & 11.9  & Y     & 9.0   & -     & 18.9  & 2.4   & 31.8  & 12.7  & Y \\
    172167* & 0.33  & -     & -     & 10.3  & -     & 1.8   & 0.1   & N     & -     & -     & 12.5  & -     & 2.3   & 3.9   & N \\
    139006 & 0.35  & 2.7   & -     & 6.3   & 7.5   & 11.0  & 13.4  & Y     & 3.3   & -     & 5.4   & -3.2  & 4.8   & 16.3  & Y \\
    95418 & 0.35  & -3.1  & -     & 0.2   & -1.6  & -     & -     & N     & -1.5  & -     & 2.7   & -0.1  & -     & -     & N \\
    13161 & 0.39  & 7.6   & -     & 8.7   & 7.3   & 13.9  & 30.2  & Y     & 8.5   & -     & 8.9   & 8.3   & 15.9  & 39.9  & Y \\
    \bottomrule
    \end{tabular}%
  \label{tab:warm}%
\end{center}

\noindent

{\em Notes:}\\
(1) The ``excess'' columns give the observed flux minus the cold-component flux minus
the stellar photospheric flux, in the units of $\sigma$.\\
(2) The objects without significant mid-IR excesses were not fitted with a two-component model.\\
For such objects, no $\chi^2_\text{one}/\chi^2_\text{two}$ and $R_\text{warm}$ values are given.\\
\revision{(*)A more detailed analysis that fully includes the IRS spectra
and identification criteria other than ours, shows that these objects very likely possess
a warm component \citep{lebreton-et-al-2013,su-et-al-2013}.
See also note (4) at Table~\ref{tab:disks_MIR_phot}.
}
\end{table*}
\subsection{Fitting parameters}

To avoid potential degeneracies and keep the procedure simple and robust,
we only include three free, independent fitting parameters for the cold component.
One of them is always the amount of dust that determines the height of the SED.
The additional two in the MBB treatment are $\lambda_0$ and $\beta$, see 
Eq.~(\ref{mod_BB}).
In the SD treatment, these are $\smin$ and $q$, see Eq.~(\ref{n(s)}).
In this case, the maximum grain size is fixed to a reasonably large value $\smax = 1\mm$;
the influence of this parameter on the fitting results is negligible.
Including the warm component  introduces two more parameters.
One is again the amount of dust, and another the radius of the warm disk $\rwarm$.
Altogether there are three free parameters for the one- and five for the two-component fit. 
The quality of the fit is characterized by

\be
 \chi^2_{\text{red}} = \frac{1}{\nu_{\text{free}}} \cdot \chi^2 = 
 \frac{1}{\nu_{\text{free}}} \cdot \sum\limits_{i = 0}^{N_{\text{data}} - 1}{\left(\frac{F_{\text{i}}^{\text{obs}} - F_{\text{i}}^{\text{model}}}{\sigma_{\text{i}}^{\text{obs}}}\right)^2}
\ee
where $\nu_{\text{free}} = N_{\text{data}} - N_{\text{parameter}}-1$
is the number of degrees of freedom.
Therefore, the model is underdetermined if the number of data points is less than five
(one-component fit) or seven (two-component fit).
This was indeed the case \revision{for HD 142091 ($\kappa$ CrB), having only 5 data 
points and thus allowing only a one-component fit to be done.
However, this fit suggests that the SED of that object does} not exhibit a warm 
excess (see Table~\ref{tab:warm}), so that the two-component fitting is not necessary.

\subsection{Iterative two-component fitting}

The fitting procedure consisted of several steps, the same for both the MBB and the SD 
method. 
At first, we made a one-component fit to determine the first-guess parameters of the cold component of 
the system, using the far-IR data only.
Since mid-IR data can be contaminated by a warm component, we discarded them at this stage.
During the second step,
the model fluxes of the cold component fit were compared with the available mid-IR data.
If there was a significant ($\ge 3\sigma$) excess at least in one of the mid-IR bands,
we added the mid-IR data and performed a two-component fit with both mid- and far-IR 
photometry to find the first-guess warm component.
In this fitting, we allowed the amount of cold dust to vary, but kept
the other two parameters of the cold component unchanged.
In the third step, we fitted the cold component again,
this time \revision{allowing the amount of warm dust to vary and keeping the radius}
of the warm component fixed. Steps 2 and 3 were repeated several times until 
the parameters of both components stopped to change.

Following \citet{ballering-et-al-2013},
we used three criteria to decide whether the object has a warm component or not.
We required that:

\medskip
\begin{enumerate}
\item[(1)]
a significant ($\ge 3\sigma$) excess in any of the IRS/22, {\it WISE}/22, MIPS/24 or IRS/31 
bands is present.
\item[(2)]
The quality of the two-component fit is much better 
than that of the one--component fit:
$\chi^2_\text{one}/\chi^2_\text{two} > 3$.
\\[-3mm]
\item[(3)]
the inferred warm dust is located outside the sublimation radius
(assuming $1300\K$ as the sublimation temperature).
\end{enumerate}

The warm component was considered real when all three criteria were met.
If one or more of them were not satisfied, the two-component fit was discarded, and
we performed a one-component fit with the mid-IR and far-IR data.


\bigskip
\section{Results}

\subsection{Systems with one and two components}

Table~\ref{tab:warm} shows that about two-thirds of the systems
(\revision{22}/34 in the MBB method and \revision{20}/34 in the SD one) reveal the second, warm 
component.
This fraction is in accord with previous studies \citep[e.g.,][]{chen-et-al-2014}.
No correlation \revision{between the presence of the warm component and the spectral type of 
the central star} is evident.
Both later- and early-type stars possess one-component disks in some cases
and two-component ones in the others.

Criteria (1) and (2) turn out to be more restrictive than (3).
There are only a few cases where (1) and (2) suggest the presence of a warm component, yet
the inferred radius of the warm dust is smaller than the sublimation distance.
\revision{This happens if the warm excess is small,
so that the parameters of a presumed warm component are very uncertain.}

\begin{figure*}[htb!]
   \centering
   \includegraphics[width=0.48\textwidth, angle=0]{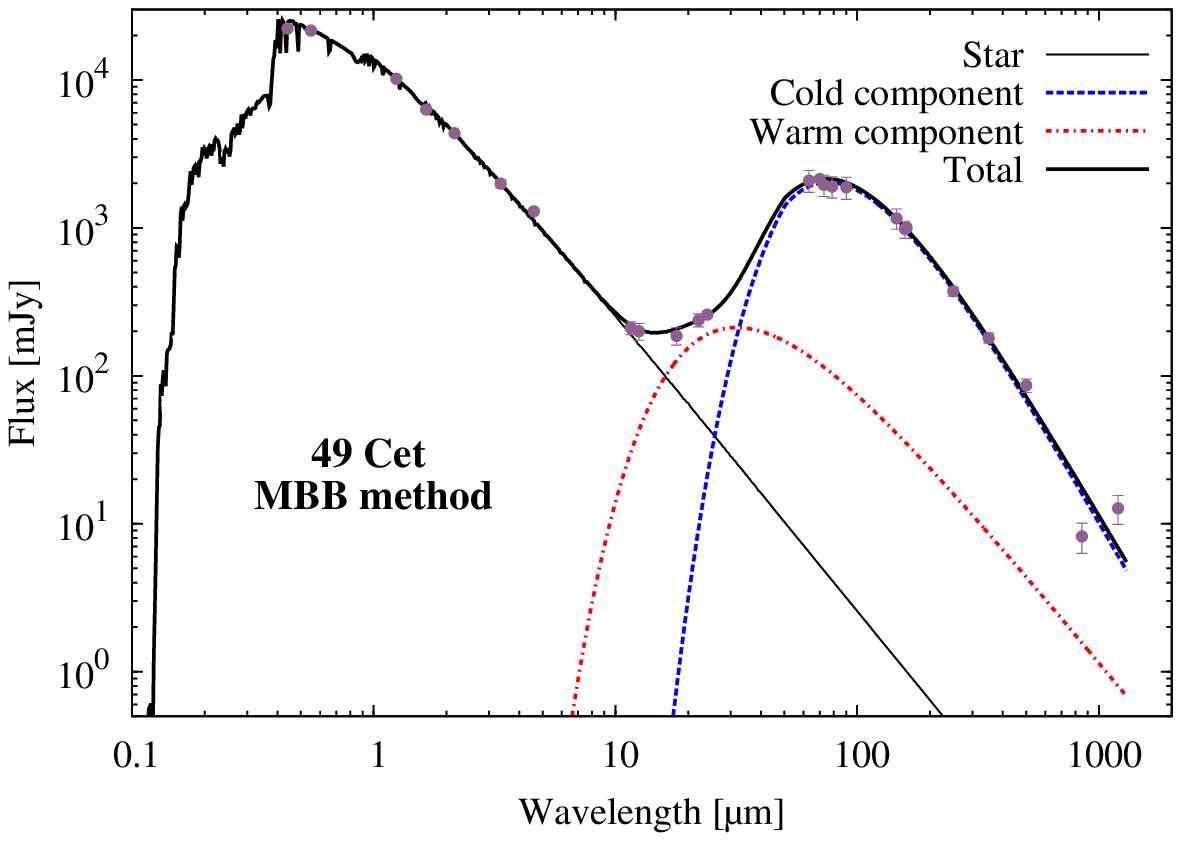}
   \includegraphics[width=0.48\textwidth, angle=0]{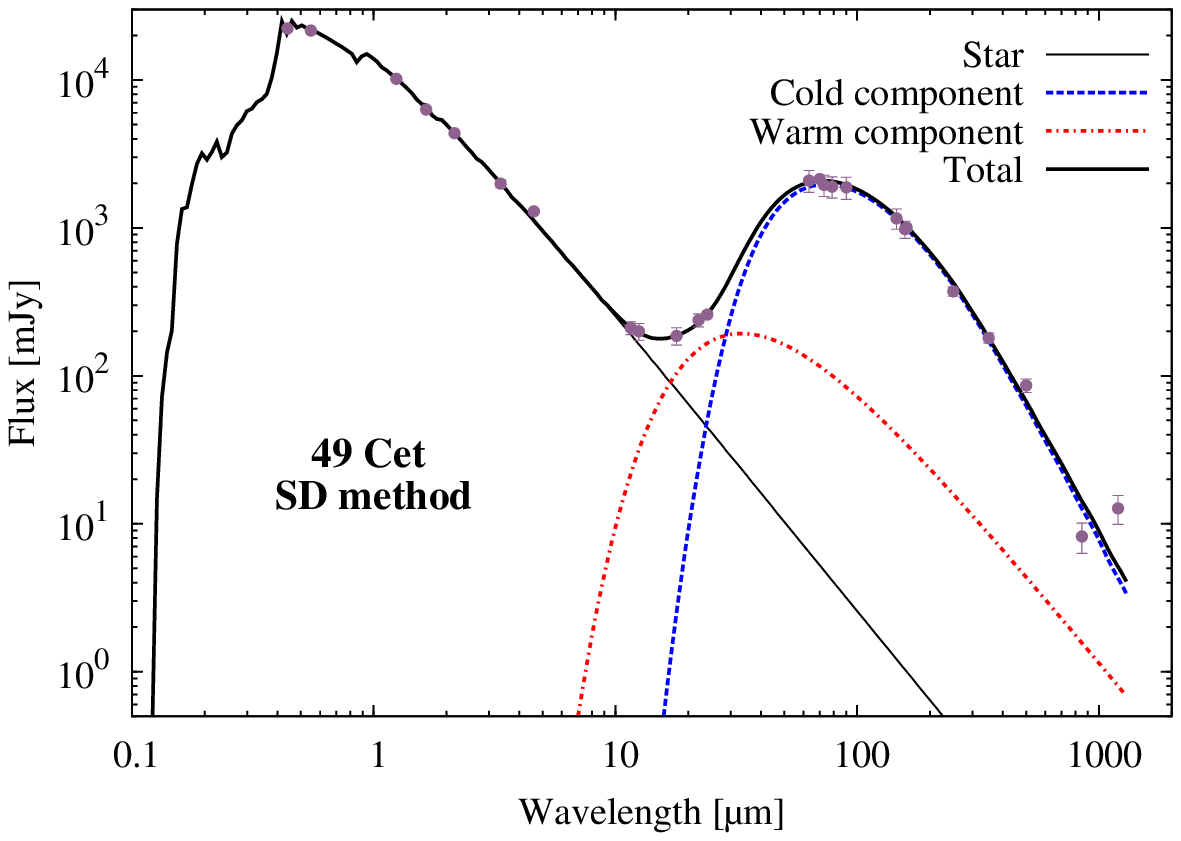}\\
   \includegraphics[width=0.48\textwidth, angle=0]{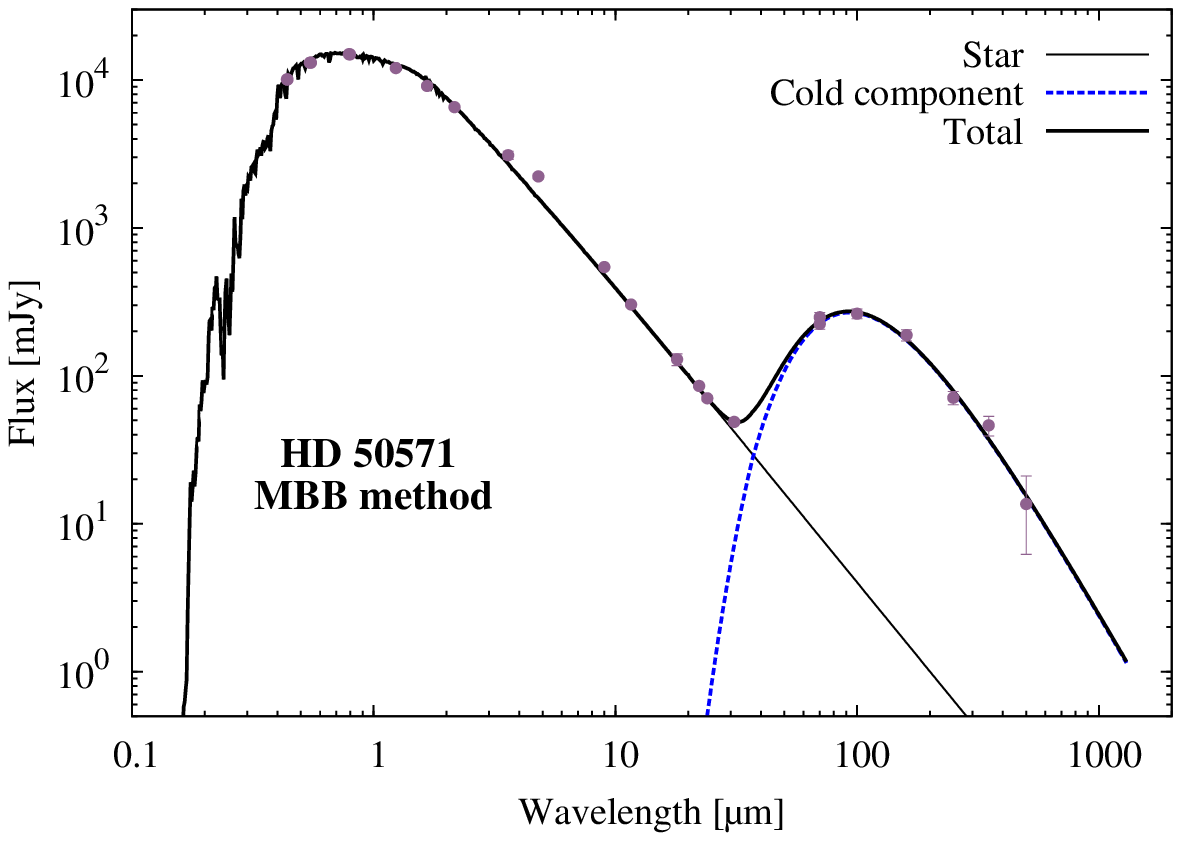}
   \includegraphics[width=0.48\textwidth, angle=0]{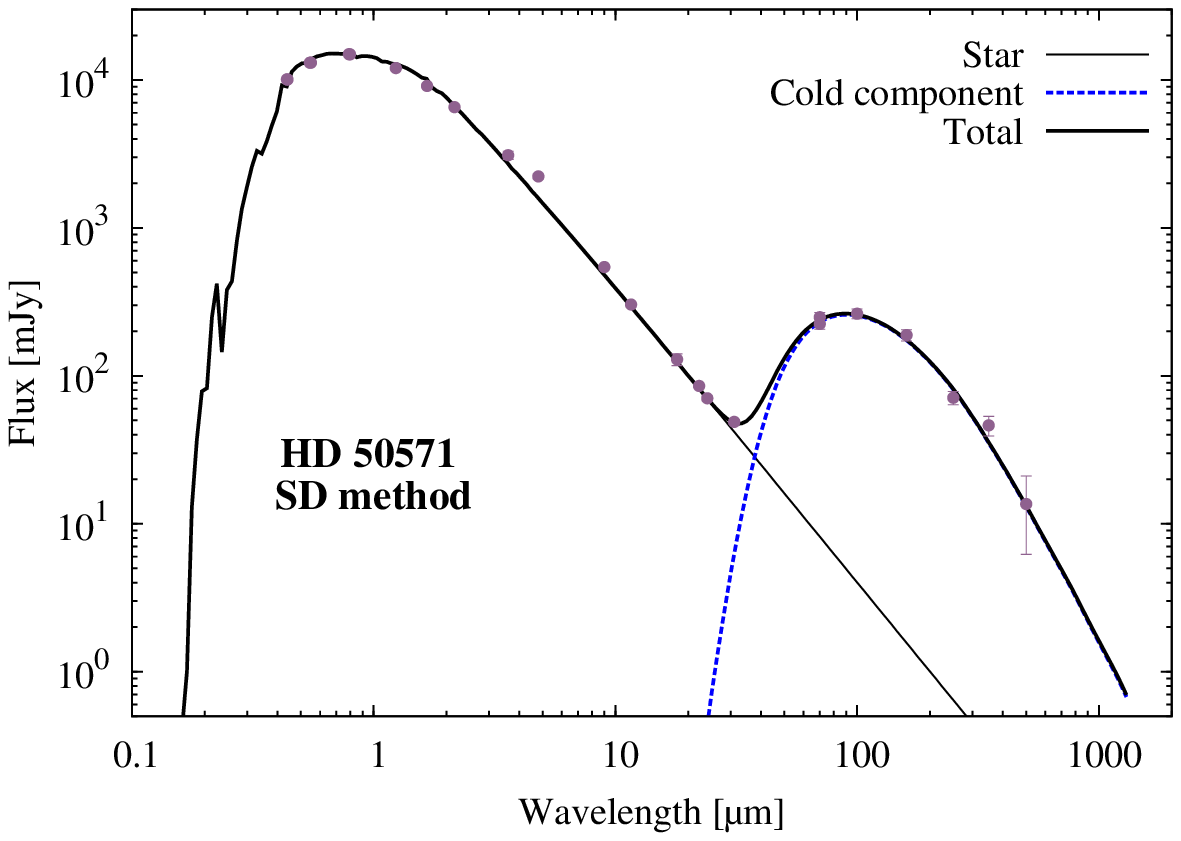}\\
   \includegraphics[width=0.48\textwidth, angle=0]{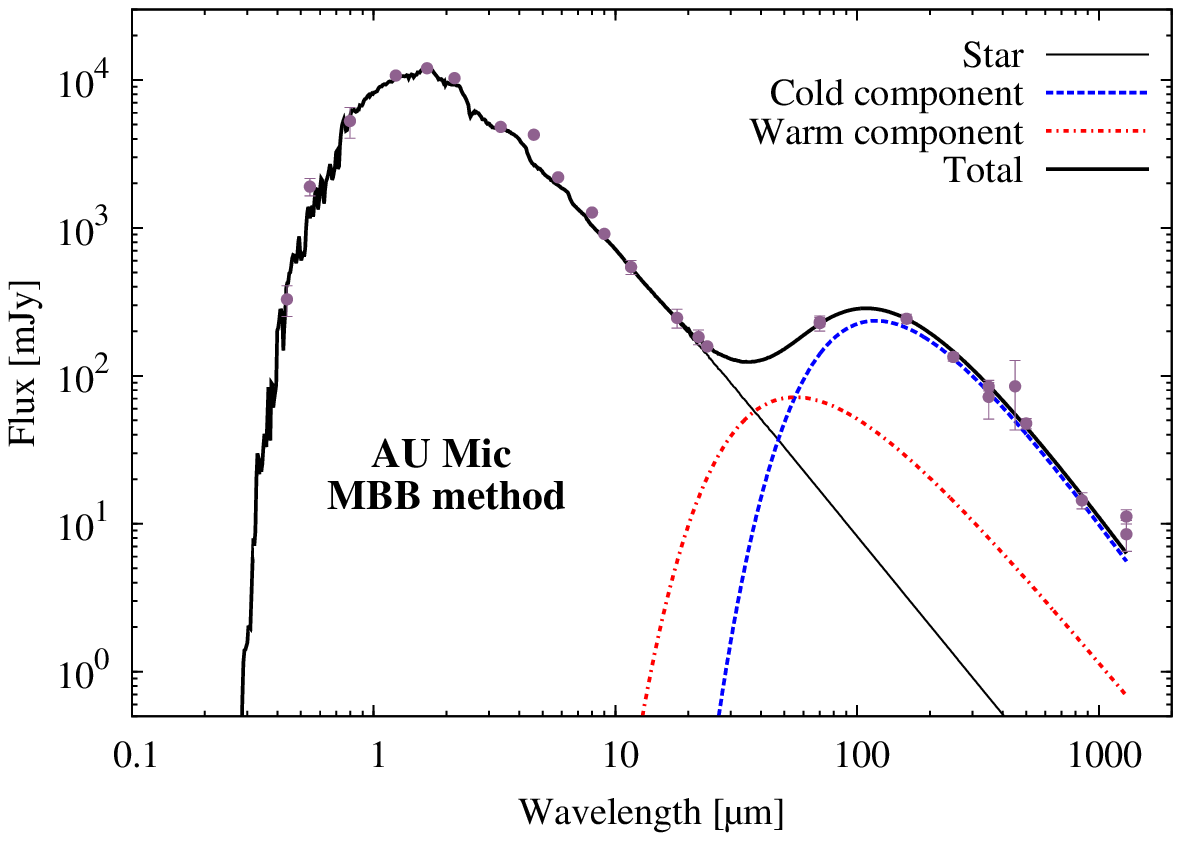}
   \includegraphics[width=0.48\textwidth, angle=0]{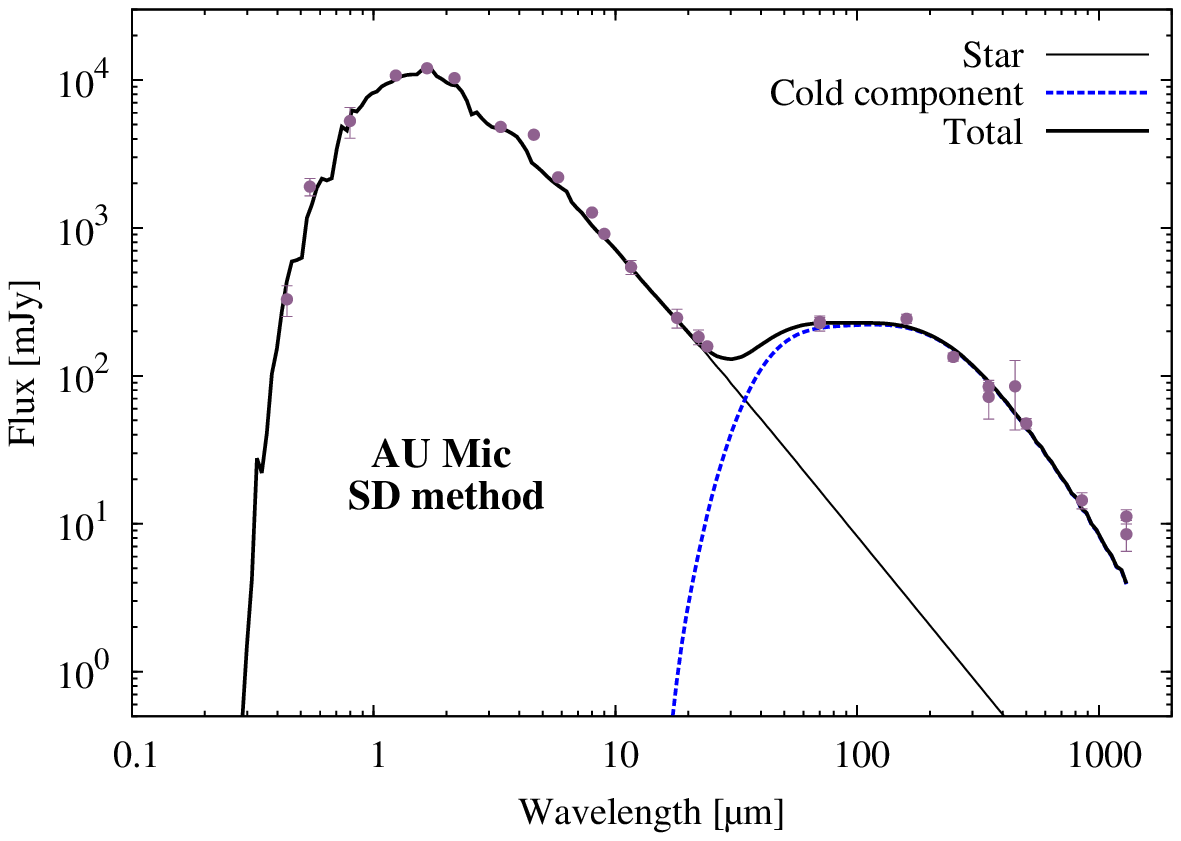}
   \caption{The SEDs for 49~Cet (top), HD~50571 (middle), and AU~Mic (bottom)
   fitted by the MBB method (left) and the SD method (right).
   Symbols with error bars are measured fluxes and their uncertainties.
   Lines are explained in the legend.
   \label{fig:SEDs}
   }
   \end{figure*}

In most of the cases, the results obtained with the two methods are similar,
albeit not identical.
The top panels in Figure~\ref{fig:SEDs} show the SEDs of 49~Cet,
a typical system with two components, fitted with both methods.
The middle panels do the same for HD~50571, which is a typical one-component system.
There are a few cases, however, where one of the methods suggests a
warm component whereas another one does not.
An example is AU~Mic. 
Bottom panels in Figure~\ref{fig:SEDs} use AU~Mic as an example
and demonstrate that the warm component may (MBB method) or may not be present (SD method).
This is because the exact shape of the cold-component SEDs, on top of which we are seeking warm excesses,
is slightly different in these two methods.

\begin{figure}[tbh!]
   \centering
   \includegraphics[width=0.45\textwidth,angle=0]{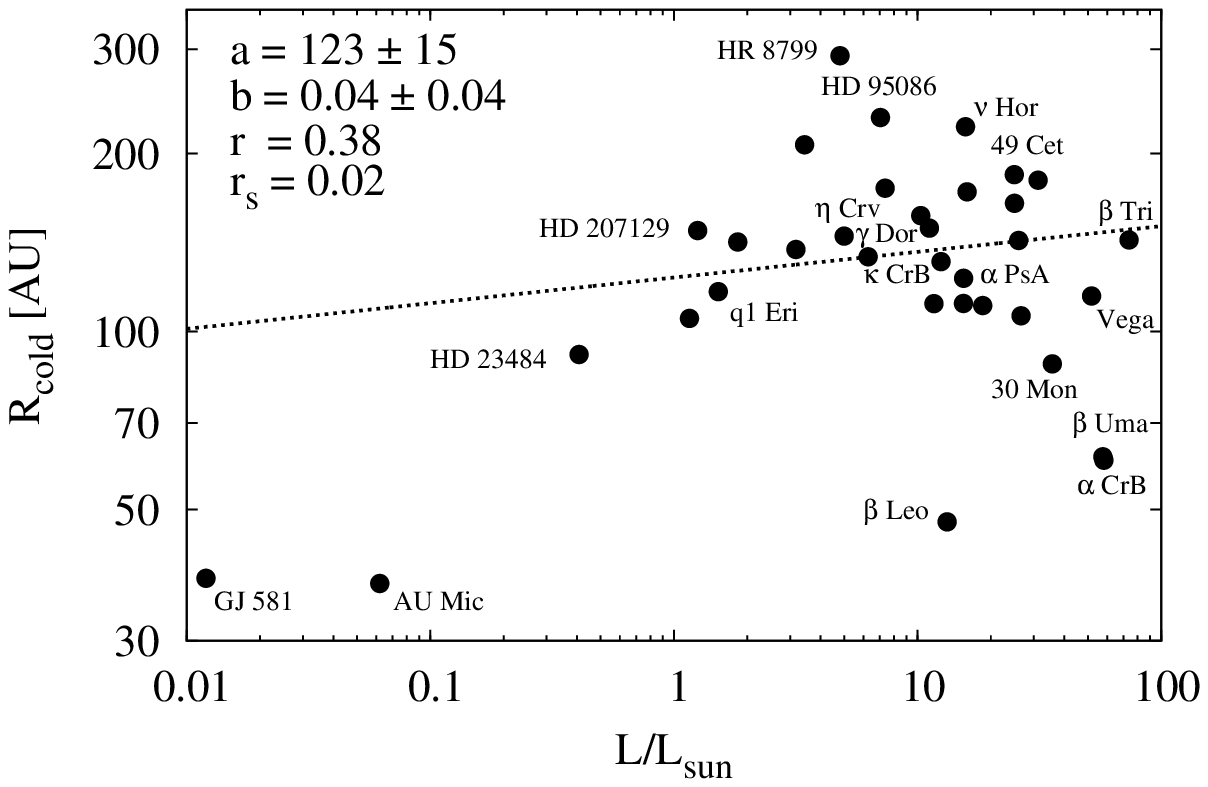}
   \caption{Disk radii versus stellar luminosities.
   The straight line is the best linear fit to the log-log data,
   equivalent to a power law fit of the form $\rdisc = a (L/L_\odot)^b$.
   The best-fit parameters $a$ and $b$  together with
   the Pearson's $r$ and Spearman's $r_s$ are indicated in the panel.
   \label{fig:rdisk}
   }
   \end{figure}

A detailed study of the incidence and properties of the warm component is deferred to a
subsequent paper. Here we concentrate on the properties of the main, cold component and thus
use the knowledge of the warm component to \revision{exclude} it from the SED, as 
described above.
The disk radii and the fitting results for the cold component
obtained both with the MBB and SD methods 
are listed in Tables~\ref{tab:fitting_MBB}--\ref{tab:fitting_SD} and plotted in
Figures~\ref{fig:rdisk}--\ref{fig:Td_smin}.
These are discussed in the following subsections.

\subsection{Disk radii}

   \begin{figure*}[tbh!]
   \centering
   \includegraphics[width=0.45\textwidth,angle=0]{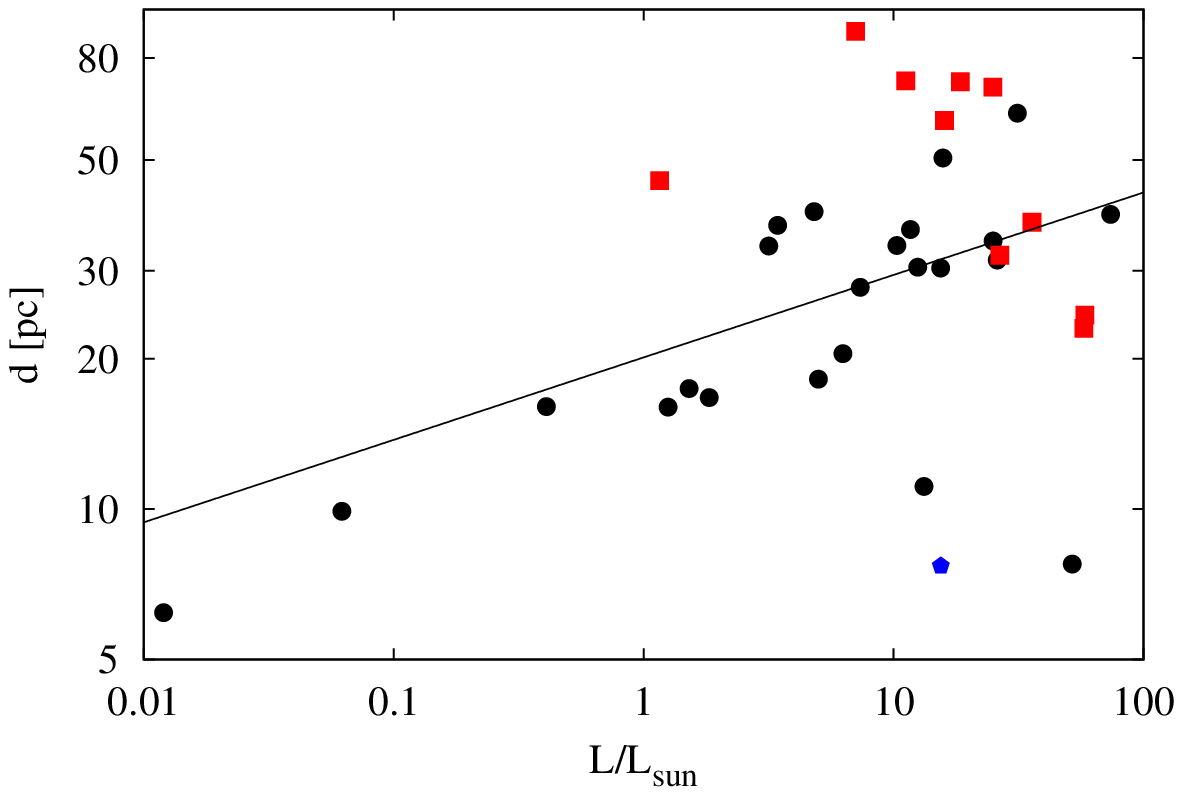}
   \includegraphics[width=0.45\textwidth,angle=0]{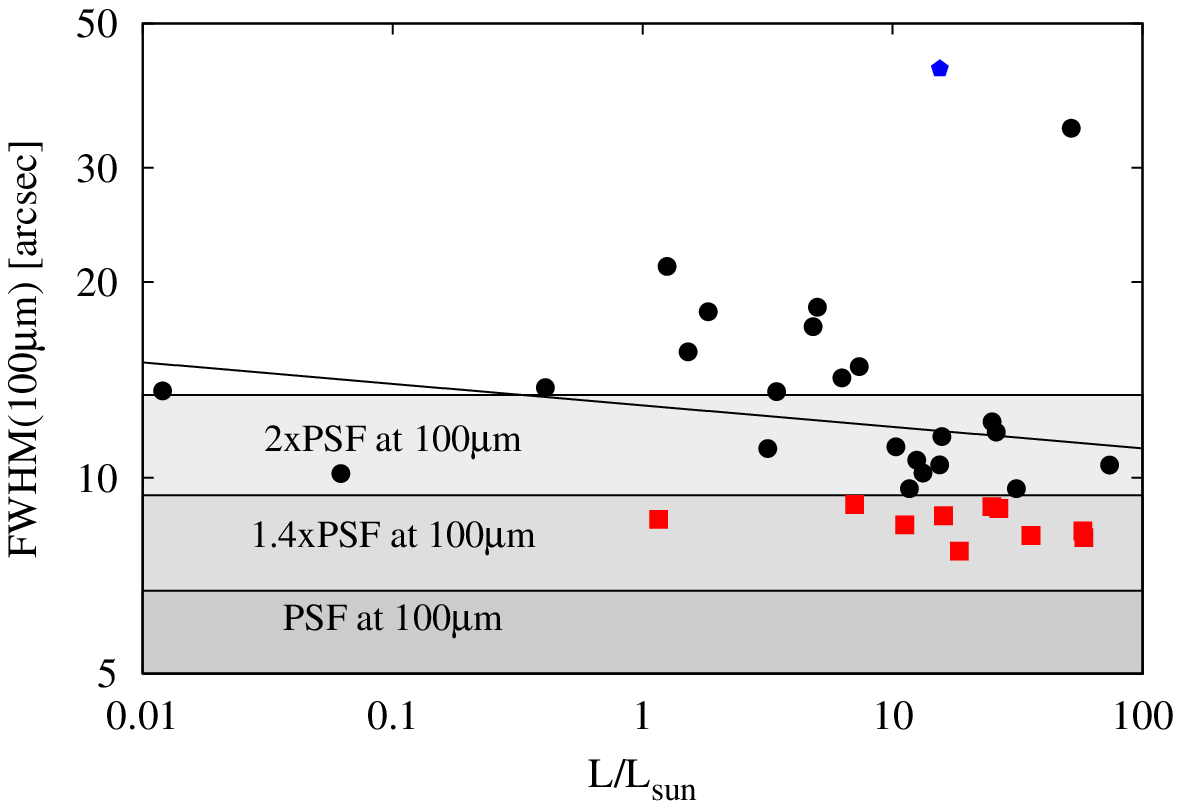}
   \vspace*{-3mm}
   \caption{Distance to targets (left) and the observed long-axis FWHM at $100\mum$ (right) 
   as a function of stellar luminosity.
   Tilted straight lines are log-log best fits through the symbols.
   For Fomalhaut (blue diamond), the FWHM at $70\mum$ instead of $100\mum$ is shown in the right panel.
   Grey-shaded areas correspond to 1.0, 1.4, and 2.0 times the beam size, showing how well the disks have 
   been resolved.
   Red squares are marginally resolved systems, i.e. those with $\text{FWHM} < 1.4 \text{PSF}$.
   \label{fig:d-L-bias}
   }
   \end{figure*}

The cold disk radii are plotted in Figure~\ref{fig:rdisk}.
They reveal a huge scatter, from $\approx 40\AU$ to $\approx 290\AU$.
At a first glance, the figure suggests a weak positive correlation with the stellar
luminosity.
However, the Pearson correlation coefficient between $\log\rdisc$ and $\log L$ is 
only $r=+0.38$.
The Spearman rank correlation coefficient (that does not require linearity and treats possible
outliers better than Pearson does) is as small as $r_s=0.02$.
To interpret these coefficients, we can calculate a two-tailed probability that the
correlation is absent. The correlation is treated as \revision{significant if this probability 
is less than 1\%; for a sample of 34 stars, this requires} $r$ or $r_s$ larger than 0.43.
Judging by $r$ and especially by $r_s$, the correlation
between $\log\rdisc$ and $\log L$ is unlikely.
Also, the best-fit trend line $\rdisc = a (L/L_\odot)^b$ has
$a=123\pm15\AU$ and $b=0.04\pm0.04$, so that the slope $b$ is statistically indistinguishable from zero.
Furthermore, the results strongly depend on a few individual stars with radii far from the
trend line. For example, excluding the two M-stars (which are the two points in the
left bottom corner of Figure~\ref{fig:rdisk}) would change the best-fit coefficients to
$a=152\pm20\AU$ and $b=0.03\pm0.05$.

\begin{table*}[htbp]
  \begin{center}
  \caption{\change{Table slightly corrected}Fitting Results for the MBB Method}
  \tabcolsep 3.5pt
    \begin{tabular}{rrcc|rrrrr}
    \toprule
    HD    & $r_\text{cold}$ & $s_\text{blow}$ & $T_\text{BB}$ & $\lambda_\text{0}/(2\pi)$ & $\beta$    & $T_\text{d}$ 
     & $T_\text{d}/T_\text{BB}$ & $\chi^2_\text{red}$ \\
    \midrule
    GJ 581 	& 38    & -     & 15    & 0.06~$\pm$~0.44  & 0.74~$\pm$~0.19  & 35~$\pm$~3     & 2.31~$\pm$~0.17  & 1.01 \\
    197481 	& 38    & -     & 22    & 0.02~$\pm$~0.19  & 0.37~$\pm$~0.08  & 38~$\pm$~1     & 1.74~$\pm$~0.05  & 2.63 \\
    23484 	& 91    & -     & 24    & 0.13~$\pm$~0.42  & 0.37~$\pm$~0.11  & 36~$\pm$~3     & 1.53~$\pm$~0.13  & 0.86 \\
    104860 	& 105   & 0.39  & 28    & 9.24~$\pm$~0.25  & 0.88~$\pm$~0.07  & 33~$\pm$~3     & 1.18~$\pm$~0.11  & 3.22 \\
    207129 	& 148   & 0.41  & 24    & 0.03~$\pm$~0.70  & 0.53~$\pm$~0.15  & 46~$\pm$~3     & 1.90~$\pm$~0.12  & 4.35 \\
    10647 	& 117   & 0.47  & 28    & 0.13~$\pm$~0.22  & 0.44~$\pm$~0.10  & 47~$\pm$~4     & 1.68~$\pm$~0.14  & 5.02 \\
    48682 	& 142   & 0.54  & 28    & 0.11~$\pm$~0.44  & 0.52~$\pm$~0.11  & 50~$\pm$~3     & 1.78~$\pm$~0.11  & 9.36 \\
    50571 	& 138   & 0.81  & 31    & 5.67~$\pm$~0.39  & 0.95~$\pm$~0.10  & 40~$\pm$~3     & 1.27~$\pm$~0.10  & 1.69 \\
    170773 	& 207   & 0.86  & 26    & 5.96~$\pm$~0.56  & 1.08~$\pm$~0.08  & 35~$\pm$~1     & 1.34~$\pm$~0.04  & 0.55 \\
    218396 	& 293   & 1.11  & 24    & 2.64~$\pm$~0.26  & 0.94~$\pm$~0.08  & 37~$\pm$~1     & 1.52~$\pm$~0.04  & 7.15 \\
    109085 	& 145   & 1.14  & 35    & 9.11~$\pm$~0.23  & 0.26~$\pm$~0.10  & 36~$\pm$~1     & 1.03~$\pm$~0.03  & 3.05 \\
    27290 	& 134   & 1.35  & 38    & 3.76~$\pm$~0.30  & 0.64~$\pm$~0.10  & 47~$\pm$~2     & 1.23~$\pm$~0.05  & 11.18 \\
    95086 	& 230   & 1.47  & 30    & 0.02~$\pm$~0.30  & 0.45~$\pm$~0.08  & 52~$\pm$~2     & 1.73~$\pm$~0.07  & 1.87 \\
    195627 	& 175   & 1.51  & 34    & 5.56~$\pm$~0.79  & 0.97~$\pm$~0.10  & 43~$\pm$~2     & 1.26~$\pm$~0.06  & 2.14 \\
    20320 	& 157   & 1.94  & 45    & 2.31~$\pm$~0.75  & 0.55~$\pm$~0.08  & 56~$\pm$~2     & 1.24~$\pm$~0.04  & 0.96 \\
    21997 	& 150   & 2.06  & 42    & 5.73~$\pm$~0.53  & 0.82~$\pm$~0.08  & 49~$\pm$~1     & 1.17~$\pm$~0.02  & 0.80 \\
    110411 	& 111   & 2.13  & 50    & 6.04~$\pm$~0.38  & 1.22~$\pm$~0.08  & 60~$\pm$~3     & 1.19~$\pm$~0.06  & 6.10 \\
    142091 	& 131   & 2.24  & 45    & 0.05~$\pm$~0.40  & 0.32~$\pm$~0.09  & 63~$\pm$~4     & 1.41~$\pm$~0.09  & 8.47 \\
    102647 	& 48    & 2.33  & 78    & 4.62~$\pm$~0.74  & 1.04~$\pm$~0.10  & 87~$\pm$~1     & 1.12~$\pm$~0.01  & 16.07 \\
    125162 	& 112   & 2.61  & 56    & 5.33~$\pm$~0.44  & 0.69~$\pm$~0.08  & 63~$\pm$~1     & 1.12~$\pm$~0.02  & 4.09 \\
    216956 	& 123   & 2.62  & 50    & 16.40 ~$\pm$~0.18  & 0.85~$\pm$~0.09  & 51~$\pm$~2     & 1.01~$\pm$~0.04  & 4.39 \\
    17848 	& 222   & 2.65  & 37    & 5.05~$\pm$~0.35  & 0.80~$\pm$~0.08  & 45~$\pm$~4     & 1.21~$\pm$~0.11  & 3.66 \\
    9672  	& 172   & 2.68  & 43    & 8.01~$\pm$~0.47  & 0.97~$\pm$~0.10  & 49~$\pm$~2     & 1.16~$\pm$~0.05  & 0.55 \\
    71722 	& 111   & 2.99  & 52    & 3.03~$\pm$~0.35  & 0.63~$\pm$~0.07  & 64~$\pm$~1     & 1.23~$\pm$~0.02  & 8.63 \\
    182681 	& 184   & 3.72  & 46    & 1.27~$\pm$~0.35  & 0.74~$\pm$~0.09  & 68~$\pm$~1     & 1.49~$\pm$~0.02  & 6.55 \\
    14055 	& 165   & 3.73  & 51    & 19.73 ~$\pm$~1.28  & 1.14~$\pm$~0.09  & 51~$\pm$~1     & 1.01~$\pm$~0.02  & 3.02 \\
    161868 	& 143   & 3.84  & 53    & 3.15~$\pm$~0.33  & 0.69~$\pm$~0.08  & 65~$\pm$~1     & 1.22~$\pm$~0.02  & 4.34 \\
    188228 	& 106   & 3.90  & 62    & 0.73~$\pm$~0.98  & 0.36~$\pm$~0.12  & 77~$\pm$~6     & 1.25~$\pm$~0.10  & 1.39 \\
    10939 	& 180   & 4.40  & 49    & 20.80 ~$\pm$~0.48  & 1.22~$\pm$~0.08  & 50~$\pm$~1     & 1.03~$\pm$~0.02  & 3.08 \\
    71155 	& 88    & 4.85  & 72    & 8.65~$\pm$~0.39  & 2.22~$\pm$~0.10  & 79~$\pm$~15    & 1.08~$\pm$~0.21  & 2.17 \\
    172167 	& 115   & 6.38  & 60    & 23.22 ~$\pm$~1.25  & 1.35~$\pm$~0.09  & 70~$\pm$~3     & 1.18~$\pm$~0.05  & 7.50 \\
    139006 	& 61    & 6.90  & 98    & 11.89 ~$\pm$~1.03  & 1.80~$\pm$~0.10  & 101 ~$\pm$~3     & 1.03~$\pm$~0.03  & 1.94 \\
    95418 	& 61    & 6.95  & 98    & 11.54 ~$\pm$~0.93  & 1.37~$\pm$~0.12  & 100 ~$\pm$~4     & 1.02~$\pm$~0.04  & 9.48 \\
    13161 	& 143   & 8.28  & 69    & 22.60 ~$\pm$~0.43  & 1.33~$\pm$~0.11  & 68~$\pm$~1     & 0.99~$\pm$~0.01  & 2.56 \\
    \bottomrule
    \end{tabular}%
  \label{tab:fitting_MBB}%
  \end{center}

  \medskip
  \noindent
  {\em Notes:}\\
  The disk radii are in AU, the temperatures in K and $\lambda_\text{0}/(2\pi)$, $s_\text{blow}$ in $\mum$.\\
\end{table*}%

The conclusion that $\rdisc$ and $L$ appear uncorrelated is not necessarily in 
contradiction with
a possible weak trend of disk radius getting larger towards more luminous stars reported by 
\citet{eiroa-et-al-2013}.
The reason is that Eiroa et al. consider the blackbody radius $\rbb$ rather than the true disk radius 
$\rdisc$.
Below we will see that the ratio of the two, $\Gamma = \rdisc/\rbb$, decreases with $L$,
which may compensate an increase in $\rbb$ towards higher $L$, leading to an $\rdisc$ uncorrelated with
the stellar luminosity.

\revision{The} above analysis should be treated with caution.
One caveat is related to luminous, F- and especially A-stars.
The more luminous the stars, the more distant they are on the average.
Therefore, their disks often have a smaller angular size, and thus are
less well resolved, as illustrated in Figure~\ref{fig:d-L-bias}.
It shows, for example, that all disks around stars with $L > 10 L_\odot$,
except for Fomalhaut and Vega, are resolved in less than two beams.
This may explain why the scatter in $\rdisc$ is larger for more luminous stars.

Another caveat is that our narrow-disk (single-radius) approximation may be rather poor
for some disks where the dust distribution is strongly extended.   
The finite disk widths are particularly apparent in the scattered light images that are most sensitive
to the small grains.
For a small sample of large, bright disks, \citet{krist-et-al-2012} find the disk widths at half-maximum brightness
to range from 20\% to 60\%  of the belt radius.
Also, multi-wavelength studies of some of the individual disks, especially of A-stars
such as Vega \citep{su-et-al-2006} and HR~8799 \citep{matthews-et-al-2013b}
reveal huge halos probably composed of small dust grains in weakly bound orbits
\citep{matthews-et-al-2013}.
For instance, the huge ($290\AU$) radius derived from the HR~8799 PACS image
may not necessarily measure the ``peak'' radius of the disk (i.e. the radius
of the dust-producing planetesimal belt).

   \begin{figure*}[htb!]
   \centering
   \hspace*{8mm}
   \includegraphics[width=0.45\textwidth,angle=0]{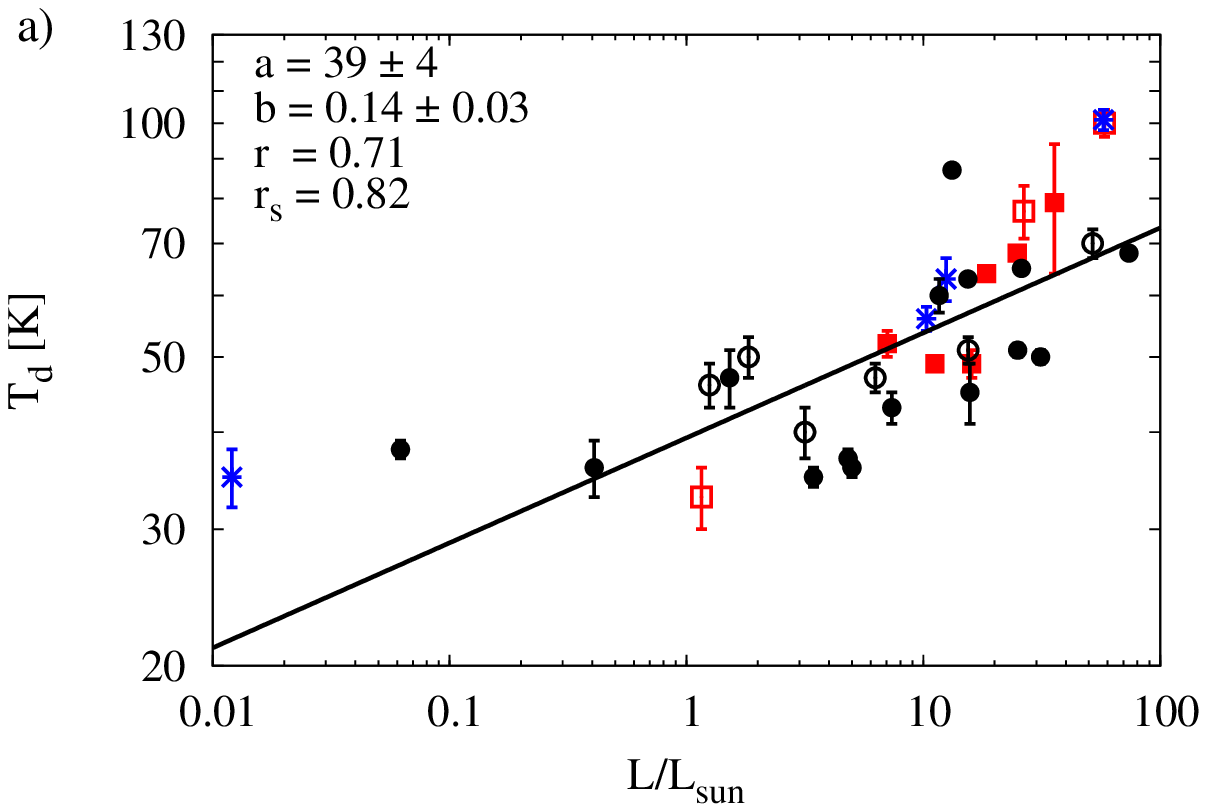}
   \includegraphics[width=0.45\textwidth,angle=0]{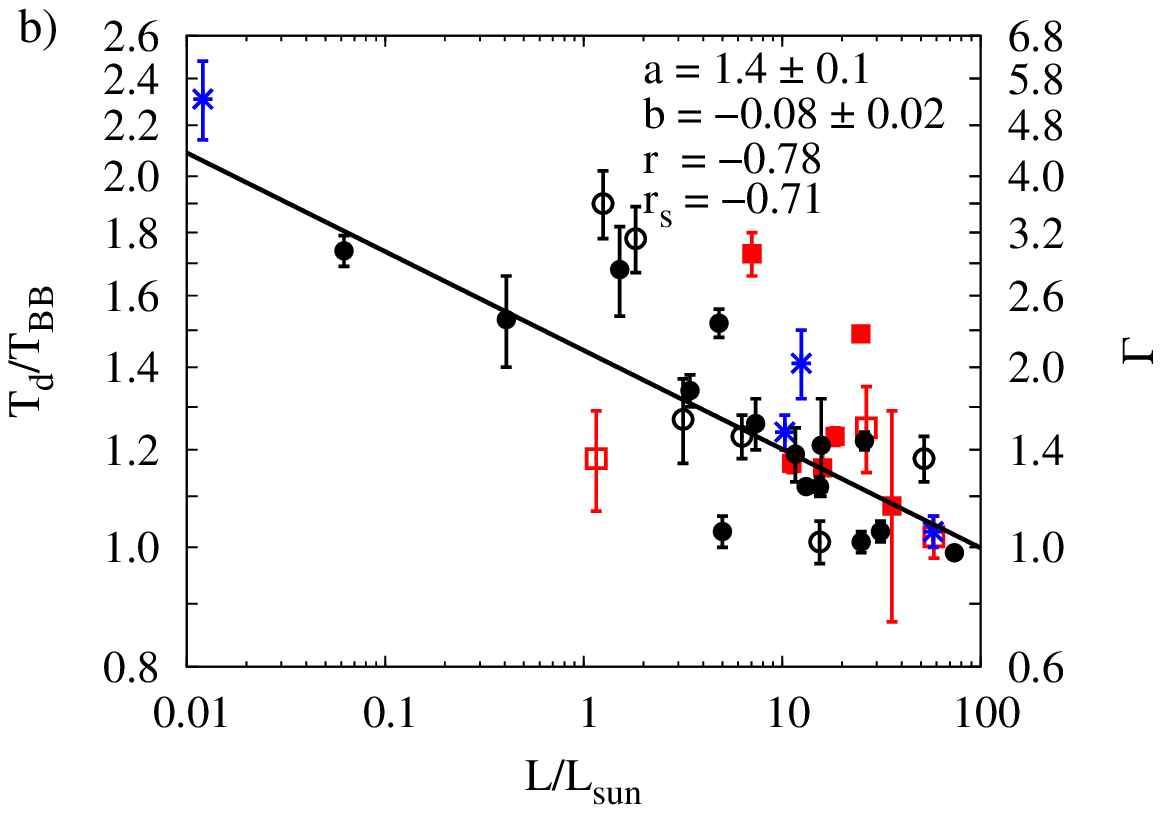}\\
   \includegraphics[width=0.45\textwidth,angle=0]{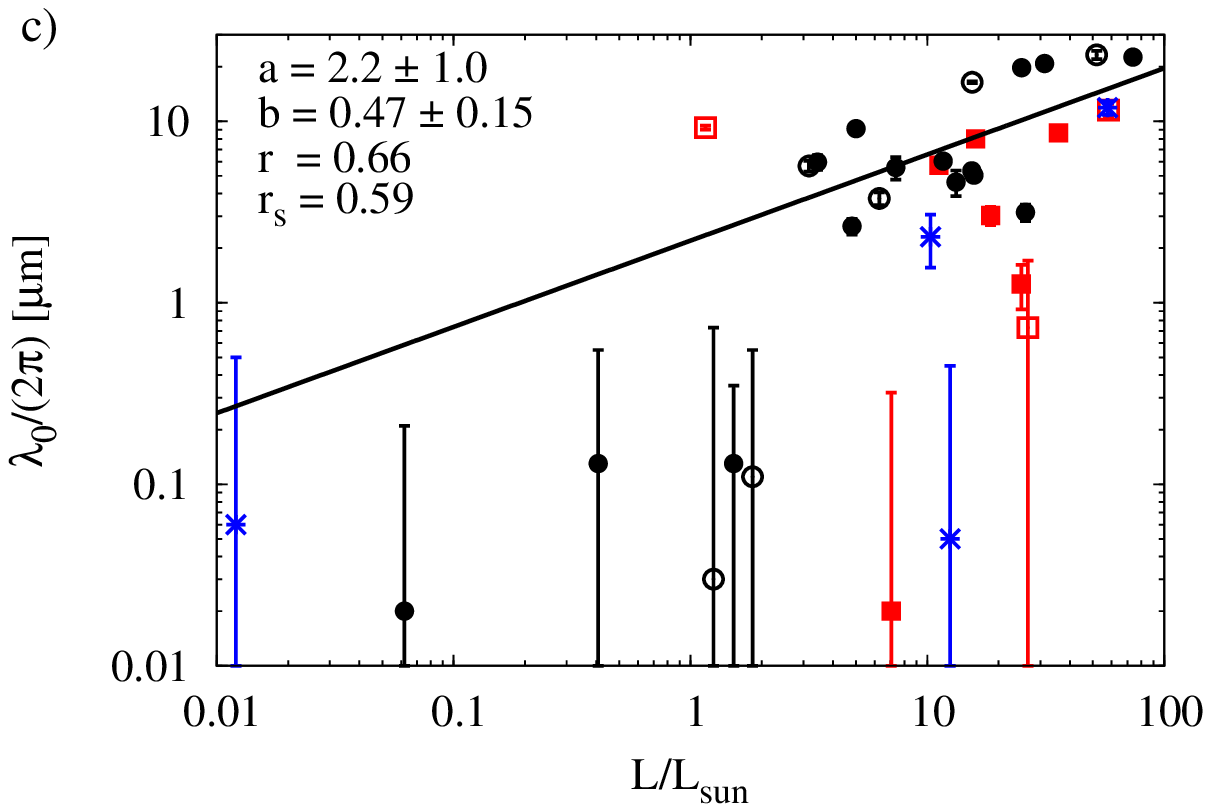}
   \includegraphics[width=0.45\textwidth,angle=0]{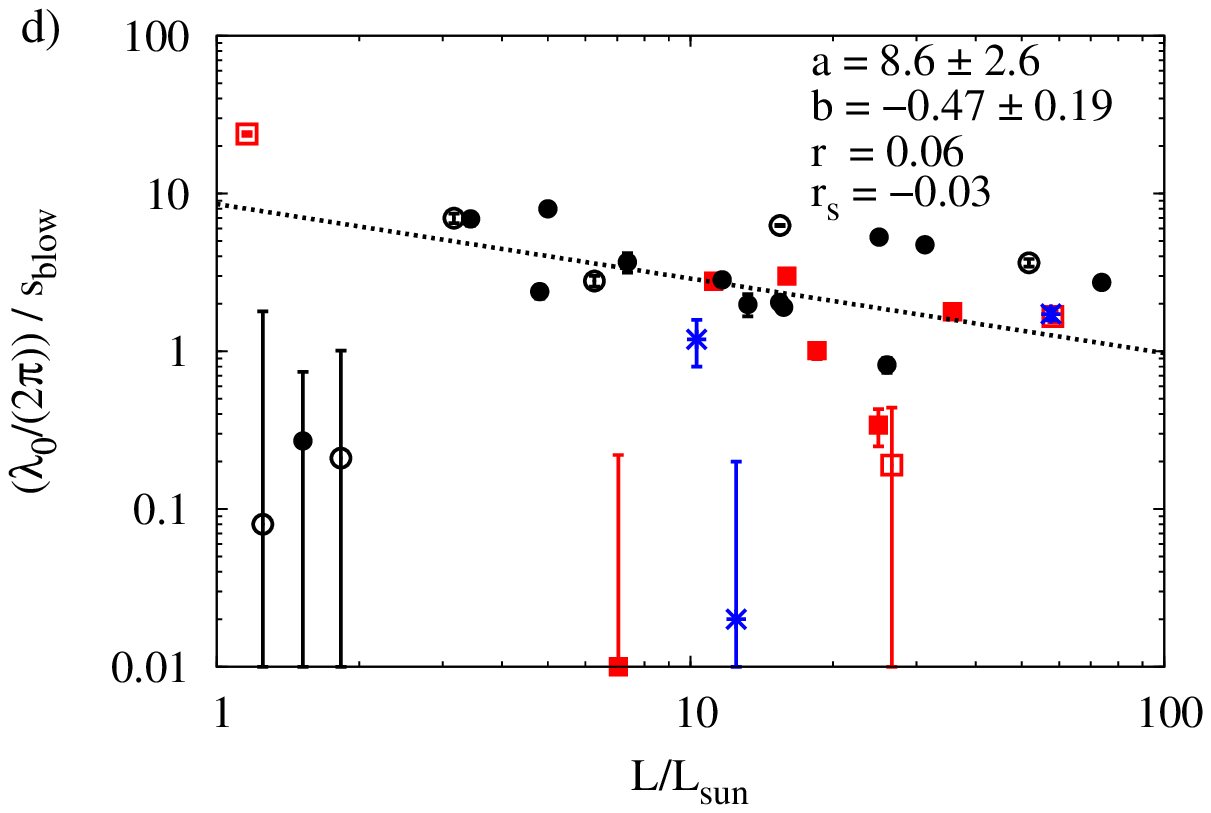}\\
   \includegraphics[width=0.45\textwidth,angle=0]{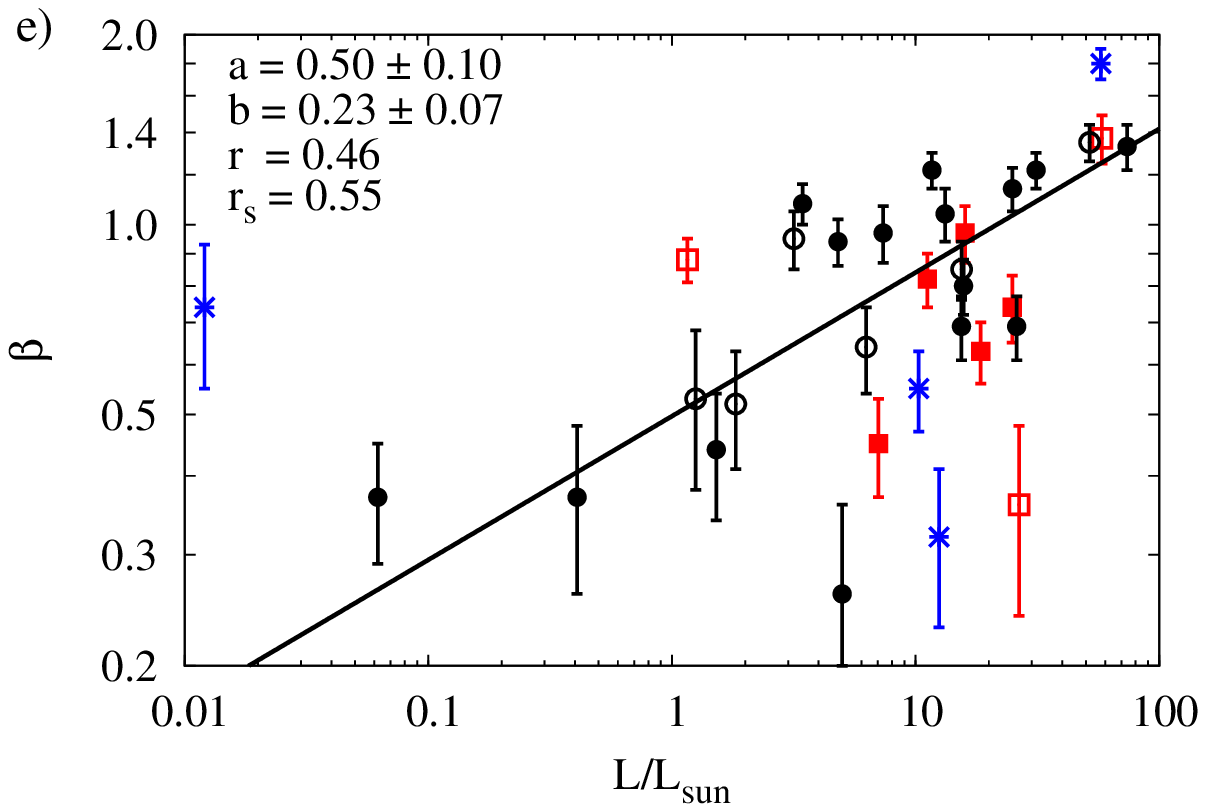}
   \caption
   {   \change{Figure slightly amended}
       Disk parameters obtained with the MBB method, plotted as functions of stellar luminosity:
       (a)~dust temperature,
       (b)~the ratio of the dust temperature to the blackbody temperature
       and the ratio $\Gamma$ of the true disk radius to its blackbody radius,
       (c)~the characteristic size $\lambda_0/(2\pi)$,
       (d)~the ratio of $\lambda_0/(2\pi)$ to the blowout size $\sblow$ for stars with $L > L_\odot$, and
       (e)~the opacity \revision{index} $\beta$.
       Open and filled symbols stand for one- and two-component disks, respectively.
       Circles represent well-resolved disks, squares indicate marginally resolved ones.
       The blue asterisks are the objects with no (sub)-mm detections.
       Lines are linear fits to the log-log data.
       The quantities $a$, $b$, $r$, and $r_s$ are as in Figure~\ref{fig:rdisk}.
       The trend lines are solid if a correlation is \revision{formally} significant
       (both $r$ and $r_s$ are greater than 0.43) and dashed otherwise.
   }
    \label{fig:Td_smin_MBB}
    
   \end{figure*}

\subsection{Dust temperatures and sizes assuming MBB}

The temperature as a function of the luminosity of the primary stars
is plotted in Figure~\ref{fig:Td_smin_MBB}a.
It shows that the dust temperature rises from 30--50$\K$ for
late- and solar-type stars to $60$--$100\K$ for the 
most luminous A-stars.
\change{Correlation coefficients in this section slightly changed}
With the Pearson's $r=0.71$ and the Spearman's $r_s=0.82$,
the trend is very significant.
This is in a good agreement with other {\it Herschel}-based studies, most notably
\citet{booth-et-al-2013} who find temperatures in the $70$--$120\K$ range
for their sample of A-stars.
Our results are also roughly consistent with {\it Spitzer} results
by \citet{morales-et-al-2011}, although
a trend of temperatures getting warmer towards A-stars was weaker there.
They find 
median values for their samples of G- and A-stars to be close to each other~---
$59\K$ and $62\K$, respectively.
We can also compare the temperatures with the [24]-[70]
color temperatures of 44 A-star disks derived by \citet{su-et-al-2006}
(see their Figure~7).
For a subsample of eight A-stars only detected at $70\mum$ but not at $24\mum$,
the upper limits on the temperature vary between $60\K$ and $120\K$,
with the median value of $79\K$.
Most recently, \citet{ballering-et-al-2013} studied the circumstellar 
environment
of more than 500 main-sequence stars using {\it Spitzer} IRS/MIPS data.
Considering the SEDs of 174 debris disks, they found a correlation
between the temperature of cold debris component and stellar temperature.
Our Figure~\ref{fig:Td_smin_MBB}a
is in excellent agreement with their Figure~5.

Figure~\ref{fig:Td_smin_MBB}b presents the \revision{ratio $\td/\tbb$,
where $\tbb \propto L^{1/4} \rdisc^{-1/2}$ is the blackbody temperature at a distance
$\rdisc$ from the star.
If the variation in $\td$ seen in Figure~\ref{fig:Td_smin_MBB}a were caused solely
by different luminosities of the central stars and differences in the radii between the 
individual disks, the ratio $\td/\tbb$ would be nearly constant for all the systems.}
Instead, we see a clear trend of $\td/\tbb$ decreasing with the increasing stellar luminosity
($r=-0.78$, $r_s=-0.71$),
in a good agreement with \citet{booth-et-al-2013}.
An equivalent quantity to  $\td/\tbb$
is $\Gamma = (\td/\tbb)^2$, which is
the ratio of the true disk radius to the radius it would have if its material were emitting
as blackbody \citep{booth-et-al-2013}.
To facilitate comparisons with the results obtained by other authors, we plot
$\Gamma$ in the second axis of Figure~\ref{fig:Td_smin_MBB}b.
The trend of  $\td/\tbb$ or $\Gamma$ decreasing with $L$
must be indicative of dust grains getting larger in disks of earlier-type 
stars, because bigger grains are colder.

\begin{table*}[htbp]
  \begin{center}
  \caption{\change{Table slightly corrected}Fitting Results for the SD Method}
  \tabcolsep 3.5pt
    \begin{tabular}{rrcc|rrrrr}
    \toprule
    HD    & $r_\text{cold}$ & $s_\text{blow}$ & $T_\text{BB}$ & $s_\text{min}$  & $q$    & $T_\text{d}$ 
    & $T_\text{d}/T_\text{BB}$ & $\chi^2_\text{red}$ \\
    \midrule   
    GJ 581 	& 38    & -     & 15    & 1.50~$\pm$~2.32  & 3.37~$\pm$~0.60  & 24~$\pm$~11    & 1.59~$\pm$~0.74  & 0.78 \\
    197481 	& 38    & -     & 22    & 0.21~$\pm$~2.27  & 3.33~$\pm$~0.43  & 42~$\pm$~12    & 1.93~$\pm$~0.55  & 5.28 \\
    23484 	& 91    & -     & 24    & 1.46~$\pm$~2.45  & 3.41~$\pm$~0.45  & 37~$\pm$~10    & 1.56~$\pm$~0.42  & 1.56 \\
    104860 	& 105   & 0.39  & 28    & 6.96~$\pm$~2.41  & 3.83~$\pm$~0.40  & 44~$\pm$~10    & 1.58~$\pm$~0.36  & 2.99 \\
    207129 	& 148   & 0.41  & 24    & 4.93~$\pm$~1.67  & 4.23~$\pm$~0.46  & 56~$\pm$~16    & 2.33~$\pm$~0.67  & 6.19 \\
    10647 	& 117   & 0.47  & 28    & 3.92~$\pm$~2.21  & 3.87~$\pm$~0.37  & 56~$\pm$~11    & 2.00~$\pm$~0.39  & 5.33 \\
    48682 	& 142   & 0.54  & 27    & 1.83~$\pm$~2.73  & 3.70~$\pm$~0.60  & 65~$\pm$~11    & 2.42~$\pm$~0.41  & 4.62 \\
    50571 	& 138   & 0.81  & 31    & 5.36~$\pm$~2.75  & 4.01~$\pm$~0.49  & 56~$\pm$~13    & 1.80~$\pm$~0.41  & 1.85 \\
    170773 	& 207   & 0.86  & 26    & 5.04~$\pm$~2.83  & 4.02~$\pm$~0.46  & 53~$\pm$~7     & 1.99~$\pm$~0.26  & 3.37 \\
    218396 	& 293   & 1.11  & 24    & 4.84~$\pm$~2.72  & 4.01~$\pm$~0.55  & 51~$\pm$~11    & 2.11~$\pm$~0.46  & 6.42 \\
    109085 	& 145   & 1.14  & 35    & 7.46~$\pm$~1.81  & 2.85~$\pm$~0.40  & 35~$\pm$~17    & 1.01~$\pm$~0.49  & 1.80 \\
    27290 	& 134   & 1.35  & 38    & 5.02~$\pm$~2.11  & 3.85~$\pm$~0.31  & 61~$\pm$~18    & 1.60~$\pm$~0.48  & 37.36 \\
    95086 	& 230   & 1.47  & 30    & 2.96~$\pm$~3.54  & 3.83~$\pm$~0.54  & 61~$\pm$~19    & 2.03~$\pm$~0.64  & 1.30 \\
    195627 	& 175   & 1.51  & 34    & 4.83~$\pm$~1.90  & 4.00~$\pm$~0.41  & 61~$\pm$~7     & 1.78~$\pm$~0.21  & 2.24 \\
    20320 	& 157   & 1.94  & 45    & 5.38~$\pm$~2.56  & 3.90~$\pm$~0.59  & 68~$\pm$~7     & 1.50~$\pm$~0.15  & 0.52 \\
    21997 	& 150   & 2.06  & 42    & 4.46~$\pm$~1.88  & 3.88~$\pm$~0.33  & 65~$\pm$~16    & 1.57~$\pm$~0.39  & 1.28 \\
    110411 	& 111   & 2.13  & 50    & 3.42~$\pm$~1.91  & 4.10~$\pm$~0.44  & 87~$\pm$~21    & 1.73~$\pm$~0.42  & 9.21 \\
    142091 	& 131   & 2.24  & 45    & 2.42~$\pm$~2.85  & 3.64~$\pm$~0.43  & 73~$\pm$~15    & 1.61~$\pm$~0.33  & 7.61 \\
    102647 	& 48    & 2.33  & 78    & 8.84~$\pm$~2.14  & 4.24~$\pm$~0.44  & 72~$\pm$~11    & 0.92~$\pm$~0.14  & 1.79 \\
    125162 	& 112   & 2.61  & 56    & 3.94~$\pm$~2.11  & 3.75~$\pm$~0.52  & 78~$\pm$~8     & 1.39~$\pm$~0.14  & 3.95 \\
    216956 	& 123   & 2.62  & 50    & 3.88~$\pm$~2.64  & 3.44~$\pm$~0.39  & 63~$\pm$~12    & 1.25~$\pm$~0.24  & 3.51 \\
    17848 	& 222   & 2.65  & 37    & 4.69~$\pm$~2.11  & 3.88~$\pm$~0.38  & 62~$\pm$~6     & 1.66~$\pm$~0.16  & 3.85 \\
    9672  	& 172   & 2.68  & 43    & 3.98~$\pm$~2.02  & 3.85~$\pm$~0.38  & 70~$\pm$~16    & 1.64~$\pm$~0.38  & 1.08 \\
    71722 	& 111   & 2.99  & 52    & 5.11~$\pm$~1.63  & 3.94~$\pm$~0.61  & 75~$\pm$~21    & 1.44~$\pm$~0.40  & 9.16 \\
    182681 	& 184   & 3.72  & 46    & 2.47~$\pm$~1.74  & 3.95~$\pm$~0.44  & 90~$\pm$~19    & 1.97~$\pm$~0.42  & 2.32 \\
    14055 	& 165   & 3.73  & 51    & 6.70~$\pm$~3.18  & 3.69~$\pm$~0.40  & 63~$\pm$~19    & 1.24~$\pm$~0.37  & 3.07 \\
    161868 	& 143   & 3.84  & 53    & 4.81~$\pm$~2.41  & 3.72~$\pm$~0.36  & 72~$\pm$~7     & 1.35~$\pm$~0.13  & 9.61 \\
    188228 	& 106   & 3.90  & 62    & 1.81~$\pm$~3.71  & 3.34~$\pm$~0.33  & 79~$\pm$~24    & 1.28~$\pm$~0.39  & 1.50 \\
    10939 	& 180   & 4.40  & 49    & 8.81~$\pm$~1.90  & 3.78~$\pm$~0.38  & 59~$\pm$~5     & 1.20~$\pm$~0.10  & 4.00 \\
    71155 	& 88    & 4.85  & 72    & 5.14~$\pm$~2.12  & 5.80~$\pm$~0.41  & 104~$\pm$~38   & 1.44~$\pm$~0.52  & 2.94 \\
    172167 	& 115   & 6.38  & 60    & 9.05~$\pm$~1.91  & 3.67~$\pm$~0.34  & 68~$\pm$~2     & 1.14~$\pm$~0.03  & 9.87 \\
    139006 	& 61    & 6.90  & 98    & 6.74~$\pm$~2.50  & 5.29~$\pm$~0.41  & 97~$\pm$~44    & 0.99~$\pm$~0.45  & 2.35 \\
    95418 	& 61    & 6.95  & 98    & 5.18~$\pm$~2.12  & 4.43~$\pm$~0.33  & 110~$\pm$~43   & 1.12~$\pm$~0.44  & 9.50 \\
    13161 	& 143   & 8.28  & 71    & 15.09 ~$\pm$~2.49  & 4.47~$\pm$~0.37  & 59~$\pm$~6   & 0.82~$\pm$~0.08  & 3.46 \\
    \bottomrule
    \end{tabular}%
  \label{tab:fitting_SD}%
  \end{center}

  \medskip
  \noindent
  {\em Notes:}\\
 The disk radii are in AU, the temperatures in K and $s_\text{blow}$  and $\smin$ in $\mum$.\\
\end{table*}%

To check this, in Figure~\ref{fig:Td_smin_MBB}c we plot
the quantity $\lambda_0/(2\pi)$, which is a proxy for the typical grain sizes
in the MBB method.
Indeed, there is a strong ($r=0.66$, $r_s=0.59$)
trend of the typical size increasing
with the stellar luminosity.
That size rises from $\la 1\mum$ for GKM-type stars
to a few $\mum$ or even more for A-type stars.

We expect that the reason for the dust sizes to increase towards more luminous stars
is that so does the radiation pressure blowout
limit. To verify this, we calculate it as \citep[see, e.g.,][]{burns-et-al-1979}:
\be
  {\frac{\sblow}{1\mum}} = 0.35 \Qpr \left(\frac{3.3 \g\cm^{-3}}{\rho} \right)
                               \left(\frac{L/L_\odot}{M/M_\odot}\right),
\label{s_blow}
\ee
where $\rho$ is the grain's bulk density,
$L$ and $M$ are stellar luminosity and mass,
and $\Qpr$ is the radiation pressure efficiency.
\revision{
We adopted $\rho=3.3\g\cm^{-3}$ and took
$L$ and $M$ from Table~\ref{tab:stars}.
Next, we assumed $\Qpr = 1$.
As long as $L \ge L_\odot$, the error introduced by this assertion 
is much smaller than the accuracy of the $\lambda_0/(2\pi)$ determination.
For instance, the astrosilicate particles with $s=\sblow$ around a $1L_\odot$ and a $10L_\odot$ star 
would have $\Qpr = 1.4$ and $\Qpr = 1.1$, respectively.
If $L < L_\odot$, the assumption $\Qpr = 1$ is no longer valid.
One has to consider $\Qpr$ as a function of the grain size and treat
the formula~(\ref{s_blow}) as an equation for $\sblow$. For low-luminosity
stars, this equation will not have a solution, meaning that radiation pressure
is too weak for the blowout limit to exist.
For this reason, we only calculated $\sblow$ for stars with
$L \ge L_\odot$.}

\change{Indent}The result is shown in Figure~\ref{fig:Td_smin_MBB}d.
We find that $\lambda_0/(2\pi)$ for nearly all the stars lies in the range
from $(1...10) s_\text{blow}$, regardless of the stellar luminosity.
Judging by the trend line, there might be a subtle trend of
the $\lambda_0/(2\pi)$ to $s_\text{blow}$ ratio decreasing towards earlier spectral classes
(the slope $-0.47$).
However, the correlation coefficients are as small as $r=+0.06$ and $r_s=-0.03$.
The Spearman probability that the ratio of $\lambda_0/(2\pi)$ to $s_\text{blow}$
is constant is \revision{87}\%.

Finally, Figure~\ref{fig:Td_smin_MBB}e depicts the MBB index $\beta$. As expected,
it lies between 0 and 2, with the majority of disks best fitted with $\beta$
between zero and unity.
\revision{This is consistent with previous results, based on sub-mm observations of debris
disks \citep[e.g.,][]{williams-andrews-2006,nilsson-et-al-2010}.}
There might be a slight trend of $\beta$ increasing with the stellar luminosity.
However, with $r=0.46$ and $r=0.55$, the trend 
\revision{is only of moderate significance.
If it is real, it may} be a consequence of the fact that $\beta$ is not completely
independent of $\lambda_0$ (see Appendix B).

\subsection{Dust temperatures and sizes assuming SD}

   \begin{figure*}[htb!]
   \centering
   \hspace*{8mm}
   \includegraphics[width=0.45\textwidth,angle=0]{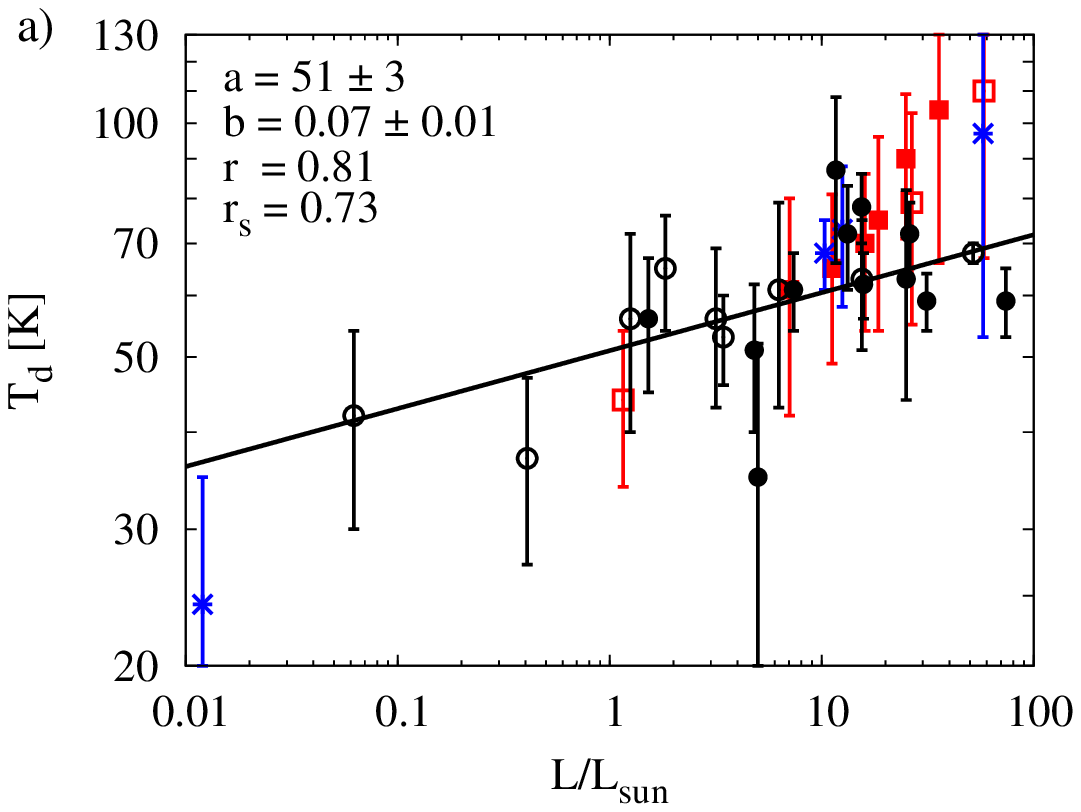}
   \includegraphics[width=0.45\textwidth,angle=0]{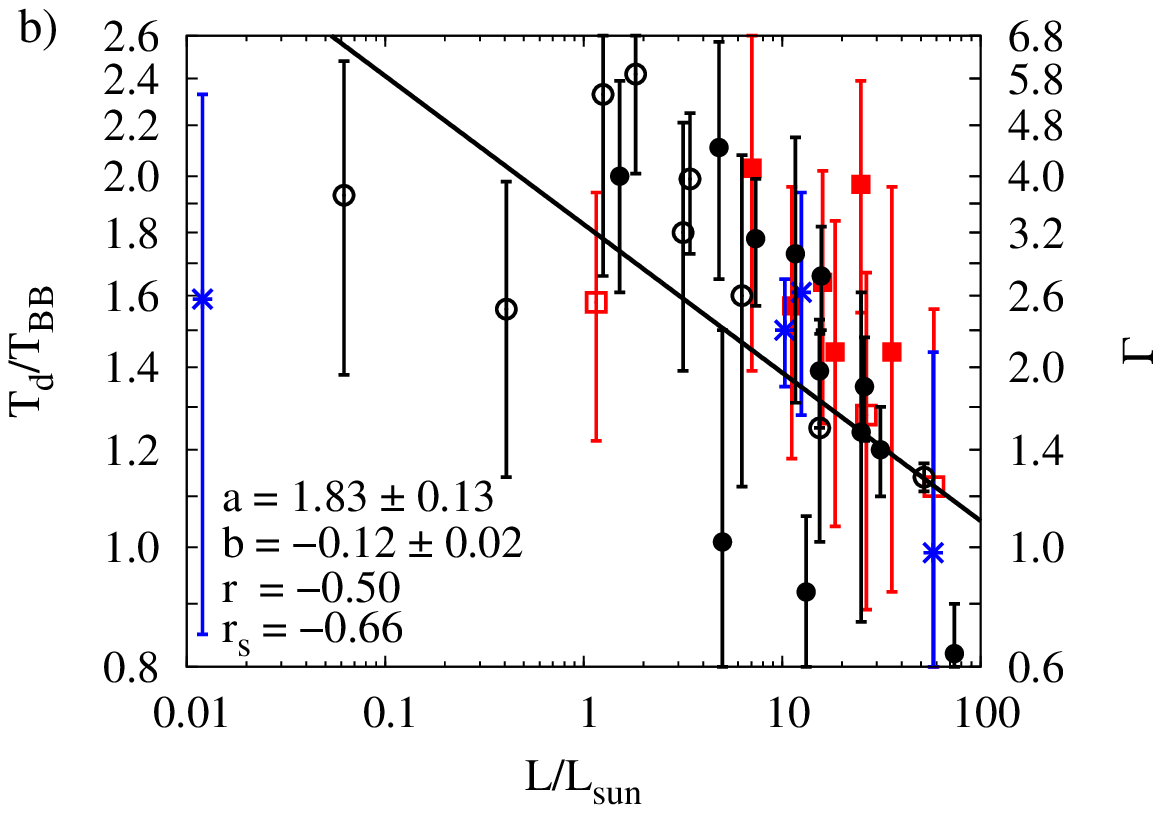}\\
   \includegraphics[width=0.45\textwidth,angle=0]{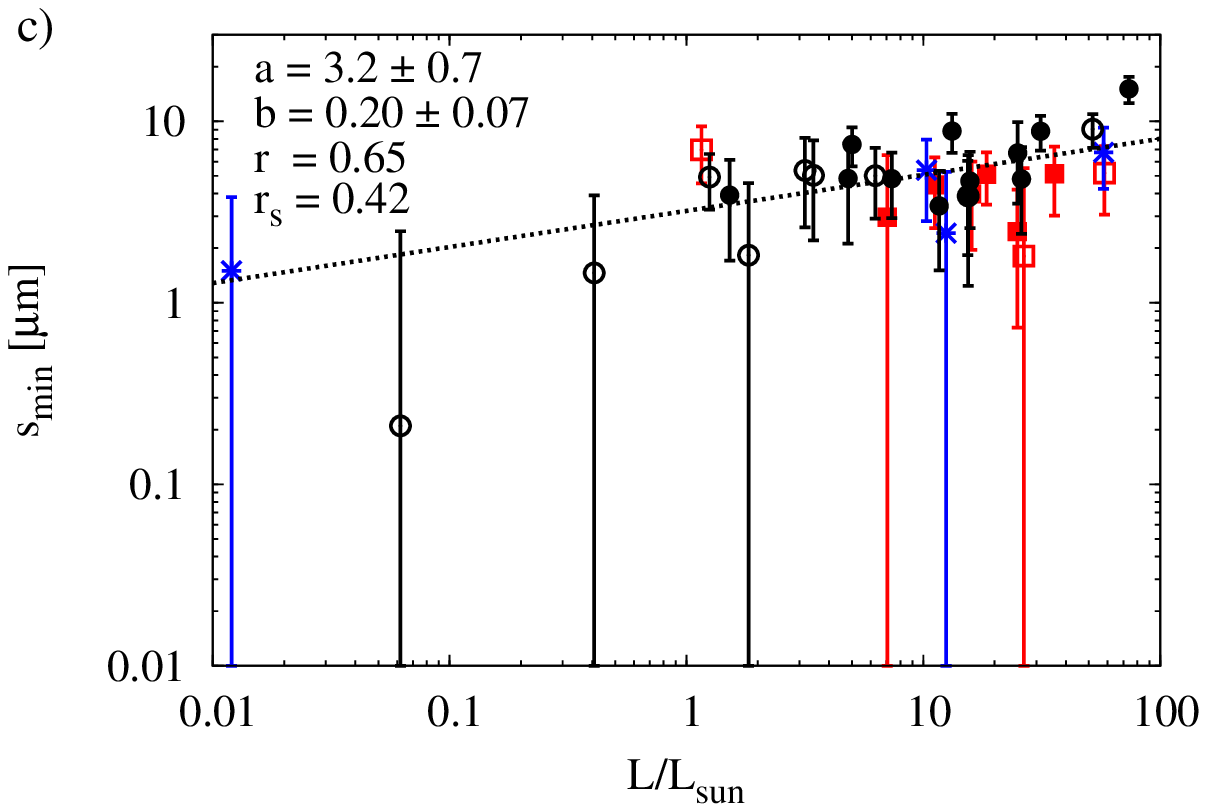}
   \includegraphics[width=0.45\textwidth,angle=0]{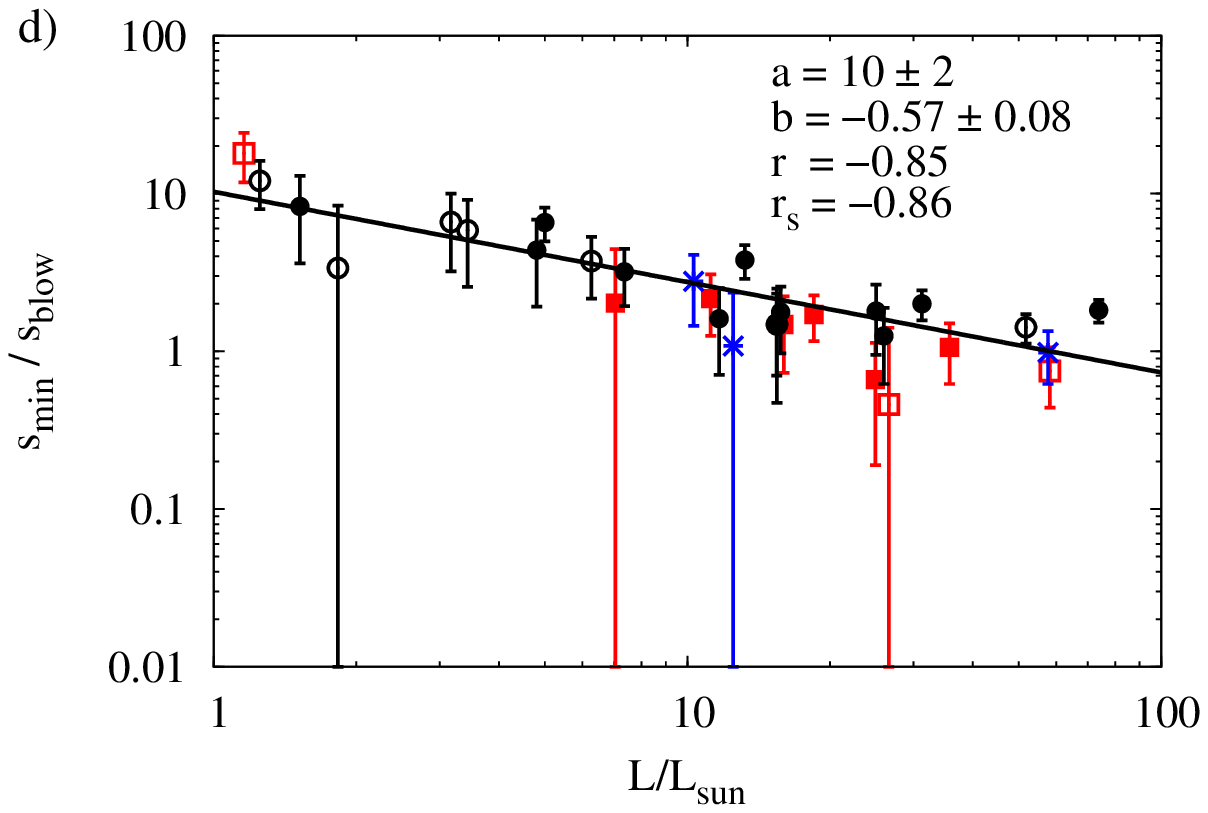}\\
   \includegraphics[width=0.45\textwidth,angle=0]{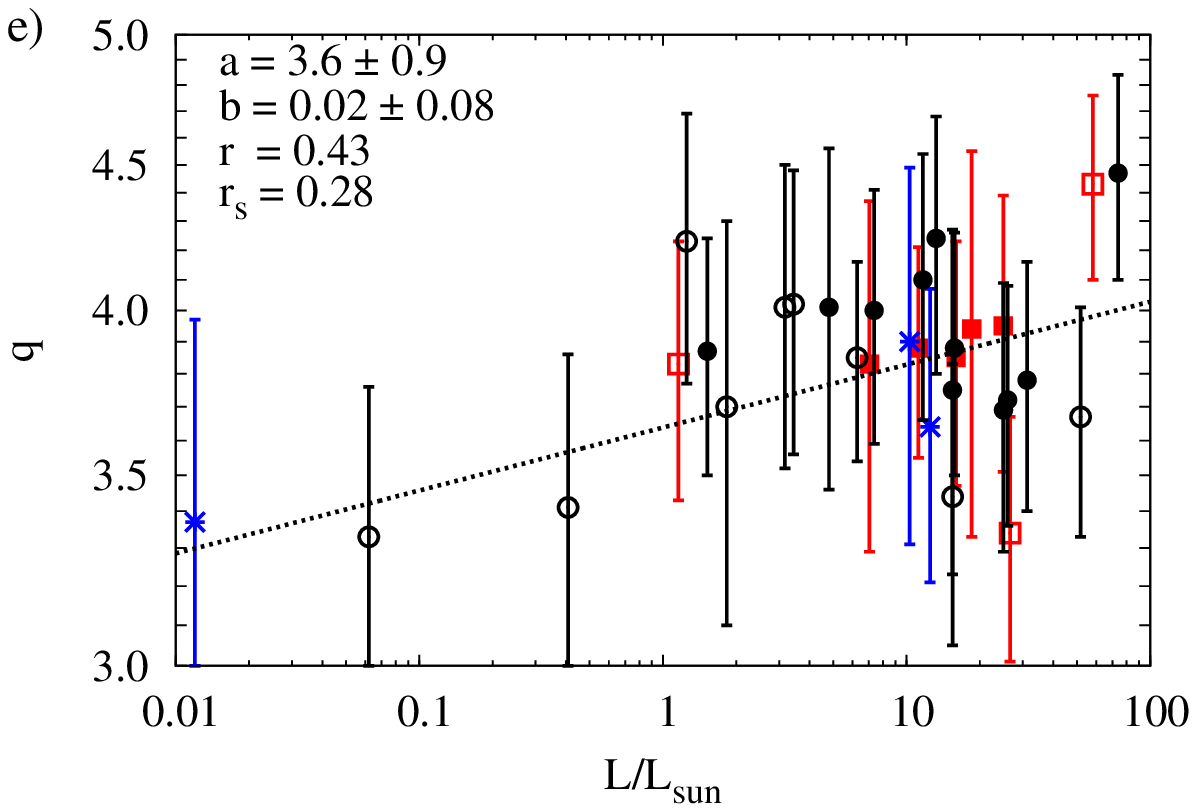}
   \caption
   {   \change{Figure slightly amended}
       Disk parameters obtained with the SD method, plotted as functions of stellar luminosity:
       (a)~dust temperature,
       (b)~the ratio of the dust temperature to the blackbody temperature
       and the ratio of the true disk radius to its blackbody radius,
       (c)~minimum size $\smin$,
       (d)~the ratio of $\smin$ and $\sblow$ for stars with $L > L_\odot$, and
       (e)~the size distribution index $q$.
       The meaning of symbols and fitting lines is the same as in Figure~\ref{fig:Td_smin_MBB}.
    \label{fig:Td_smin}
   }
   \end{figure*}

We now present the results obtained with the second method of the SED fitting,
which assumed a size distribution of particles and their particular emitting
properties (here: astrosilicate).
Figure~\ref{fig:Td_smin} is organized in a similar way as Figure~\ref{fig:Td_smin_MBB},
with the only difference that we now plot
$\smin$ instead of $\lambda_0/(2\pi)$ and
$q$ instead of $\beta$.

The temperature plots, i.e. $\td$ and $\td/\tbb$,
readily show some quantitative differences to the MBB method.
These are already visible at the level of individual stars.
For instance, one of the two M-stars in our sample, GJ~581,
reveals a smaller $\smin$ than earlier-type stars, consistent
with the general trend. However, the dust temperature 
($\td$ in Figure~\ref{fig:Td_smin}a,
$\td/\tbb$ in Figure~\ref{fig:Td_smin}b)
is much lower that one might expect.
This might seem strange, as smaller grains are usually hotter, not colder, than larger ones.
However, this can be easily explained with absorption properties of the astrosilicate,
whose absorption efficiency in the visible (i.e. where the stellar light is absorbed)
drops drastically at small grain sizes. As a result, \revision{small}
particles become colder again \citep{krivov-et-al-2008}.
\revision{For M-stars such as GJ~581, this happens for grains smaller than a few microns.}

Notwithstanding some individual cases, the temperatures derived with
the SD method show the same qualitative trends as those obtained with the MBB.
The dust temperature grows with luminosity (Figure~\ref{fig:Td_smin}a),
the ratio of the dust temperature to the blackbody temperature decreases (Figure~\ref{fig:Td_smin}b), 
and the minimum grain size gets larger towards more luminous stars (Figure~\ref{fig:Td_smin}c).
Judging by the Pearson's correlation coefficient as indicated in the panels,
all of these trends are significant.
Spearman's ranking suggests the same, except that the $\smin$--$L$ correlation has
$r=0.42$ and is only moderately significant at a \revision{1}\% level.
However, quantitative differences from the MBB results are not to be overlooked.
The increase of temperature  towards larger $L/L_\odot$ is now weaker,
the decrease of $\td/\tbb$ stronger,
and $\smin$ (Figure~\ref{fig:Td_smin}c)
increases with $L/L_\odot$ more gently than $\lambda_0/(2\pi)$ does.
As a consequence, the inferred sizes are more consistent
with $\smin/\sblow$ decreasing towards more luminous stars
from $\sim 10$ for solar-type stars to
nearly unity for A-stars (Figure~\ref{fig:Td_smin}d).
Interestingly, this trend is formally the strongest of all correlations found in our paper
($r=-0.85$, \revision{$r_s=-0.86$,} significance $<10^{-8})$!
This needs to be explained, and we will return to this in Sect. 5.

Finally, the slope $q$ lies between 3.3 and 4.5 for nearly all the stars
(Figure~\ref{fig:Td_smin}e), being close to the
``canonical'' value 3.5 \citep{dohnanyi-1969}.
Like $\beta$ in the MBB method, it seems to reveal
a trend of $q$ increasing with the stellar luminosity,
with a marginal significance 
($r=0.43$, $r_s=0.28$).
However, the apparent trend is primarily caused by three individual stars of subsolar luminosity.
Excluding these would lead to $r=0.28$, suggesting that the trend may not be real.
Also, similar to the MBB method, this might trace back to the shape of the $\chi^2$-isolines
discussed in Appendix B.

\subsection{MBB versus SD}

Why are the results of the MBB and SD methods somewhat different, and which of
these are more trustworthy?
The main reason for the differences is that both methods yield SEDs of different shape,
and that shape responds differently to the variation of fitting parameters.
An SED generated with the MBB method consists of a Planck curve at $\lambda < \lambda_0$ and
a steeper Rayleigh-Jeans tail at $\lambda > \lambda_0$, separated by an artificial
knee at $\lambda_0$.
Thus the entire modeled SED can only be narrower than a Planck curve.
In contrast, the SD method assumes a continuous distribution of particle sizes (and thus
temperatures) and is able to yield smooth SEDs that are broader than a single-temperature
blackbody curve (Figure~\ref{fig:SEDs}, bottom panels).
Furthermore, with this method we are at liberty to vary the dust
composition and structure by varying $\Qabs(\lambda,s)$, which makes the fitting
even more flexible.
Altogether, we expect the SD method to be superior to the MBB one.
Nevertheless, we deem it reasonable to present results obtained with both techniques.
This is because the MBB approximation is simple, transparent, and may be more 
\revision{appropriate} for stars with relatively sparse SEDs.
Next, the MBB results can be directly compared
to those derived by other authors, many of which employ this technique.
Finally, a comparison between the MBB and SD results allows us to better judge how
significant our findings are. If one or another trend is evident with both methods,
this can be treated as an additional argument that the trend may be real.

\subsection{Origin of scatter}

Most of the plots presented above reveal an appreciable scatter
amongst the data points. What is the origin of this scatter?
Some of it may reflect real differences between individual systems of any spectral class, but 
much of it likely comes from uncertainties of numerous parameters and measurements 
themselves. These include not only a random, but also a systematic component.
This makes it difficult to find out, for instance, whether a somewhat larger scatter
at higher luminosities reflects a higher degree of dissimilarity of debris disks around more 
luminous stars or is caused by larger uncertainties associated with such stars.

   \begin{figure}[htb!]
   \centering
   \includegraphics[width=0.45\textwidth,angle=0]{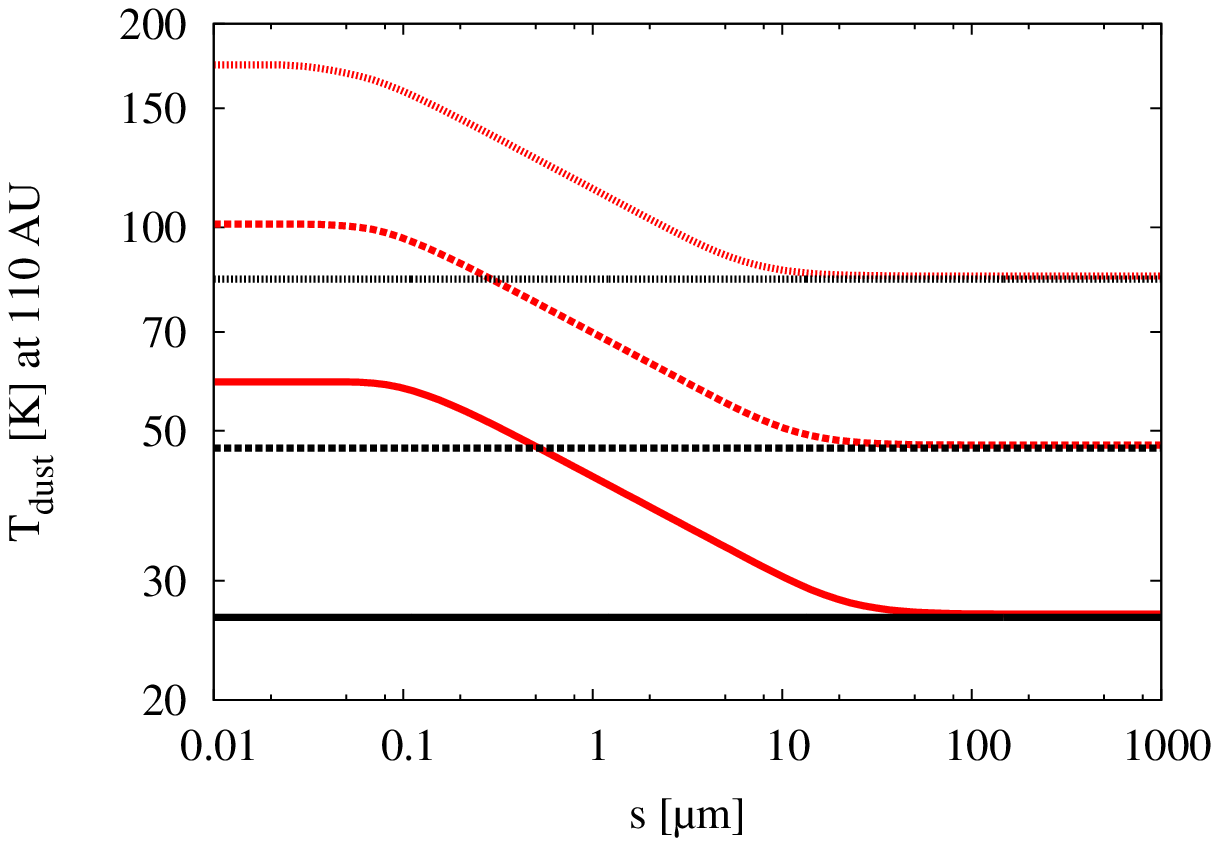}\\
   \includegraphics[width=0.45\textwidth,angle=0]{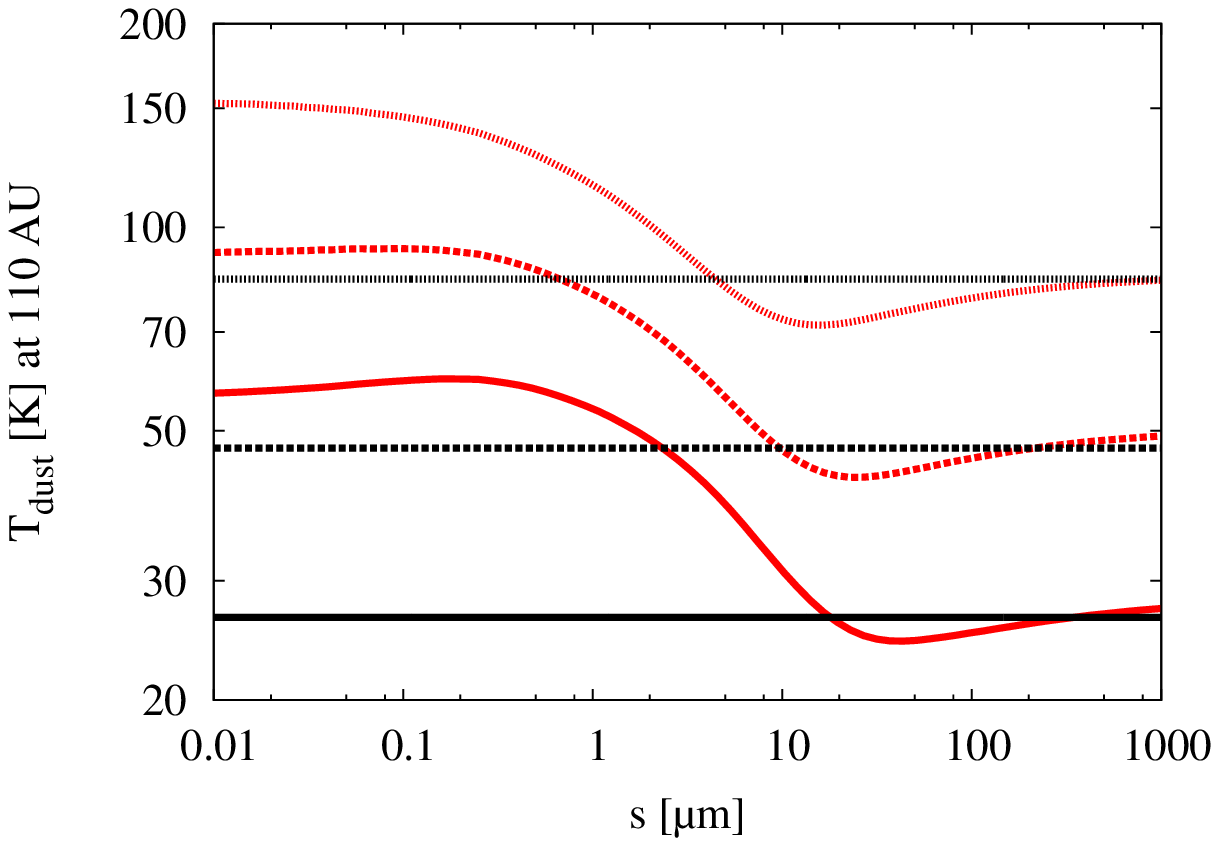}
   \caption{Equilibrium temperature of  dust grains (red lines) and blackbody temperature (black)
   against grain size. Top: MBB dust grains, bottom: astrosilicate particles.
   Solid lines: $L_* = L_\odot$,
   dashed: $L_* = 10 L_\odot$,
   dotted: $L_* = 100 L_\odot$.
   \label{fig:Td_of_s}
   }
   \end{figure}

To illustrate this, let us consider A-stars.
Modeling of their disks uncovers three serious problems.
The first one is a poorer accuracy of disk
radius determination for A-stars, as discussed in Sect. 4.1.
The second problem is that in the case of A-stars the dust temperature turns out to be 
particularly sensitive 
to the set of photometry points used for the fitting (even including or excluding one single point
can matter), and to the fitting method (MBB versus SD).
This is because the IR excesses in the SEDs of A-stars peak at shorter wavelengths than those 
of later-type stars.
Often the SED maximum lies between $\sim 30\mum$ (where the IRS 
spectra end) and  $60$--$70\mum$ (where the MIPS and PACS data start), i.e. at wavelengths 
that have never been probed by any IR instruments.
The third problem is that $\smin$ is also very sensitive to the dust temperature,
especially for A-stars.
This can be understood by looking at Figure~\ref{fig:Td_of_s},
where we plotted the temperature of grains calculated
with two methods. In one case (upper panel), we used the ``MBB grains'',
whose absorption efficiency is given by
Eq.~(\ref{Qabs_MBB}). In another case (lower panel), we assumed the
absorption efficiency of astrosilicate.
For the explanations of the resulting temperature behavior the reader is referred
to \citet{krivov-et-al-2008},
where also $Q_\textrm{abs} (\lambda)$ curves for different-sized grains are plotted.
It is seen that for luminous stars, the grains with radii between several microns
and several tens of microns have temperatures close to $\tbb$.
For astrosilicate grains, $\td$ may even be below $\tbb$.
Thus a small change in $\td$ for those stars would imply a large change in $\smin$.
Therefore, for A-stars we are facing three challenges at the same time:
it is difficult to accurately measure the disk radii from the images,
it is difficult to derive $\td$ from the {\it Herschel} photometry,
and it is difficult then to derive $\smin$ from $\td$.
This makes the results of disks of A-stars intrinsically more uncertain than
those for later-type stars.


\newpage
\section{Discussion}

\subsection{Disk radii and dust sizes}

One thing that we have learned is that the radii of the resolved debris disks
appear uncorrelated with the luminosity of their host stars, and that
there is a large scatter of radii at any given luminosity. This confirms that the
temperature-dependent processes such as ice lines do not play the dominant role
in setting the debris disk dimensions \citep{ballering-et-al-2013}.
However, a conclusion that the planetesimal
formation is not directly related to such processes would be premature.
This is because the location of debris disks depends not only on the location 
of zones where planetesimals preferentially form, but also on the subsequent
complex dynamical and collisional evolution of these planetesimals.
It is thus possible that planetesimal formation mechanisms are temperature-driven,
for instance are largely related to location of ice lines
\citep[e.g.][]{dodson-robinson-et-al-2009b,ros-johansen-2013},
but the emerging debris disks are shaped by processes that are temperature-independent.
The latter may include, for instance, shaping of planetesimal populations by planets,
stirring by various mechanisms, and long-term collisional depletion.

Next, we have confirmed that the dust temperature in Kuiper belt-like disks is higher
around more luminous stars, as found previously
\citep[e.g.][]{ballering-et-al-2013,booth-et-al-2013,eiroa-et-al-2013,chen-et-al-2014}.
However, the temperature rise with the stellar luminosity is gentler than the one we could expect
if the sizes of dust grains were the same of all the stars, as first pointed out by
\citet{booth-et-al-2013}. Instead, the grains in disks of more luminous stars must be
colder, and thus bigger. We attribute this to direct radiation pressure that
swiftly eliminates all grains with sizes $s < \sblow$, where the blowout limit $\sblow$
 is larger at higher stellar luminosities. We have shown that the ratio of the dominant grain
size to the blowout size lies between roughly one and ten at all luminosities.
This confirms theoretical predictions and provides direct evidence that radiation pressure
does play the leading role in setting the size distribution of debris disks.

\subsection{Dust properties}

In Section 4.4, we noted that the SD method uncovered a
significant trend of $\smin/\sblow$ decreasing with $L$.
A similar trend, albeit at a low significance level, was also seen in the MBB results.
This correlation certainly needs to be verified in the future studies over larger
samples. However, should this trend be real, it requires explanation.

One possibility is to attribute it to the dust properties.
Our modeling assumed compact, spherical astrosilicate particles.
However, the reality is expected to be more complex: 
e.g. different mixtures of ices,
silicates, and organics, and different packing factors, around stars of different
spectral types.
Indeed, detailed models of individual debris disks often favor more realistic dust
compositions \citep[e.g.][]{lebreton-et-al-2012,donaldson-et-al-2013}.
This is not any surprise, since the material in debris disks should inherit such a complex 
chemistry and morphology from the preceding, protoplanetary phase 
\citep{dutrey-et-al-2013}.
Another piece of evidence for rich composition of debris material comes from the analysis of 
the oldest debris disk systems, namely from the spectra of debris-polluted white dwarfs
\citep{zuckerman-et-al-2003,gaensicke-et-al-2006,farihi-et-al-2009}.

Changing the grain properties (chemical composition and porosity and thus 
$Q_\textrm{abs}$, $Q_\textrm{pr}$, and $\rho$) may change the results considerably.
For instance, if dust grains are icy, the blowout size changes compared to that
of pure silicate in a non-trivial way
\citep[see, e.g., Figure~2 in][]{reidemeister-et-al-2011}.
For G-type stars, \revision{$\sblow$ remains} nearly unchanged, but for A-stars, it 
decreases considerably.
\revision{However, the icy grains of a given size at a given distance from the star
are colder than silicate ones, and so, we may expect that $\smin$ derived from the SED fitting
will be smaller.
This effect could make the ratio $\smin/\sblow$ more uniform
than Figure~\ref{fig:Td_smin}d suggests.
However, further modeling, which is outside the scope of this paper, is required to find out
how exactly the ratio $\smin/\sblow$ will be affected.}

The grains can also be moderately or even very porous. This, too, would
change $\sblow$ \citep[see, e.g., Figures 7--8 in][]{kirchschlager-wolf-2013}.
Porosity moderately increases $\sblow$ for G-stars.
For A-stars, $\sblow$ will strongly increase compared to compact silicate grains.
\revision{However, $\smin$ will be affected as well, since porous grains have lower 
absorption in the visible
\citep[see, e.g., Figure 3 in][]{kirchschlager-wolf-2013}
and thus a lower temperature than the compact grains.
Modeling is needed here, too, to describe quantitatively the changes in
the ratio $\smin/\sblow$ caused by porosity.}

\subsection{Stirring level}

Besides dust properties, the trends of dust sizes around stars of different luminosities
can also be affected by the following effect.
Assume that, for whatever reasons, the disks around early-type
stars are more strongly stirred than those of later-type ones.
One can indeed expect this, as more luminous stars are more massive,
and may have had more massive protoplanetary disks \citep[e.g.][]{williams-cieza-2011},
which may have been
able to build larger stirring planetesimals \citep[e.g.][]{kenyon-bromley-2008,kobayashi-et-al-2011}.

If this is true, this will modify the inferred size distribution,
shifting the dominant dust
size to smaller values, which are closer to $\sblow$
\citep[e.g.][]{thebault-wu-2008,krivov-et-al-2013}. This could possibly explain
what is seen in Figure~\ref{fig:Td_smin}d.

What is more, the alleged higher stirring level in disks of early-type stars
should also affect the radial distribution of material,
by making them more extended beyond the location of their parent
belts \citep[e.g.][]{thebault-wu-2008,krivov-et-al-2013}.
Conversely, the disks of later-type
stars may tend to stretch inward from the location of the birth
belts. 
In fact, this seems to be supported by observations.
Several disks of A-stars are known to have pronounced halos
and rather sharp inner edges
(examples: Vega,
\citeauthor{su-et-al-2005} \citeyear{su-et-al-2005},
\citeauthor{mueller-et-al-2009} \citeyear{mueller-et-al-2009};
HR~8799,
\citeauthor{su-et-al-2009} \citeyear{su-et-al-2009},
\citeauthor{matthews-et-al-2013b} \citeyear{matthews-et-al-2013b}),
whereas some disks of late-type stars have sharper outer edges
and a leakage of dust into the cavities
(like that of
HD~107146, \citeauthor{ertel-et-al-2011} \citeyear{ertel-et-al-2011},
and
HD 207129, \citeauthor{loehne-et-al-2011} \citeyear{loehne-et-al-2011}).
In that case, the disk radii retrieved from the resolved images
would be \revision{overestimated} around early-type stars and/or \revision{underestimated}
around late-type ones.  In other words, the disks of A-type stars may be smaller than we 
think, while those around FGKM-type stars may be larger. To compensate for this, we would 
need to make the grains in disks of early-type stars cooler (and thus larger),
and grains in disks of later-type stars warmer (and thus smaller).
\revision{Especially for the early-type stars the effect can be significant, i.e.,
a slight overestimation of their disk radii would lead to a strong underestimation of the 
grain sizes. This is because the temperature of disks of A-stars 
is close to the blackbody one, so that a substantial increase of the grain size is needed to 
decrease their temperature only slightly.}

\change{New subsection}
\subsection{The smallest collisional fragments}

All of the previous debris disk models tacitly assumed that collisions produce fragments of
all sizes down to macromolecular ones, and that the minimum size $\smin$ is set by one or 
another physical process that swiftly eliminates the smallest debris from the system.
However, \citet{krijt-kama-2014} recently realized that tiny fragments below a certain 
size may not be even produced. This is because a fraction of the impact energy
available at any collisional event, $\eta$, 
has to be spent to create surfaces of the collisionally produced grains. This provides a 
constraint on the minimum possible grain size which
can be calculated as a function of stellar mass and luminosity, disk radius, 
the fraction $\eta$ just described, the surface tension of the disk material $\gamma$,
and the typical collisional velocity in the units of the local Keplerian speed $f$
(Equation 7 in their paper). The constraint is the strongest (i.e., the minimum size is the 
largest) for the least luminous stars, the largest disks,  
and/or the disks with the lowest dynamical excitation.
Regrettably, most of these parameters, especially $\eta$, $\gamma$, and $f$, are not or 
only poorly known. Nevertheless, we made estimates with the default values assumed by  
\citet{krijt-kama-2014} to find that the minimum fragment size set by the surface energy 
constraint can be on the order of $\sim 10 \sblow$ for solar-type stars, decreasing to
below $\sblow$ for the A-type ones. This is astonishingly close to our results plotted 
in Figure~\ref{fig:Td_smin}d, suggesting this effect as another potential explanation for 
the trends of grain sizes deduced in this work.


\section{Conclusions}

In this paper, we considered a sample of 34 selected debris disks around AFGKM-type stars,
trying to find correlations between the disk parameters and stellar luminosity.
Since these disks are well-resolved, we can measure the disk radii,
thus breaking the degeneracy between the grain size and dust location in 
the SED modeling.
We employ two different methods of the SED fitting:
using the modified blackbody approximation in which the typical grain size is
given by $\lambda_0/(2\pi)$,
and fitting with astrosilicate particles having a size distribution,
which gives the cutoff size $\smin$.
Our key findings are as follows:

{\em Disk radii.}
The disk radii were found to have
a large dispersion for host stars of any spectral class, but no significant
trend with the stellar luminosity is seen.
This finding does not necessarily speak against suggestions that
formation of planetesimals (including those that are large enough to build planets
by core accretion) is largely related to
location of ice lines. It is still possible that planetesimal formation is driven
by temperature-dependent processes, but the final dimensions of debris disks are
determined by the subsequent collisional evolution of planetesimal belts
or their alteration by planetary perturbers, i.e. by processes
that are obviously temperature-indepenent.

{\em Dust temperature.}
The dust temperature systematically increases towards earlier spectral types.
However, the ratio of the dust temperature $\td$ to the blackbody temperature $\tbb$
at the disk radius decreases with the stellar luminosity.

{\em Minimum size of dust grains.}
The decrease of $\td/\tbb$ is explained by an
increase of the typical grain sizes towards more luminous stars.
Such a trend is expected, because radiation pressure exerted on the dust grains should 
get stronger towards earlier spectral types, and we do confirm it with our analysis.
From M- to A-stars, both $\lambda_0/(2\pi)$ and $\smin$
rise roughly from $\sim 0.1\mum$ to $\sim 10\mum$.
Larger grains are colder than smaller ones and have temperatures closer to $\tbb$.
For spectral classes earlier than K where radiation pressure is strong enough
for the blowout limit $\sblow$ to exist, the typical grain size is found to lie between
one and ten times $\sblow$.
This agrees very well with theoretical predictions.
The ratio $\smin/\sblow$ appears to decrease slightly towards A-stars.
Should the latter be confirmed, it might indicate, for instance, that earlier-type stars
have disks that are more excited dynamically and thus are more active collisionally than
those of later-type stars.

{\em Size distribution index.}
\revision{The spectral index of the dust opacity $\beta$ in the modified blackbody treatment is in the range of 0.3 to 2.0
for all the disks in the sample, in accord with previous determinations
based on sub-mm data.}
The index $q$ of the size distribution varies from 3.3 to 4.6.
\revision{This is roughly consistent with what is predicted by 
detailed models of collisional cascade.}


\begin{acknowledgements}
We thank  Amy Bonsor, Grant Kennedy, Torsten L\"ohne,
and Christian Vitense for stimulating discussions.
NP is grateful to Steve Ertel for
useful advice on various aspects of the SED fitting with the thermal annealing 
algorithm.
Insightful and constructive comments of the anonymous referee greatly helped to 
improve the paper.
NP and AVK acknowledge support by the DFG through grant Kr~2164/10-1.
JPM, BM, CE are partly supported by Spanish grant AYA 2011-26202.
This work was also partly supported by the grant OTKA K101393 of the Hungarian Scientific Research Fund,
and the LP2014-6/2014 Lend\"ulet Young  Researchers’ Program.
\end{acknowledgements}


\newcommand{\AAp}      {A\& A}
\newcommand{\AApR}     {Astron. Astrophys. Rev.}
\newcommand{\AApS}    {AApS}
\newcommand{\AApSS}    {AApSS}
\newcommand{\AApT}     {Astron. Astrophys. Trans.}
\newcommand{\AdvSR}    {Adv. Space Res.}
\newcommand{\AJ}       {AJ}
\newcommand{\AN}       {AN}
\newcommand{\AO}       {App. Optics}
\newcommand{\ApJ}      {ApJ}
\newcommand{\ApJL}     {ApJL}
\newcommand{\ApJS}     {ApJS}
\newcommand{\ApSS}     {Astrophys. Space Sci.}
\newcommand{\ARAA}     {ARA\& A}
\newcommand{\ARevEPS}  {Ann. Rev. Earth Planet. Sci.}
\newcommand{\BAAS}     {BAAS}
\newcommand{\CelMech}  {Celest. Mech. Dynam. Astron.}
\newcommand{\EMP}      {Earth, Moon and Planets}
\newcommand{\EPS}      {Earth, Planets and Space}
\newcommand{\GRL}      {Geophys. Res. Lett.}
\newcommand{\JGR}      {J. Geophys. Res.}
\newcommand{\JOSAA}    {J. Opt. Soc. Am. A}
\newcommand{\MemSAI}   {Mem. Societa Astronomica Italiana}
\newcommand{\MNRAS}    {MNRAS}
\newcommand{\PASJ}     {PASJ}
\newcommand{\PASP}     {PASP}
\newcommand{\PSS}      {Planet. Space Sci.}
\newcommand{\RAA}      {Research in Astron. Astrophys.}
\newcommand{\SolPhys}  {Sol. Phys.}
\newcommand{\SolSysRes}{Sol. Sys. Res.}
\newcommand{\SSR}      {Space Sci. Rev.}



\begin{thebibliography}{109}
\expandafter\ifx\csname natexlab\endcsname\relax\def\natexlab#1{#1}\fi

\bibitem[{{Acke} {et~al.}(2012){Acke}, {Min}, {Dominik}, {Vandenbussche},
  {Sibthorpe}, {Waelkens}, {Olofsson}, {Degroote}, {Smolders}, {Pantin},
  {Barlow}, {Blommaert}, {Brandeker}, {De Meester}, {Dent}, {Exter}, {Di
  Francesco}, {Fridlund}, {Gear}, {Glauser}, {Greaves}, {Harvey}, {Henning},
  {Hogerheijde}, {Holland}, {Huygen}, {Ivison}, {Jean}, {Liseau}, {Naylor},
  {Pilbratt}, {Polehampton}, {Regibo}, {Royer}, {Sicilia-Aguilar}, \&
  {Swinyard}}]{acke-et-al-2012}
{Acke}, B., {Min}, M., {Dominik}, C., {et~al.} 2012, \AAp, 540, A125

\bibitem[{{Allende Prieto} {et~al.}(1999){Allende Prieto}, {Garc{\'{\i}}a
  L{\'o}pez}, {Lambert}, \& {Gustafsson}}]{allende-prieto-et-al-1999}
{Allende Prieto}, C., {Garc{\'{\i}}a L{\'o}pez}, R.~J., {Lambert}, D.~L., \&
  {Gustafsson}, B. 1999, \ApJ, 527, 879

\bibitem[{Backman \& Paresce(1993)}]{backman-paresce-1993}
Backman, D., \& Paresce, F. 1993, in Protostars and Planets III, ed. E.~H. Levy
  \& J.~I. Lunine (Univ. of Arizona Press), 1253--1304

\bibitem[{{Backman} {et~al.}(2009){Backman}, {Marengo}, {Stapelfeldt}, {Su},
  {Wilner}, {Dowell}, {Watson}, {Stansberry}, {Rieke}, {Megeath}, {Fazio}, \&
  {Werner}}]{backman-et-al-2009}
{Backman}, D., {Marengo}, M., {Stapelfeldt}, K., {et~al.} 2009, \ApJ, 690, 1522

\bibitem[{{Ballering} {et~al.}(2013){Ballering}, {Rieke}, {Su}, \&
  {Montiel}}]{ballering-et-al-2013}
{Ballering}, N.~P., {Rieke}, G.~H., {Su}, K.~Y.~L., \& {Montiel}, E. 2013,
  \ApJ, 775, 55

\bibitem[{{Balog} {et~al.}(2013){Balog}, {M{\"u}ller}, {Nielbock}, {Altieri},
  {Klaas}, {Blommaert}, {Linz}, {Lutz}, {Mo{\'o}r}, {Billot}, {Sauvage}, \&
  {Okumura}}]{balog-et-al-2013}
{Balog}, Z., {M{\"u}ller}, T., {Nielbock}, M., {et~al.} 2013, Experimental
  Astronomy, arXiv:1309.6099

\bibitem[{{Bessell}(1979)}]{bessell-1979}
{Bessell}, M.~S. 1979, \PASP, 91, 589

\bibitem[{{Boley} {et~al.}(2012){Boley}, {Payne}, {Corder}, {Dent}, {Ford}, \&
  {Shabram}}]{boley-et-al-2012}
{Boley}, A.~C., {Payne}, M.~J., {Corder}, S., {et~al.} 2012, \ApJ, 750, L21

\bibitem[{{Bonsor} {et~al.}(2013){Bonsor}, {Kennedy}, {Crepp}, {Johnson},
  {Wyatt}, {Sibthorpe}, \& {Su}}]{bonsor-et-al-2013b}
{Bonsor}, A., {Kennedy}, G.~M., {Crepp}, J.~R., {et~al.} 2013, \mnras, 431,
  3025

\bibitem[{{Booth} {et~al.}(2013){Booth}, {Kennedy}, {Sibthorpe}, {Matthews},
  {Wyatt}, {Duch{\^e}ne}, {Kavelaars}, {Rodriguez}, {Greaves}, {Koning},
  {Vican}, {Rieke}, {Su}, {Moro-Mart{\'{\i}}n}, \& {Kalas}}]{booth-et-al-2013}
{Booth}, M., {Kennedy}, G., {Sibthorpe}, B., {et~al.} 2013, \MNRAS, 428, 1263

\bibitem[{{Broekhoven-Fiene} {et~al.}(2013){Broekhoven-Fiene}, {Matthews},
  {Kennedy}, {Booth}, {Sibthorpe}, {Lawler}, {Kavelaars}, {Wyatt}, {Qi},
  {Koning}, {Su}, {Rieke}, {Wilner}, \& {Greaves}}]{broekhovenfiene-et-al-2013}
{Broekhoven-Fiene}, H., {Matthews}, B.~C., {Kennedy}, G.~M., {et~al.} 2013,
  \ApJ, 762, 52

\bibitem[{{Brott} \& {Hauschildt}(2005)}]{brott-hauschildt-2005}
{Brott}, I., \& {Hauschildt}, P.~H. 2005, in ESA SP, Vol. 576, The
  Three-Dimensional Universe with Gaia, ed. {C.~Turon, K.~S.~O'Flaherty, \&
  M.~A.~C.~Perryman}, 565

\bibitem[{Burns {et~al.}(1979)Burns, Lamy, \& Soter}]{burns-et-al-1979}
Burns, J.~A., Lamy, P.~L., \& Soter, S. 1979, Icarus, 40, 1

\bibitem[{{Castelli} \& {Kurucz}(2004)}]{castelli-kurucz-2004}
{Castelli}, F., \& {Kurucz}, R.~L. 2004, ArXiv:astro-ph/0405087

\bibitem[{{Chen} {et~al.}(2014){Chen}, {Mittal}, {Kuchner}, {Forrest}, {Lisse},
  {Manoj}, {Sargent}, \& {Watson}}]{chen-et-al-2014}
{Chen}, C.~H., {Mittal}, T., {Kuchner}, M., {et~al.} 2014, \ApJS, 211, 25

\bibitem[{{Chen} {et~al.}(2012){Chen}, {Pecaut}, {Mamajek}, {Su}, \&
  {Bitner}}]{chen-et-al-2012}
{Chen}, C.~H., {Pecaut}, M., {Mamajek}, E.~E., {Su}, K.~Y.~L., \& {Bitner}, M.
  2012, \apj, 756, 133

\bibitem[{{Churcher} {et~al.}(2011){Churcher}, {Wyatt}, \&
  {Smith}}]{churcher-et-al-2011}
{Churcher}, L., {Wyatt}, M., \& {Smith}, R. 2011, \MNRAS, 410, 2

\bibitem[{{Cohen} {et~al.}(2003){Cohen}, {Wheaton}, \&
  {Megeath}}]{cohen-et-al-2003}
{Cohen}, M., {Wheaton}, W.~A., \& {Megeath}, S.~T. 2003, \AJ, 126, 1090

\bibitem[{{Cutri} {et~al.}(2003){Cutri}, {Skrutskie}, {van Dyk}, {Beichman},
  {Carpenter}, {Chester}, {Cambresy}, {Evans}, {Fowler}, {Gizis}, {Howard},
  {Huchra}, {Jarrett}, {Kopan}, {Kirkpatrick}, {Light}, {Marsh}, {McCallon},
  {Schneider}, {Stiening}, {Sykes}, {Weinberg}, {Wheaton}, {Wheelock}, \&
  {Zacarias}}]{cutri-et-al-2003}
{Cutri}, R.~M., {Skrutskie}, M.~F., {van Dyk}, S., {et~al.} 2003, {2MASS All
  Sky Catalog of point sources.}

\bibitem[{{Dodson-Robinson} {et~al.}(2009){Dodson-Robinson}, {Willacy},
  {Bodenheimer}, {Turner}, \& {Beichman}}]{dodson-robinson-et-al-2009b}
{Dodson-Robinson}, S.~E., {Willacy}, K., {Bodenheimer}, P., {Turner}, N.~J., \&
  {Beichman}, C.~A. 2009, Icarus, 200, 672

\bibitem[{Dohnanyi(1969)}]{dohnanyi-1969}
Dohnanyi, J.~S. 1969, \JGR, 74, 2531

\bibitem[{{Donaldson} {et~al.}(2013){Donaldson}, {Lebreton}, {Roberge},
  {Augereau}, \& {Krivov}}]{donaldson-et-al-2013}
{Donaldson}, J.~K., {Lebreton}, J., {Roberge}, A., {Augereau}, J.-C., \&
  {Krivov}, A.~V. 2013, \ApJ, 772, 17

\bibitem[{{Draine}(2003)}]{draine-2003}
{Draine}, B.~T. 2003, \ARAA, 41, 241

\bibitem[{{Duch{\^e}ne} {et~al.}(2014){Duch{\^e}ne}, {Arriaga}, {Wyatt},
  {Kennedy}, {Sibthorpe}, {Lisse}, {Holland}, {Wisniewski}, {Clampin}, {Kalas},
  {Pinte}, {Wilner}, {Booth}, {Horner}, {Matthews}, \&
  {Greaves}}]{duchene-et-al-2014}
{Duch{\^e}ne}, G., {Arriaga}, P., {Wyatt}, M., {et~al.} 2014, \ApJ, 784, 148

\bibitem[{{Dutrey} {et~al.}(2014){Dutrey}, {Semenov}, Chapillon, {X}, {Y}, \&
  {Z}}]{dutrey-et-al-2013}
{Dutrey}, A., {Semenov}, D., Chapillon, E., {et~al.} 2014, in {Protostars and
  Planets VI}, ed. H.~Beuther, R.~Klessen, C.~Dullemond, \& T.~Henning (Univ.
  of Arizona Press), in press, arXiv:1402.3503

\bibitem[{{Eiroa} {et~al.}(2013){Eiroa}, {Marshall}, {Mora}, {Montesinos},
  {Absil}, {Augereau}, {Bayo}, {Bryden}, {Danchi}, {del Burgo}, {Ertel},
  {Fridlund}, {Heras}, {Krivov}, {Launhardt}, {Liseau}, {L{\"o}hne},
  {Maldonado}, {Pilbratt}, {Roberge}, {Rodmann}, {Sanz-Forcada}, {Solano},
  {Stapelfeldt}, {Th{\'e}bault}, {Wolf}, {Ardila}, {Ar{\'e}valo}, {Beichmann},
  {Faramaz}, {Gonz{\'a}lez-Garc{\'{\i}}a}, {Guti{\'e}rrez}, {Lebreton},
  {Mart{\'{\i}}nez-Arn{\'a}iz}, {Meeus}, {Montes}, {Olofsson}, {Su}, {White},
  {Barrado}, {Fukagawa}, {Gr{\"u}n}, {Kamp}, {Lorente}, {Morbidelli},
  {M{\"u}ller}, {Mutschke}, {Nakagawa}, {Ribas}, \&
  {Walker}}]{eiroa-et-al-2013}
{Eiroa}, C., {Marshall}, J.~P., {Mora}, A., {et~al.} 2013, \AAp, 555, A11

\bibitem[{{Ertel} {et~al.}(2011){Ertel}, {Wolf}, {Metchev}, {Schneider},
  {Carpenter}, {Meyer}, {Hillenbrand}, \& {Silverstone}}]{ertel-et-al-2011}
{Ertel}, S., {Wolf}, S., {Metchev}, S., {et~al.} 2011, \AAp, 533, A132

\bibitem[{{Ertel} {et~al.}(2014){Ertel}, {Marshall}, {Augereau}, {Krivov},
  {L{\"o}hne}, {Eiroa}, {Mora}, {del Burgo}, {Montesinos}, {Bryden}, {Danchi},
  {Kirchschlager}, {Liseau}, {Maldonado}, {Pilbratt}, {Sch{\"u}ppler},
  {Th{\'e}bault}, {White}, \& {Wolf}}]{ertel-et-al-2013}
{Ertel}, S., {Marshall}, J.~P., {Augereau}, J.-C., {et~al.} 2014, \AAp, 561,
  A114

\bibitem[{{Farihi} {et~al.}(2009){Farihi}, {Jura}, \&
  {Zuckerman}}]{farihi-et-al-2009}
{Farihi}, J., {Jura}, M., \& {Zuckerman}, B. 2009, \ApJ, 694, 805

\bibitem[{{Frisch} {et~al.}(2011){Frisch}, {Redfield}, \&
  {Slavin}}]{frisch-et-al-2011}
{Frisch}, P.~C., {Redfield}, S., \& {Slavin}, J.~D. 2011, \ARAA, 49, 237

\bibitem[{{G{\"a}nsicke} {et~al.}(2006){G{\"a}nsicke}, {Marsh}, {Southworth},
  \& {Rebassa-Mansergas}}]{gaensicke-et-al-2006}
{G{\"a}nsicke}, B.~T., {Marsh}, T.~R., {Southworth}, J., \&
  {Rebassa-Mansergas}, A. 2006, Science, 314, 1908

\bibitem[{{Golimowski} {et~al.}(2006){Golimowski}, {Ardila}, {Krist},
  {Clampin}, {Ford}, {Illingworth}, {Bartko}, {Ben{\'{\i}}tez}, {Blakeslee},
  {Bouwens}, {Bradley}, {Broadhurst}, {Brown}, {Burrows}, {Cheng}, {Cross},
  {Demarco}, {Feldman}, {Franx}, {Goto}, {Gronwall}, {Hartig}, {Holden},
  {Homeier}, {Infante}, {Jee}, {Kimble}, {Lesser}, {Martel}, {Mei},
  {Menanteau}, {Meurer}, {Miley}, {Motta}, {Postman}, {Rosati}, {Sirianni},
  {Sparks}, {Tran}, {Tsvetanov}, {White}, {Zheng}, \&
  {Zirm}}]{golimowski-et-al-2006}
{Golimowski}, D.~A., {Ardila}, D.~R., {Krist}, J.~E., {et~al.} 2006, \AJ, 131,
  3109

\bibitem[{{Gray}(1992)}]{gray-1992}
{Gray}, D.~F. 1992, {The observation and analysis of stellar photospheres.},
  ed. {Gray, D.~F.}

\bibitem[{{Gray} {et~al.}(2006){Gray}, {Corbally}, {Garrison}, {McFadden},
  {Bubar}, {McGahee}, {O'Donoghue}, \& {Knox}}]{gray-et-al-2006}
{Gray}, R.~O., {Corbally}, C.~J., {Garrison}, R.~F., {et~al.} 2006, \AJ, 132,
  161

\bibitem[{{Gray} {et~al.}(2003){Gray}, {Corbally}, {Garrison}, {McFadden}, \&
  {Robinson}}]{gray-et-al-2003}
{Gray}, R.~O., {Corbally}, C.~J., {Garrison}, R.~F., {McFadden}, M.~T., \&
  {Robinson}, P.~E. 2003, \AJ, 126, 2048

\bibitem[{{Griffin} {et~al.}(2010){Griffin}, {Abergel}, {Abreu}, {Ade},
  {Andr{\'e}}, {Augueres}, {Babbedge}, {Bae}, {Baillie}, {Baluteau}, {Barlow},
  {Bendo}, {Benielli}, {Bock}, {Bonhomme}, {Brisbin}, {Brockley-Blatt},
  {Caldwell}, {Cara}, {Castro-Rodriguez}, {Cerulli}, {Chanial}, {Chen},
  {Clark}, {Clements}, {Clerc}, {Coker}, {Communal}, {Conversi}, {Cox},
  {Crumb}, {Cunningham}, {Daly}, {Davis}, {de Antoni}, {Delderfield}, {Devin},
  {di Giorgio}, {Didschuns}, {Dohlen}, {Donati}, {Dowell}, {Dowell}, {Duband},
  {Dumaye}, {Emery}, {Ferlet}, {Ferrand}, {Fontignie}, {Fox}, {Franceschini},
  {Frerking}, {Fulton}, {Garcia}, {Gastaud}, {Gear}, {Glenn}, {Goizel},
  {Griffin}, {Grundy}, {Guest}, {Guillemet}, {Hargrave}, {Harwit}, {Hastings},
  {Hatziminaoglou}, {Herman}, {Hinde}, {Hristov}, {Huang}, {Imhof}, {Isaak},
  {Israelsson}, {Ivison}, {Jennings}, {Kiernan}, {King}, {Lange}, {Latter},
  {Laurent}, {Laurent}, {Leeks}, {Lellouch}, {Levenson}, {Li}, {Li},
  {Lilienthal}, {Lim}, {Liu}, {Lu}, {Madden}, {Mainetti}, {Marliani}, {McKay},
  {Mercier}, {Molinari}, {Morris}, {Moseley}, {Mulder}, {Mur}, {Naylor},
  {Nguyen}, {O'Halloran}, {Oliver}, {Olofsson}, {Olofsson}, {Orfei}, {Page},
  {Pain}, {Panuzzo}, {Papageorgiou}, {Parks}, {Parr-Burman}, {Pearce},
  {Pearson}, {P{\'e}rez-Fournon}, {Pinsard}, {Pisano}, {Podosek}, {Pohlen},
  {Polehampton}, {Pouliquen}, {Rigopoulou}, {Rizzo}, {Roseboom}, {Roussel},
  {Rowan-Robinson}, {Rownd}, {Saraceno}, {Sauvage}, {Savage}, {Savini},
  {Sawyer}, {Scharmberg}, {Schmitt}, {Schneider}, {Schulz}, {Schwartz},
  {Shafer}, {Shupe}, {Sibthorpe}, {Sidher}, {Smith}, {Smith}, {Smith},
  {Spencer}, {Stobie}, {Sudiwala}, {Sukhatme}, {Surace}, {Stevens}, {Swinyard},
  {Trichas}, {Tourette}, {Triou}, {Tseng}, {Tucker}, {Turner}, {Vaccari},
  {Valtchanov}, {Vigroux}, {Virique}, {Voellmer}, {Walker}, {Ward}, {Waskett},
  {Weilert}, {Wesson}, {White}, {Whitehouse}, {Wilson}, {Winter}, {Woodcraft},
  {Wright}, {Xu}, {Zavagno}, {Zemcov}, {Zhang}, \&
  {Zonca}}]{griffin-et-al-2010}
{Griffin}, M.~J., {Abergel}, A., {Abreu}, A., {et~al.} 2010, \AAp, 518, L3

\bibitem[{{Holmberg} {et~al.}(2009){Holmberg}, {Nordstr{\"o}m}, \&
  {Andersen}}]{holmberg-et-al-2009}
{Holmberg}, J., {Nordstr{\"o}m}, B., \& {Andersen}, J. 2009, \AAp, 501, 941

\bibitem[{{Ishihara} {et~al.}(2010){Ishihara}, {Onaka}, {Kataza}, {Salama},
  {Alfageme}, {Cassatella}, {Cox}, {Garc{\'{\i}}a-Lario}, {Stephenson},
  {Cohen}, {Fujishiro}, {Fujiwara}, {Hasegawa}, {Ita}, {Kim}, {Matsuhara},
  {Murakami}, {M{\"u}ller}, {Nakagawa}, {Ohyama}, {Oyabu}, {Pyo}, {Sakon},
  {Shibai}, {Takita}, {Tanab{\'e}}, {Uemizu}, {Ueno}, {Usui}, {Wada},
  {Watarai}, {Yamamura}, \& {Yamauchi}}]{ishihara-et-al-2010}
{Ishihara}, D., {Onaka}, T., {Kataza}, H., {et~al.} 2010, \aap, 514, A1

\bibitem[{{Johansen} {et~al.}(2014){Johansen}, {Blum}, {Tanaka}, {Ormel},
  {Bizzarro}, \& {Rickman}}]{johansen-et-al-2014}
{Johansen}, A., {Blum}, J., {Tanaka}, H., {et~al.} 2014, in {Protostars and
  Planets VI}, ed. H.~Beuther, R.~Klessen, C.~Dullemond, \& T.~Henning (Univ.
  of Arizona Press), in press, arXiv:1402.1344

\bibitem[{{Kennedy} \& {Wyatt}(2010)}]{kennedy-wyatt-2010}
{Kennedy}, G.~M., \& {Wyatt}, M.~C. 2010, \MNRAS, 405, 1253

\bibitem[{{Kennedy} {et~al.}(2012){Kennedy}, {Wyatt}, {Sibthorpe}, {Phillips},
  {Matthews}, \& {Greaves}}]{kennedy-et-al-2012b}
{Kennedy}, G.~M., {Wyatt}, M.~C., {Sibthorpe}, B., {et~al.} 2012, \MNRAS, 426,
  2115

\bibitem[{{Kenyon} \& {Bromley}(2008)}]{kenyon-bromley-2008}
{Kenyon}, S.~J., \& {Bromley}, B.~C. 2008, \ApJS, 179, 451

\bibitem[{{Kirchschlager} \& {Wolf}(2013)}]{kirchschlager-wolf-2013}
{Kirchschlager}, F., \& {Wolf}, S. 2013, \aap, 552, A54

\bibitem[{{Kobayashi} {et~al.}(2011){Kobayashi}, {Tanaka}, \&
  {Krivov}}]{kobayashi-et-al-2011}
{Kobayashi}, H., {Tanaka}, H., \& {Krivov}, A.~V. 2011, \ApJ, 738, 35

\revision{
\bibitem[{{Krijt} \& {Kama}(2014)}]{krijt-kama-2014}
{Krijt}, S., \& {Kama}, M. 2014, \AAp, 566, L2
}

\bibitem[{{Krist} {et~al.}(2012){Krist}, {Stapelfeldt}, {Bryden}, \&
  {Plavchan}}]{krist-et-al-2012}
{Krist}, J.~E., {Stapelfeldt}, K.~R., {Bryden}, G., \& {Plavchan}, P. 2012,
  \AJ, 144, 45

\bibitem[{Krivov {et~al.}(2006)Krivov, L\"ohne, \&
  Srem\v{c}evi\'c}]{krivov-et-al-2006}
Krivov, A.~V., L\"ohne, T., \& Srem\v{c}evi\'c, M. 2006, \AAp, 455, 509

\bibitem[{Krivov {et~al.}(2008)Krivov, M\"uller, L\"ohne, \&
  Mutschke}]{krivov-et-al-2008}
Krivov, A.~V., M\"uller, S., L\"ohne, T., \& Mutschke, H. 2008, \ApJ, 687, 608

\bibitem[{{Krivov} {et~al.}(2013){Krivov}, {Eiroa}, {L{\"o}hne}, {Marshall},
  {Montesinos}, {del Burgo}, {Absil}, {Ardila}, {Augereau}, {Bayo}, {Bryden},
  {Danchi}, {Ertel}, {Lebreton}, {Liseau}, {Mora}, {Mustill}, {Mutschke},
  {Neuh{\"a}user}, {Pilbratt}, {Roberge}, {Schmidt}, {Stapelfeldt},
  {Th{\'e}bault}, {Vitense}, {White}, \& {Wolf}}]{krivov-et-al-2013}
{Krivov}, A.~V., {Eiroa}, C., {L{\"o}hne}, T., {et~al.} 2013, \ApJ, 772, 32

\bibitem[{{Kuchner} \& {Stark}(2010)}]{kuchner-stark-2010}
{Kuchner}, M.~J., \& {Stark}, C.~C. 2010, \AJ, 140, 1007

\revision{
\bibitem[{{Lebouteiller} {et~al.}(2011){Lebouteiller}, {Barry}, {Spoon},
  {Bernard-Salas}, {Sloan}, {Houck}, \& {Weedman}}]{lebouteiller-et-al-2011}
{Lebouteiller}, V., {Barry}, D.~J., {Spoon}, H.~W.~W., {et~al.} 2011, \ApJS,
  196, 8
}

\bibitem[{{Lebreton} {et~al.}(2012){Lebreton}, {Augereau}, {Thi}, {Roberge},
  {Donaldson}, {Schneider}, {Maddison}, {M{\'e}nard}, {Riviere-Marichalar},
  {Mathews}, {Kamp}, {Pinte}, {Dent}, {Barrado}, {Duch{\^e}ne}, {Gonzalez},
  {Grady}, {Meeus}, {Pantin}, {Williams}, \& {Woitke}}]{lebreton-et-al-2012}
{Lebreton}, J., {Augereau}, J.-C., {Thi}, W.-F., {et~al.} 2012, \AAp, 539, A17

\revision{
\bibitem[{{Lebreton} {et~al.}(2013){Lebreton}, {van Lieshout}, {Augereau},
  {Absil}, {Mennesson}, {Kama}, {Dominik}, {Bonsor}, {Vandeportal}, {Beust},
  {Defr{\`e}re}, {Ertel}, {Faramaz}, {Hinz}, {Kral}, {Lagrange}, {Liu}, \&
  {Th{\'e}bault}}]{lebreton-et-al-2013}
{Lebreton}, J., {van Lieshout}, R., {Augereau}, J.-C., {et~al.} 2013, \AAp,
  555, A146
}

\bibitem[{{Lestrade} {et~al.}(2012){Lestrade}, {Matthews}, {Sibthorpe},
  {Kennedy}, {Wyatt}, {Bryden}, {Greaves}, {Thilliez}, {Moro-Mart{\'{\i}}n},
  {Booth}, {Dent}, {Duch{\^e}ne}, {Harvey}, {Horner}, {Kalas}, {Kavelaars},
  {Phillips}, {Rodriguez}, {Su}, \& {Wilner}}]{lestrade-et-al-2012}
{Lestrade}, J.-F., {Matthews}, B.~C., {Sibthorpe}, B., {et~al.} 2012, \AAp,
  548, A86

\bibitem[{{Liseau} {et~al.}(2008){Liseau}, {Risacher}, {Brandeker}, {Eiroa},
  {Fridlund}, {Nilsson}, {Olofsson}, {Pilbratt}, \&
  {Th{\'e}bault}}]{liseau-et-al-2008}
{Liseau}, R., {Risacher}, C., {Brandeker}, A., {et~al.} 2008, \AAp, 480, L47

\bibitem[{{Liseau} {et~al.}(2010){Liseau}, {Eiroa}, {Fedele}, {Augereau},
  {Olofsson}, {Gonz{\'a}lez}, {Maldonado}, {Montesinos}, {Mora}, {Absil},
  {Ardila}, {Barrado}, {Bayo}, {Beichman}, {Bryden}, {Danchi}, {Del Burgo},
  {Ertel}, {Fridlund}, {Heras}, {Krivov}, {Launhardt}, {Lebreton}, {L{\"o}hne},
  {Marshall}, {Meeus}, {M{\"u}ller}, {Pilbratt}, {Roberge}, {Rodmann},
  {Solano}, {Stapelfeldt}, {Th{\'e}bault}, {White}, \&
  {Wolf}}]{liseau-et-al-2010}
{Liseau}, R., {Eiroa}, C., {Fedele}, D., {et~al.} 2010, \AAp, 518, L132

\bibitem[{L\"ohne {et~al.}(2008)L\"ohne, Krivov, \&
  Rodmann}]{loehne-et-al-2007}
L\"ohne, T., Krivov, A.~V., \& Rodmann, J. 2008, \ApJ, 673, 1123

\bibitem[{L\"ohne {et~al.}(2012)L\"ohne, Augereau, Ertel, Marshall, Eiroa,
  Mora, Absil, Stapelfeldt, Th\'ebault, {del Burgo}, Danchi, Krivov, Lebreton,
  Letawe, Magain, Maldonado, Montesinos, Pilbratt, White, \&
  Wolf}]{loehne-et-al-2011}
L\"ohne, T., Augereau, J.-C., Ertel, S., {et~al.} 2012, \AAp, 537, A110

\bibitem[{{Marshall} {et~al.}(2011){Marshall}, {L{\"o}hne}, {Montesinos},
  {Krivov}, {Eiroa}, {Absil}, {Bryden}, {Maldonado}, {Mora}, {Sanz-Forcada},
  {Ardila}, {Augereau}, {Bayo}, {Del Burgo}, {Danchi}, {Ertel}, {Fedele},
  {Fridlund}, {Lebreton}, {Gonz{\'a}lez-Garc{\'{\i}}a}, {Liseau}, {Meeus},
  {M{\"u}ller}, {Pilbratt}, {Roberge}, {Stapelfeldt}, {Th{\'e}bault}, {White},
  \& {Wolf}}]{marshall-et-al-2011}
{Marshall}, J.~P., {L{\"o}hne}, T., {Montesinos}, B., {et~al.} 2011, \AAp, 529,
  A117

\bibitem[{{Matthews} {et~al.}(2014{\natexlab{a}}){Matthews}, {Kennedy},
  {Sibthorpe}, {Booth}, {Wyatt}, {Broekhoven-Fiene}, {Macintosh}, \&
  {Marois}}]{matthews-et-al-2013b}
{Matthews}, B., {Kennedy}, G., {Sibthorpe}, B., {et~al.} 2014{\natexlab{a}},
  \apj, 780, 97

\bibitem[{{Matthews} {et~al.}(2014{\natexlab{b}}){Matthews}, {Krivov}, {Wyatt},
  {Bryden}, \& {Eiroa}}]{matthews-et-al-2013}
{Matthews}, B.~C., {Krivov}, A.~V., {Wyatt}, M.~C., {Bryden}, G., \& {Eiroa},
  C. 2014{\natexlab{b}}, in {Protostars and Planets VI}, ed. H.~Beuther,
  R.~Klessen, C.~Dullemond, \& T.~Henning (U. Arizona Press), in press,
  arXiv:1401.0743

\bibitem[{{Matthews} {et~al.}(2010){Matthews}, {Sibthorpe}, {Kennedy},
  {Phillips}, {Churcher}, {Duch{\^e}ne}, {Greaves}, {Lestrade}, {Moro-Martin},
  {Wyatt}, {Bastien}, {Biggs}, {Bouvier}, {Butner}, {Dent}, {di Francesco},
  {Eisl{\"o}ffel}, {Graham}, {Harvey}, {Hauschildt}, {Holland}, {Horner},
  {Ibar}, {Ivison}, {Johnstone}, {Kalas}, {Kavelaars}, {Rodriguez}, {Udry},
  {van der Werf}, {Wilner}, \& {Zuckerman}}]{matthews-et-al-2010}
{Matthews}, B.~C., {Sibthorpe}, B., {Kennedy}, G., {et~al.} 2010, \AAp, 518,
  L135

\bibitem[{{Mo{\'o}r} {et~al.}(2011){Mo{\'o}r}, {{\'A}brah{\'a}m}, {Juh{\'a}sz},
  {Kiss}, {Pascucci}, {K{\'o}sp{\'a}l}, {Apai}, {Henning}, {Csengeri}, \&
  {Grady}}]{moor-et-al-2011}
{Mo{\'o}r}, A., {{\'A}brah{\'a}m}, P., {Juh{\'a}sz}, A., {et~al.} 2011, \ApJL,
  740, L7

\bibitem[{{Mo{\'o}r} {et~al.}(2013{\natexlab{a}}){Mo{\'o}r}, {{\'A}brah{\'a}m},
  {K{\'o}sp{\'a}l}, {Szab{\'o}}, {Apai}, {Balog}, {Csengeri}, {Grady},
  {Henning}, {Juh{\'a}sz}, {Kiss}, {Pascucci}, {Szul{\'a}gyi}, \&
  {Vavrek}}]{moor-et-al-2013}
{Mo{\'o}r}, A., {{\'A}brah{\'a}m}, P., {K{\'o}sp{\'a}l}, {\'A}., {et~al.}
  2013{\natexlab{a}}, \ApJL, 775, L51

\bibitem[{{Mo{\'o}r} {et~al.}(2013{\natexlab{b}}){Mo{\'o}r}, {Juh{\'a}sz},
  {K{\'o}sp{\'a}l}, {{\'A}brah{\'a}m}, {Apai}, {Csengeri}, {Grady}, {Henning},
  {Hughes}, {Kiss}, {Pascucci}, {Schmalzl}, \&
  {Gab{\'a}nyi}}]{moor-et-al-2013b}
{Mo{\'o}r}, A., {Juh{\'a}sz}, A., {K{\'o}sp{\'a}l}, {\'A}., {et~al.}
  2013{\natexlab{b}}, \ApJL, 777, L25

\bibitem[{{Morales} {et~al.}(2013){Morales}, {Bryden}, {Werner}, \&
  {Stapelfeldt}}]{morales-et-al-2013}
{Morales}, F.~Y., {Bryden}, G., {Werner}, M.~W., \& {Stapelfeldt}, K.~R. 2013,
  \ApJ, 776, 111

\bibitem[{{Morales} {et~al.}(2011){Morales}, {Rieke}, {Werner}, {Bryden},
  {Stapelfeldt}, \& {Su}}]{morales-et-al-2011}
{Morales}, F.~Y., {Rieke}, G.~H., {Werner}, M.~W., {et~al.} 2011, \ApJL, 730,
  L29

\bibitem[{{Morales} {et~al.}(2009){Morales}, {Werner}, {Bryden}, {Plavchan},
  {Stapelfeldt}, {Rieke}, {Su}, {Beichman}, {Chen}, {Grogan}, {Kenyon},
  {Moro-Martin}, \& {Wolf}}]{morales-et-al-2009}
{Morales}, F.~Y., {Werner}, M.~W., {Bryden}, G., {et~al.} 2009, \ApJ, 699, 1067

\bibitem[{M\"uller {et~al.}(2010)M\"uller, L\"ohne, \&
  Krivov}]{mueller-et-al-2009}
M\"uller, S., L\"ohne, T., \& Krivov, A.~V. 2010, \ApJ, 708, 1728

\bibitem[{{Mustill} \& {Wyatt}(2009)}]{mustill-wyatt-2009}
{Mustill}, A.~J., \& {Wyatt}, M.~C. 2009, \MNRAS, 399, 1403

\bibitem[{{Najita} \& {Williams}(2005)}]{najita-williams-2005}
{Najita}, J., \& {Williams}, J.~P. 2005, \ApJ, 635, 625

\bibitem[{{Nilsson} {et~al.}(2010){Nilsson}, {Liseau}, {Brandeker}, {Olofsson},
  {Pilbratt}, {Risacher}, {Rodmann}, {Augereau}, {Bergman}, {Eiroa},
  {Fridlund}, {Th{\'e}bault}, \& {White}}]{nilsson-et-al-2010}
{Nilsson}, R., {Liseau}, R., {Brandeker}, A., {et~al.} 2010, \AAp, 518, A40

\bibitem[{{Ott}(2010)}]{ott-2010}
{Ott}, S. 2010, in ASP Conf. Series, Vol. 434, Astronomical Data Analysis
  Software and Systems XIX, ed. Y.~{Mizumoto}, K.-I. {Morita}, \& M.~{Ohishi},
  139

\bibitem[{{Paunzen} {et~al.}(2006){Paunzen}, {Schnell}, \&
  {Maitzen}}]{paunzen-et-al-2006}
{Paunzen}, E., {Schnell}, A., \& {Maitzen}, H.~M. 2006, \aap, 458, 293

\bibitem[{{Pilbratt} {et~al.}(2010){Pilbratt}, {Riedinger}, {Passvogel},
  {Crone}, {Doyle}, {Gageur}, {Heras}, {Jewell}, {Metcalfe}, {Ott}, \&
  {Schmidt}}]{pilbratt-et-al-2010}
{Pilbratt}, G.~L., {Riedinger}, J.~R., {Passvogel}, T., {et~al.} 2010, \AAp,
  518, L1

\bibitem[{{Plavchan} {et~al.}(2005){Plavchan}, {Jura}, \&
  {Lipscy}}]{plavchan-et-al-2005}
{Plavchan}, P., {Jura}, M., \& {Lipscy}, S.~J. 2005, \ApJ, 631, 1161

\bibitem[{{Poglitsch} {et~al.}(2010){Poglitsch}, {Waelkens}, {Geis},
  {Feuchtgruber}, {Vandenbussche}, {Rodriguez}, {Krause}, {Renotte}, {van
  Hoof}, {Saraceno}, {Cepa}, {Kerschbaum}, {Agn{\`e}se}, {Ali}, {Altieri},
  {Andreani}, {Augueres}, {Balog}, {Barl}, {Bauer}, {Belbachir}, {Benedettini},
  {Billot}, {Boulade}, {Bischof}, {Blommaert}, {Callut}, {Cara}, {Cerulli},
  {Cesarsky}, {Contursi}, {Creten}, {De Meester}, {Doublier}, {Doumayrou},
  {Duband}, {Exter}, {Genzel}, {Gillis}, {Gr{\"o}zinger}, {Henning},
  {Herreros}, {Huygen}, {Inguscio}, {Jakob}, {Jamar}, {Jean}, {de Jong},
  {Katterloher}, {Kiss}, {Klaas}, {Lemke}, {Lutz}, {Madden}, {Marquet},
  {Martignac}, {Mazy}, {Merken}, {Montfort}, {Morbidelli}, {M{\"u}ller},
  {Nielbock}, {Okumura}, {Orfei}, {Ottensamer}, {Pezzuto}, {Popesso},
  {Putzeys}, {Regibo}, {Reveret}, {Royer}, {Sauvage}, {Schreiber}, {Stegmaier},
  {Schmitt}, {Schubert}, {Sturm}, {Thiel}, {Tofani}, {Vavrek}, {Wetzstein},
  {Wieprecht}, \& {Wiezorrek}}]{poglitsch-et-al-2010}
{Poglitsch}, A., {Waelkens}, C., {Geis}, N., {et~al.} 2010, \AAp, 518, L2

\bibitem[{Press {et~al.}(2002)Press, Teukolsky, Vetterling, \&
  Flannery}]{numerical-recipes-2002}
Press, W.~H., Teukolsky, S.~A., Vetterling, W.~T., \& Flannery, B.~P. 2002,
  Numerical Recipes in C++ (Cambridge University Press), 1032 pp.

\bibitem[{{Raymond} {et~al.}(2011){Raymond}, {Armitage}, {Moro-Mart{\'{\i}}n},
  {Booth}, {Wyatt}, {Armstrong}, {Mandell}, {Selsis}, \&
  {West}}]{raymond-et-al-2011}
{Raymond}, S.~N., {Armitage}, P.~J., {Moro-Mart{\'{\i}}n}, A., {et~al.} 2011,
  \AAp, 530, A62

\bibitem[{Reidemeister {et~al.}(2011)Reidemeister, Krivov, Stark, Augereau,
  L\"ohne, \& M\"uller}]{reidemeister-et-al-2011}
Reidemeister, M., Krivov, A.~V., Stark, C.~C., {et~al.} 2011, \AAp, 527, A57

\bibitem[{{Reis} {et~al.}(2011){Reis}, {Corradi}, {de Avillez}, \&
  {Santos}}]{reis-et-al-2011}
{Reis}, W., {Corradi}, W., {de Avillez}, M.~A., \& {Santos}, F.~P. 2011, \ApJ,
  734, 8

\bibitem[{{Rieke} \& {Lebofsky}(1985)}]{rieke-lebofsky-1985}
{Rieke}, G.~H., \& {Lebofsky}, M.~J. 1985, \ApJ, 288, 618

\bibitem[{{Roberge} {et~al.}(2006){Roberge}, {Feldman}, {Weinberger},
  {Deleuil}, \& {Bouret}}]{roberge-et-al-2006}
{Roberge}, A., {Feldman}, P.~D., {Weinberger}, A.~J., {Deleuil}, M., \&
  {Bouret}, J.-C. 2006, Nature, 441, 724

\bibitem[{{Roberge} {et~al.}(2013){Roberge}, {Kamp}, {Montesinos}, {Dent},
  {Meeus}, {Donaldson}, {Olofsson}, {Mo{\'o}r}, {Augereau}, {Howard}, {Eiroa},
  {Thi}, {Ardila}, {Sandell}, \& {Woitke}}]{roberge-et-al-2013}
{Roberge}, A., {Kamp}, I., {Montesinos}, B., {et~al.} 2013, \ApJ, 771, 69

\bibitem[{{Rodriguez} \& {Zuckerman}(2012)}]{rodriguez-zuckerman-2012}
{Rodriguez}, D.~R., \& {Zuckerman}, B. 2012, \apj, 745, 147

\bibitem[{{Ros} \& {Johansen}(2013)}]{ros-johansen-2013}
{Ros}, K., \& {Johansen}, A. 2013, \AAp, 552, A137

\bibitem[{{Sheret} {et~al.}(2004){Sheret}, {Dent}, \&
  {Wyatt}}]{sheret-et-al-2004}
{Sheret}, I., {Dent}, W.~R.~F., \& {Wyatt}, M.~C. 2004, \MNRAS, 348, 1282

\bibitem[{{Sibthorpe} {et~al.}(2010){Sibthorpe}, {Vandenbussche}, {Greaves},
  {Pantin}, {Olofsson}, {Acke}, {Barlow}, {Blommaert}, {Bouwman}, {Brandeker},
  {Cohen}, {De Meester}, {Dent}, {di Francesco}, {Dominik}, {Fridlund}, {Gear},
  {Glauser}, {Gomez}, {Hargrave}, {Harvey}, {Henning}, {Heras}, {Hogerheijde},
  {Holland}, {Ivison}, {Leeks}, {Lim}, {Liseau}, {Matthews}, {Naylor},
  {Pilbratt}, {Polehampton}, {Regibo}, {Royer}, {Sicilia-Aguilar}, {Swinyard},
  {Waelkens}, {Walker}, \& {Wesson}}]{sibthorpe-et-al-2010}
{Sibthorpe}, B., {Vandenbussche}, B., {Greaves}, J.~S., {et~al.} 2010, \AAp,
  518, L130

\bibitem[{{Stock} {et~al.}(2010){Stock}, {Su}, {Liu}, {Hinz}, {Rieke},
  {Marengo}, {Stapelfeldt}, {Hines}, \& {Trilling}}]{stock-et-al-2010}
{Stock}, N.~D., {Su}, K.~Y.~L., {Liu}, W., {et~al.} 2010, \apj, 724, 1238

\bibitem[{{Su} {et~al.}(2005){Su}, {Rieke}, {Misselt}, {Stansberry},
  {Moro-Mart\'in}, {Stapelfeldt}, {Werner}, {Trilling}, {Bendo}, {Gordon},
  {Hines}, {Wyatt}, {Holland}, {Marengo}, {Megeath}, \&
  {Fazio}}]{su-et-al-2005}
{Su}, K.~Y.~L., {Rieke}, G.~H., {Misselt}, K.~A., {et~al.} 2005, \ApJ, 628, 487

\bibitem[{{Su} {et~al.}(2006){Su}, {Rieke}, {Stansberry}, {Bryden},
  {Stapelfeldt}, {Trilling}, {Muzerolle}, {Beichman}, {Moro-Martin}, {Hines},
  \& {Werner}}]{su-et-al-2006}
{Su}, K.~Y.~L., {Rieke}, G.~H., {Stansberry}, J.~A., {et~al.} 2006, \ApJ, 653,
  675

\bibitem[{{Su} {et~al.}(2009){Su}, {Rieke}, {Stapelfeldt}, {Malhotra},
  {Bryden}, {Smith}, {Misselt}, {Moro-Martin}, \& {Williams}}]{su-et-al-2009}
{Su}, K.~Y.~L., {Rieke}, G.~H., {Stapelfeldt}, K.~R., {et~al.} 2009, \ApJ, 705,
  314

\revision{
\bibitem[{{Su} {et~al.}(2013){Su}, {Rieke}, {Malhotra}, {Stapelfeldt},
  {Hughes}, {Bonsor}, {Wilner}, {Balog}, {Watson}, {Werner}, \&
  {Misselt}}]{su-et-al-2013}
{Su}, K.~Y.~L., {Rieke}, G.~H., {Malhotra}, R., {et~al.} 2013, \ApJ, 763, 118
}

\bibitem[{{Th{\' e}bault} \& {Augereau}(2007)}]{thebault-augereau-2007}
{Th{\' e}bault}, P., \& {Augereau}, J.-C. 2007, \AAp, 472, 169

\bibitem[{{Th{\' e}bault} \& {Wu}(2008)}]{thebault-wu-2008}
{Th{\' e}bault}, P., \& {Wu}, Y. 2008, \AAp, 481, 713

\bibitem[{{Torres}(2010)}]{torres-2010}
{Torres}, G. 2010, \AJ, 140, 1158

\bibitem[{{Trilling} {et~al.}(2007){Trilling}, {Stansberry}, {Stapelfeldt},
  {Rieke}, {Su}, {Gray}, {Corbally}, {Bryden}, {Chen}, {Boden}, \&
  {Beichman}}]{trilling-et-al-2007}
{Trilling}, D.~E., {Stansberry}, J.~A., {Stapelfeldt}, K.~R., {et~al.} 2007,
  \ApJ, 658, 1289

\bibitem[{{Vitense} {et~al.}(2010){Vitense}, {Krivov}, \&
  {L\"ohne}}]{vitense-et-al-2010}
{Vitense}, C., {Krivov}, A.~V., \& {L\"ohne}, T. 2010, \AAp, 520, A32

\bibitem[{{von Braun} {et~al.}(2011){von Braun}, {Boyajian}, {Kane}, {van
  Belle}, {Ciardi}, {L{\'o}pez-Morales}, {McAlister}, {Henry}, {Jao}, {Riedel},
  {Subasavage}, {Schaefer}, {ten Brummelaar}, {Ridgway}, {Sturmann},
  {Sturmann}, {Mazingue}, {Turner}, {Farrington}, {Goldfinger}, \&
  {Boden}}]{braun-et-al-2011}
{von Braun}, K., {Boyajian}, T.~S., {Kane}, S.~R., {et~al.} 2011, \ApJL, 729,
  L26

\bibitem[{{Wahhaj} {et~al.}(2007){Wahhaj}, {Koerner}, \&
  {Sargent}}]{wahhaj-et-al-2007}
{Wahhaj}, Z., {Koerner}, D.~W., \& {Sargent}, A.~I. 2007, \ApJ, 661, 368

\bibitem[{{Williams} \& {Andrews}(2006)}]{williams-andrews-2006}
{Williams}, J.~P., \& {Andrews}, S.~M. 2006, \ApJ, 653, 1480

\bibitem[{{Williams} \& {Cieza}(2011)}]{williams-cieza-2011}
{Williams}, J.~P., \& {Cieza}, L.~A. 2011, \ARAA, 49, 67

\bibitem[{{Wright} {et~al.}(2010){Wright}, {Eisenhardt}, {Mainzer}, {Ressler},
  {Cutri}, {Jarrett}, {Kirkpatrick}, {Padgett}, {McMillan}, {Skrutskie},
  {Stanford}, {Cohen}, {Walker}, {Mather}, {Leisawitz}, {Gautier}, {McLean},
  {Benford}, {Lonsdale}, {Blain}, {Mendez}, {Irace}, {Duval}, {Liu}, {Royer},
  {Heinrichsen}, {Howard}, {Shannon}, {Kendall}, {Walsh}, {Larsen}, {Cardon},
  {Schick}, {Schwalm}, {Abid}, {Fabinsky}, {Naes}, \&
  {Tsai}}]{wright-et-al-2010}
{Wright}, E.~L., {Eisenhardt}, P.~R.~M., {Mainzer}, A.~K., {et~al.} 2010, \AJ,
  140, 1868

\bibitem[{{Wyatt}(2008)}]{wyatt-2008}
{Wyatt}, M.~C. 2008, \ARAA, 46, 339

\bibitem[{{Wyatt} {et~al.}(2011){Wyatt}, Clarke, \& {Booth}}]{wyatt-et-al-2011}
{Wyatt}, M.~C., Clarke, C.~J., \& {Booth}, M. 2011, \CelMech, 111, 1

\bibitem[{{Wyatt} {et~al.}(2012){Wyatt}, {Kennedy}, {Sibthorpe}, {Moro-Martin},
  {Lestrade}, {Ivison}, {Matthews}, {Udry}, {Greaves}, {Kalas}, {Lawler}, {Su},
  {Rieke}, {Booth}, {Bryden}, {Horner}, {Kavelaars}, \&
  {Wilner}}]{wyatt-et-al-2012}
{Wyatt}, M.~C., {Kennedy}, G., {Sibthorpe}, B., {et~al.} 2012, \MNRAS, 424,
  1206

\bibitem[{{Zuckerman} {et~al.}(2003){Zuckerman}, {Koester}, {Reid}, \&
  {H{\"u}nsch}}]{zuckerman-et-al-2003}
{Zuckerman}, B., {Koester}, D., {Reid}, I.~N., \& {H{\"u}nsch}, M. 2003, \ApJ,
  596, 477

\end{thebibliography}



\appendix

\section{A. Photometry of the systems}

Table~\ref{tab:disks_MIR_phot} lists the
mid-IR photometry data for all the systems in our sample.
Table~\ref{tab:disks_FIR_phot} does the same for
the far-IR and sub-millimeter photometry.

\begin{turnpage}
 \begin{table}[htb!!]
 \caption{\change{More photometry points added}Mid-IR Photometry Used in Creating the Target SEDs}
 {\scriptsize
 \begin{tabular}{rllllllll}
 \hline
  Target (HD) & \multicolumn{8}{c}{} \\
\hline
           &\multicolumn{7}{c}{Flux [mJy]} \\
Wavelength 	 	& 11.56$\mu$m 		& 16$\mu$m		& 18$\mu$m 		& 22$\mu$m		& 22.09$\mu$m 		& 24$\mu$m		& 31$\mu$m		& Note\\ 
Instrument 		& b 			& a 			& c 			& a 			& b			& d 			& a			&     \\
\hline\hline
GJ 581			& 213~$\pm$~19		& 108~$\pm$~1		& \ldots		& 58.8~$\pm$~1.0	& \ldots		& \ldots		& 31.4~$\pm$~1.3 	& (3) \\
197481			& 543~$\pm$~60		& \ldots		& 246~$\pm$~36		& 183~$\pm$~21		& \ldots		& 158~$\pm$~3		& \ldots		&     \\
23484			& 314.4~$\pm$~5.2	& 172~$\pm$~6		& \ldots		& 90.3~$\pm$~8.4	& 90.4~$\pm$~2.0	& 80.47~$\pm$~5.09	& 56.6~$\pm$~15.4	& (2) \\
104860			& 81.1~$\pm$~1.1	& 44.49~$\pm$~2.93	& \ldots		& 24.73~$\pm$~3.53	& 23.3~$\pm$~1.1	& 21.54~$\pm$~2.91	& 13.36~$\pm$~3.10	&     \\
207129			& 635~$\pm$~8		& 371~$\pm$~5		& 262~$\pm$~31		& 197~$\pm$~4		& 196~$\pm$~5		& 155~$\pm$~3		& 114~$\pm$~5		& (2) \\
10647			& 600.2~$\pm$~7.7	& \ldots		& 315.3~$\pm$~39.7	& \ldots		& 218.1~$\pm$~4.0	& 184.8~$\pm$~3.8	& 185.4~$\pm$~7.3	&     \\
48682			& 774~$\pm$~11		& 425~$\pm$~12		& 458~$\pm$~17		& 240~$\pm$~5		& 246~$\pm$~5		& 193~$\pm$~4		& 152~$\pm$~4  		& (2) \\
50571			& 303~$\pm$~4		& 152~$\pm$~4	 	& 129~$\pm$~12		& 82.7~$\pm$~5.0	& 85.4~$\pm$~1.9	& 70.4~$\pm$~2.8	& 48.0~$\pm$~4.0	& (2) \\
170773			& 262~$\pm$~3		& 136~$\pm$~5		& \ldots		& 78.4~$\pm$~5.7	& 76.4~$\pm$~2.2	& 65.3~$\pm$~2.6	& 54.6~$\pm$~6.0	& (2) \\
218396			& 258~$\pm$~4		& 140~$\pm$~2		& 119~$\pm$~80		& 91.3~$\pm$~2.3	& 94.7~$\pm$~2.8	& 86.6~$\pm$~1.7	& 68.0~$\pm$~2.3	& (2) \\
109085			& 1460~$\pm$~70		& \ldots		& 820~$\pm$~20		& \ldots		& 680~$\pm$~40		& 590~$\pm$~20		& \ldots		& (4) \\
27290			& 1269.12~$\pm$~9.35	& 704.8~$\pm$~12.7	& \ldots		& 263.3~$\pm$~8.9	& 367.76~$\pm$~7.11	& 315.6~$\pm$~3.2	& \ldots		&     \\
95086			& 62.7~$\pm$~3.0	& \ldots		& \ldots		& 46.7~$\pm$~7.3	& 51.6~$\pm$~3.3	& 45.6~$\pm$~2.0	& 96.7~$\pm$~10.0	& (1) \\
195627			& 761~$\pm$~10		& 439.6~$\pm$~8.3	& 325~$\pm$~13		& 248.5~$\pm$~6.7	& 237~$\pm$~5		& 186~$\pm$~7		& 159.46~$\pm$~4.85	&     \\
20320			& \ldots		& 333.5~$\pm$~6.6	& 263.5~$\pm$~56.0	& 186.8~$\pm$~5.0	& \ldots		& 162.6~$\pm$~4.2	& 109.18~$\pm$~2.94	&     \\
21997			& 106.6~$\pm$~5.2	& \ldots		& \ldots		& \ldots		& 57.2~$\pm$~3.7	& 55.1~$\pm$~2.2	& 92.3~$\pm$~10.3	& (1) \\
110411			& 247.4~$\pm$~7.2	& \ldots		& 204.2~$\pm$~18.3	& 162.8~$\pm$~6.0	& \ldots		& 149.7~$\pm$~3.9	& \ldots		&     \\
142091			& \ldots		& \ldots		& \ldots		& \ldots		& 890.4~$\pm$~13.1	& 800.1~$\pm$~8.0	& \ldots		&     \\
102647			& \ldots		& 2989~$\pm$~48		& \ldots		& 1724~$\pm$~30		& \ldots		& 1647~$\pm$~33		& \ldots		&     \\
125162			& \ldots		& 442~$\pm$~5		& 415.7~$\pm$~15.6	& 310~$\pm$~6		& \ldots		& 270.8~$\pm$~2.3	& 276~$\pm$~6		& (2) \\
216956			& \ldots		& 6947~$\pm$~112	& 5338~$\pm$~81.8	& 3940~$\pm$~64		& \ldots		& 3502~$\pm$~64		& \ldots		& (4) \\
17848			& 315~$\pm$~4		& 177.3~$\pm$~3.6	& 105~$\pm$~10		& 99.82~$\pm$~2.75	& 102~$\pm$~2		& 88.68~$\pm$~2.44	& 68.26~$\pm$~1.87	&     \\
9672			& 211~$\pm$~21		& \ldots		& \ldots		& \ldots		& 238~$\pm$~24		& 259~$\pm$~10		& 426~$\pm$~1		& (1) \\
71722			& 141~$\pm$~2		& 85.67~$\pm$~1.83	& \ldots		& 65.16~$\pm$~1.81	& 63.4~$\pm$~2.1	& 62.53~$\pm$~1.73	& 76.70~$\pm$~1.94	&     \\
182681			& 180~$\pm$~2		& \ldots		& \ldots		& \ldots		& 123~$\pm$~3		& \ldots		& \ldots		&     \\
14055			& \ldots		& 443~$\pm$~11		& 372.2~$\pm$~22.1	& 315~$\pm$~6		& \ldots		& 282.66~$\pm$~6.63	& 317~$\pm$~6		& (2) \\
161868			& 1160~$\pm$~15		& 643.6~$\pm$~12.6	& \ldots		& 463.3~$\pm$~11.4	& 474~$\pm$~9		& 438.5~$\pm$~10.9	& \ldots		&     \\
188228			& \ldots		& 385~$\pm$~12		& 296.6~$\pm$~31.0	& 206.8~$\pm$~7.3	& \ldots		& 170.75~$\pm$~0.55	& 128~$\pm$~14		& (2) \\
10939			& 322~$\pm$~4		& 177~$\pm$~4		& \ldots		& 116~$\pm$~5		& 121~$\pm$~3		& 104~$\pm$~2		& 114~$\pm$~5		& (2) \\
71155			& \ldots		& 455~$\pm$~10		& 398.4~$\pm$~19.5	& 337~$\pm$~9		& \ldots		& 307~$\pm$~3		& 296~$\pm$~4		& (2) \\
172167			& \ldots		& \ldots		& \ldots		& \ldots		& \ldots		& 8900~$\pm$~89		& \ldots		& (4) \\
139006			& \ldots		& 2331~$\pm$~39		& 1823~$\pm$~38		& 1375~$\pm$~25		& \ldots		& 1261.63~$\pm$~15.46	& 964.16~$\pm$~7.07	&     \\
95418			& 3283~$\pm$~27		& 1770~$\pm$~42		& 1546~$\pm$~21		& 1060~$\pm$~31		& 1167~$\pm$~25		& 1026~$\pm$~14		& 743~$\pm$~26		& (2) \\
13161			&\ldots			& 1530~$\pm$~34		& 1141~$\pm$~31		& 896~$\pm$~19		& \ldots		& 791.5~$\pm$~16.0	& 669~$\pm$~24		& (2) \\
 \end{tabular}
}
 \label{tab:disks_MIR_phot} 
 
 \noindent
 
 {\em Notes:}\\[0mm]
All values and error bars are taken from the literature as they were published.
For instance, {\it WISE} error bars for different objects may or may not include the calibration uncertainty.\\
(1) For these stars, more photometry points extracted from the IRS spectra have been published.
Although not given in the table, these were included in the SED fitting.\\
\revision{(2) For these stars, we took the IRS spectra from CASSIS \citep{lebouteiller-et-al-2011}.
The spectra of objects taken in mapping mode were reduced as in \citet{moor-et-al-2013,moor-et-al-2013b}.
For those observed in low-resolution staring mode we made additional steps 
(outlier detection, module stitching) as described in Moor et al. (in prep.).
In both cases, we then took averages in the wavelength ranges
15--17, 21--23, and 30-32 microns to extract the photometry points.\\
(3) This object was observed in high resolution staring mode. Since no CASSIS spectrum is available, we
used its Spitzer Science Center IRS Enhanced Product, performed
the usual post-processing steps, and then extracted the photometry points.\\
(4) For these stars, IRS spectra in different observing modes have been taken
\citep{lebreton-et-al-2013,su-et-al-2013,chen-et-al-2014,duchene-et-al-2014}.
However, the resulting fluxes are not given in the literature, except for two photometry points
for HD~216956. We decided not to reduce the data by ourselves
because of the large spatial extent and brightness of the objects (saturation effects).
}

{\em Instruments:} (a) {\it Spitzer}/IRS; (b) {\it WISE};  (c) {\it AKARI}; (d){\it Spitzer}/MIPS.

{\em References:} WISE from \cite{wright-et-al-2010, moor-et-al-2013, moor-et-al-2013b, bonsor-et-al-2013b};
{\it Spitzer}/IRS from \cite{moor-et-al-2013, moor-et-al-2013b, roberge-et-al-2013, chen-et-al-2014};
AKARI from \cite{ishihara-et-al-2010};
{\it Spitzer}/MIPS from \cite{su-et-al-2006,trilling-et-al-2007,chen-et-al-2012, eiroa-et-al-2013, moor-et-al-2013, moor-et-al-2013b, bonsor-et-al-2013b}.
 
 \end{table}
\end{turnpage}

\begin{turnpage}
 
\begin{table}[htb!!]
\caption{Far-IR and Sub-mm Photometry Used in Creating the Target SEDs}
{\scriptsize
\begin{tabular}{rllllllllll}
\hline
Target (HD)& \multicolumn{10}{c}{} \\
\hline
           &\multicolumn{10}{c}{Flux [mJy]} \\
Wavelength & 70$\mu$m 		& 70$\mu$m 		& 100$\mu$m 		& 160$\mu$m 		& 250$\mu$m 		& 350$\mu$m 		& 450$\mu$m 		& 500$\mu$m 		& 850$\mu$m 		& 870$\mu$m \\ 
Instrument & a 			& b 			& b 			& b 			& c 			& c 			& d 			& c 			& d 			& e \\
\hline\hline
GJ 581	& 20.0~$\pm$~5.3 	& 18.9~$\pm$~1.4 	& 21.5~$\pm$~1.5 	& 22.2~$\pm$~5.0 	& $<$~24.0 		& $<$~26.0 		& \ldots 		& $<$~27.0 		& \ldots 		& \ldots 		\\
197481  & 227~$\pm$~27          & 231.3~$\pm$~16.3	& \ldots		& 243~$\pm$~17		& 134~$\pm$~8		& 84.4~$\pm$~5.4	& 85~$\pm$~42	 	& 47.6~$\pm$~3.8	& 14.4~$\pm$~1.8	& \ldots		\\
23484	& 99.1~$\pm$~8.4 	& 74.5~$\pm$~3.8 	& 91.3~$\pm$~4.7 	& 91.9~$\pm$~4.9 	& 53.0~$\pm$~10.4 	& 32.2~$\pm$~8.9 	& \ldots 		& $<$~21.6 		& \ldots 		& \ldots 		\\ 
104860	& 183.1~$\pm$~7.4 	& \ldots 		& 277.0~$\pm$~3.5 	& 243.4~$\pm$~5.2 	& \ldots 		& 50.1~$\pm$~15.0	& 47.0~$\pm$~14.0 	& \ldots 		& 6.8~$\pm$~1.2 	& \ldots 		\\
207129	& 278.2~$\pm$~21.5 	& 284.0~$\pm$~1.5 	& 311~$\pm$~1	 	& 211.0~$\pm$~1.5 	& 113~$\pm$~18	 	& 44.3~$\pm$~9.0 	& \ldots 		& 25.9~$\pm$~8.0 	& \ldots 		&  5.1~$\pm$~2.7 	\\
10647	& 863.4~$\pm$~58.7 	& 896.2~$\pm$~26.9 	& 897.1~$\pm$~26.9 	& 635.9~$\pm$~31.8 	& 312.3~$\pm$~25.6 	& 179.9~$\pm$~14.6 	& \ldots 		& 78.4~$\pm$~9.8 	& \ldots 		& 39.4~$\pm$~4.1 	\\
48682	& 262.8~$\pm$~18.3 	& 264.0~$\pm$~4.1 	& 252.3~$\pm$~3.2 	& 182.1~$\pm$~3.8 	& 90~$\pm$~15	 	& 25~$\pm$~8	 	& \ldots 		& $<$~24.0 		& 5.5~$\pm$~1.1 	& \ldots 		\\ 
50571	& 248.8~$\pm$~18.7 	& 223.9~$\pm$~17.4 	& 262.9~$\pm$~19.7 	& 188.5~$\pm$~16.7 	& 71.1~$\pm$~7.3 	& 46.2~$\pm$~7.0 	& \ldots 		& 13.6~$\pm$~7.4	& \ldots 		& \ldots 		\\
170773	& 787.9~$\pm$~56.0 	& 806.7~$\pm$~56.8 	& 1109.9~$\pm$~78.3 	& 875.2~$\pm$~61.5 	& 379.1~$\pm$~21.6 	& 167.8~$\pm$~11.2 	& \ldots 		& 73.9~$\pm$~7.3 	& \ldots 		& 18.0~$\pm$~5.4 	\\
218396	& 610.0~$\pm$~31.0 	& 537~$\pm$~15	 	& 687~$\pm$~20	 	& 570~$\pm$~50	 	& 309~$\pm$~30 		& 163~$\pm$~30 		& \ldots 		& 74~$\pm$~30 		& 10.3~$\pm$~1.8 	& \ldots 		\\
109085  & 198~$\pm$~7		& 230~$\pm$~13		& 252~$\pm$~16		& 231~$\pm$~13		& \ldots		& \ldots		& 58~$\pm$~10		& \ldots		& 15.5~$\pm$~1.4	& \ldots		\\
27290	& 170.7~$\pm$~8.1 	& 171.0~$\pm$~8.7 	& 148.4~$\pm$~7.7 	& 134.3~$\pm$~14.1 	& 52.5~$\pm$~6.5 	& 23.5~$\pm$~8.0 	& \ldots 		& $<$~16.7 		& \ldots 		& \ldots 		\\
95086	& 654.6~$\pm$~44.4 	& 690.1~$\pm$~48.6 	& 675.1~$\pm$~47.6 	& 462.4~$\pm$~32.7 	& 213.4~$\pm$~12.9 	& 120.3~$\pm$~8.7 	& \ldots 		& 63.6~$\pm$~10.2 	& \ldots 		& 41.3~$\pm$~18.4 	\\
195627	& 609.0~$\pm$~60.9 	& 630.0~$\pm$~44.5 	& 607.9~$\pm$~43.5 	& 405.4~$\pm$~29.3 	& 145.9~$\pm$~14.1 	& 70.9~$\pm$~7.7 	& \ldots 		& 34.1~$\pm$~7.5 	& \ldots 		& 13.0~$\pm$~7.0 	\\
20320	& 103.0~$\pm$~8.0 	& 93.9~$\pm$~5.8 	& 84.1~$\pm$~5.9 	& 42.1~$\pm$~0.8 	& \ldots		& \ldots		& \ldots 		& \ldots		& \ldots 		& \ldots 		\\
21997	& 663.7~$\pm$~46.0	& 697.6~$\pm$~49.2 	& 665.4~$\pm$~47.5 	& 410.8~$\pm$~30.0 	& 151.4~$\pm$~11.0 	& 66.7~$\pm$~9.5 	& \ldots		& 33.1~$\pm$~9.4 	& 8.3~$\pm$~2.3 	& \ldots 		\\
110411	& 248.0~$\pm$~2.2 	& 230.1~$\pm$~4.3 	& 154.2~$\pm$~7.0 	& 67.3~$\pm$~7.0 	& 37.9~$\pm$~0.8 	& 22.7~$\pm$~0.5 	& \ldots 		& 20.3~$\pm$~0.4 	& \ldots 		& \ldots 		\\
142091	& 426.2~$\pm$~22.3 	& \ldots 		& 335.0~$\pm$~16.0 	& 192.0~$\pm$~10.0 	& \ldots  		& \ldots  		& \ldots 		& \ldots  		& \ldots 		& \ldots		\\
102647	& 743.0~$\pm$~52.0 	& \ldots 		& 480.0~$\pm$~30.0 	& 215.0~$\pm$~32.0 	& 51.0~$\pm$~12.0 	& $<$~39.0 		& $<$~50.0 		& $<$~15.0 		& $~<$~6.0 		& \ldots 		\\
125162	& 364.7~$\pm$~3.9 	& 345.3~$\pm$~17.3 	& 272.1~$\pm$~15.4 	& 142.4~$\pm$~12.1 	& 50.7~$\pm$~5.1 	& 21.3~$\pm$~5.3 	& \ldots 		& 4.2~$\pm$~4.9 	& \ldots 		& \ldots 		\\
216956	& 9057.1~$\pm$~736.4 	& 10800~$\pm$~900 	& \ldots 		& 6200~$\pm$~600 	& 2700~$\pm$~300 	& 1100~$\pm$~100 	& 595~$\pm$~35	 	& 500~$\pm$~50	 	& 97~$\pm$~5	 	& \ldots 		\\
17848	& \ldots 		& 213.8~$\pm$~17.1 	& 210.8~$\pm$~18.1 	& 138.5~$\pm$~11.4 	& 50.9~$\pm$~5.9 	& 28.8~$\pm$~6.4 	& \ldots 		& 6.8~$\pm$~10.1	& \ldots 		& \ldots 		\\
9672	& \ldots 		& 2142.0~$\pm$~58.0 	& \ldots	 	& 1004.0~$\pm$~53.0 	& 372.0~$\pm$~27.0 	& 180.0~$\pm$~14.0 	& \ldots 		& 86.0~$\pm$~9.0 	& 8.2~$\pm$~1.9 	& \ldots 		\\
71722	& 155~$\pm$~4		& \ldots		& 120.5~$\pm$~4.1	& 46.9~$\pm$~8.7	& \ldots		& \ldots		& \ldots		& \ldots		& \ldots 		& \ldots		\\
182681	& \ldots 		& 607.8~$\pm$~42.8 	& 463.2~$\pm$~33.3 	& 243.0~$\pm$~18.2 	& 84.3~$\pm$~7.2 	& 30.1~$\pm$~5.4 	& \ldots 		& 8.2~$\pm$~7.3		& \ldots 		& \ldots 		\\
14055	& 787.8~$\pm$~157.6 	& 777.6~$\pm$~38.8 	& 718.8~$\pm$~35.5 	& 444.3~$\pm$~10.4 	& 186.6~$\pm$~13.6 	& 78.1~$\pm$~7.2 	& \ldots 		& 21.8~$\pm$~5.1 	& 5.5~$\pm$~1.8		& \ldots 		\\
161868	& 1085.2~$\pm$~217.0 	& 1219.4~$\pm$~85.5 	& 1044.5~$\pm$~73.4 	& 587.8~$\pm$~44.4 	& 177.9~$\pm$~12.8 	& 98.3~$\pm$~10.1 	& \ldots 		& 57.0~$\pm$~11.6	& \ldots 		& 12.8~$\pm$~5.2 	\\
188228	& 69.0~$\pm$~6.1 	& 63.1~$\pm$~4.8 	& 41.9~$\pm$~4.6 	& 23.6~$\pm$~3.2 	& 5.0~$\pm$~3.9 	& 0.7~$\pm$~4.8 	& \ldots		& 0.0~$\pm$~4.7 	& \ldots 		& \ldots 		\\
10939	& 384.6~$\pm$~16.5 	& 396.3~$\pm$~28.3 	& 403.7~$\pm$~28.8 	& 277.9~$\pm$~20.7 	& 94.9~$\pm$~7.7 	& 43.5~$\pm$~7.0 	& \ldots		& 4.0~$\pm$~6.7	 	& \ldots 		& \ldots 		\\
71155	& 211.7~$\pm$~2.8 	& 206.2~$\pm$~10.5 	& 86.5~$\pm$~5.9 	& 25.5~$\pm$~0.7 	& 7.0~$\pm$~3.4 	& 4.2~$\pm$~4.5 	& \ldots 		& $<$~13.8 		& \ldots 		& \ldots 		\\ 
172167	& 11416.1~$\pm$~2283.2 	& 10120~$\pm$~1180 	& \ldots 		& 4610~$\pm$~900 	& 1680~$\pm$~260 	& 610~$\pm$~100 	& \ldots 		& 210~$\pm$~40	 	& 45.7~$\pm$~5.4 	& \ldots 		\\
139006	& 542.0~$\pm$~80.7 	& 515.0~$\pm$~25.2 	& 235.1~$\pm$~12.6 	& 67.6~$\pm$~2.4 	& \ldots 		& \ldots 		& \ldots 		& \ldots 		& \ldots 		& \ldots 		\\
95418	& 421.13~$\pm$~84.23 	& 393.0~$\pm$~19.4 	& 189.2~$\pm$~9.6 	& 58.3~$\pm$~10.5 	& 18.3~$\pm$~3.9 	& 14.1~$\pm$~4.3 	& \ldots 		& 9.1~$\pm$~4.7 	& \ldots 		& \ldots 		\\
13161	& 643.0~$\pm$~51.0 	& 641.2~$\pm$~31.5 	& 481.1~$\pm$~23.4 	& 263.6~$\pm$~0.3 	& 87.1~$\pm$~7.3 	& 34.6~$\pm$~5.6 	& \ldots 		& 5.1~$\pm$~4.9 	& \ldots 		& \ldots 		\\
\hline 
\end{tabular}
} 
\label{tab:disks_FIR_phot}

\noindent
{\em Notes:}\\[0mm]
All values and error bars are taken from the literature as they were published.
This explains, for instance, why non-detections are sometimes given as measured values ($4.2 \pm 4.9$)
and sometimes as upper limits ($<15.0$).

{\em Instruments:}\\[0mm]
(a) {\it Spitzer}/MIPS; (b) {\it Herschel}/PACS; (c) {\it Herschel}/SPIRE; (d) JCMT/SCUBA; (e) APEX/LABOCA.

{\em References:}\\[0mm]
{\it Spitzer}/MIPS from \cite{su-et-al-2006,trilling-et-al-2007,eiroa-et-al-2013, moor-et-al-2013, moor-et-al-2013b, bonsor-et-al-2013b};\\
{\it Herschel}/PACS and /SPIRE see 
Table~\ref{tab:list};\\
JCMT/SCUBA from \cite{sheret-et-al-2004,najita-williams-2005,williams-andrews-2006};\\
APEX/LABOCA from \cite{nilsson-et-al-2010, liseau-et-al-2008}.

\end{table}
\end{turnpage}

\clearpage

\section{B. $\chi^2$-maps}

   \begin{figure*}[htb!]
   \centering
   \includegraphics[width=0.33\textwidth, angle=0]{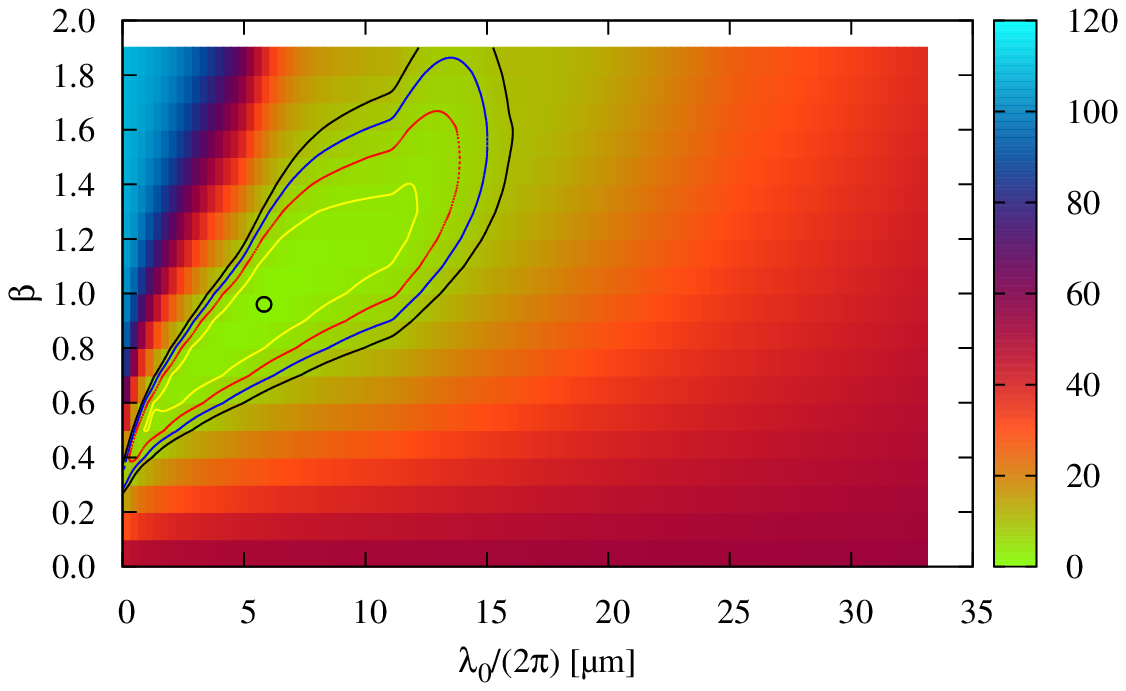}
   \hspace*{-5mm}
   \includegraphics[width=0.33\textwidth, angle=0]{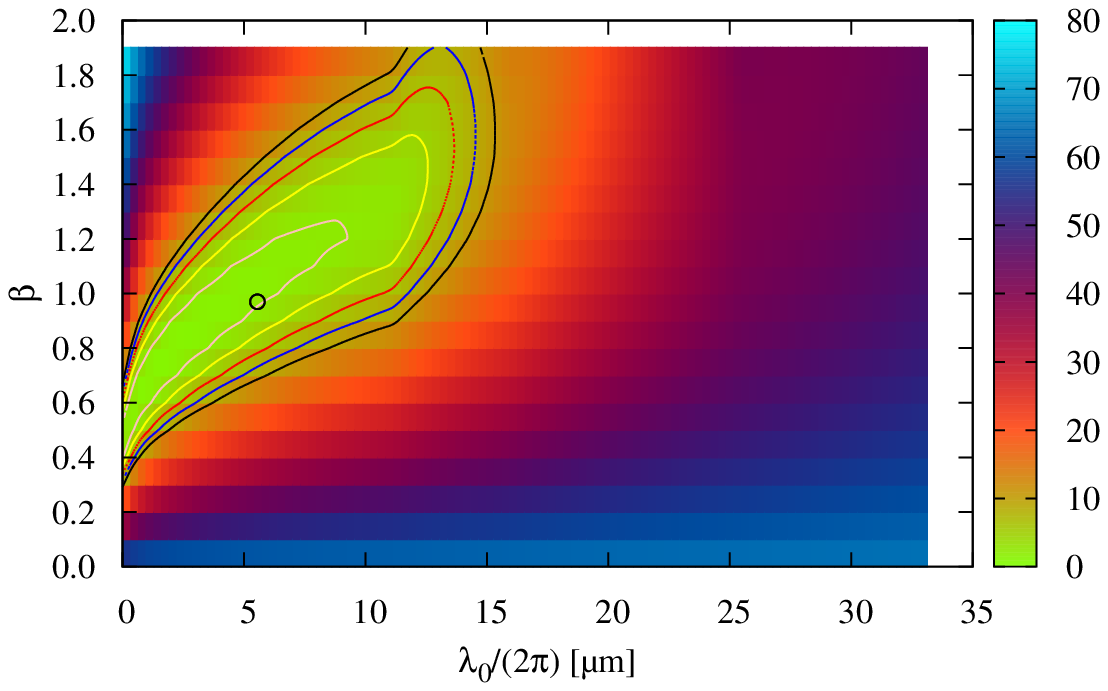}
   \hspace*{-5mm}
   \includegraphics[width=0.33\textwidth, angle=0]{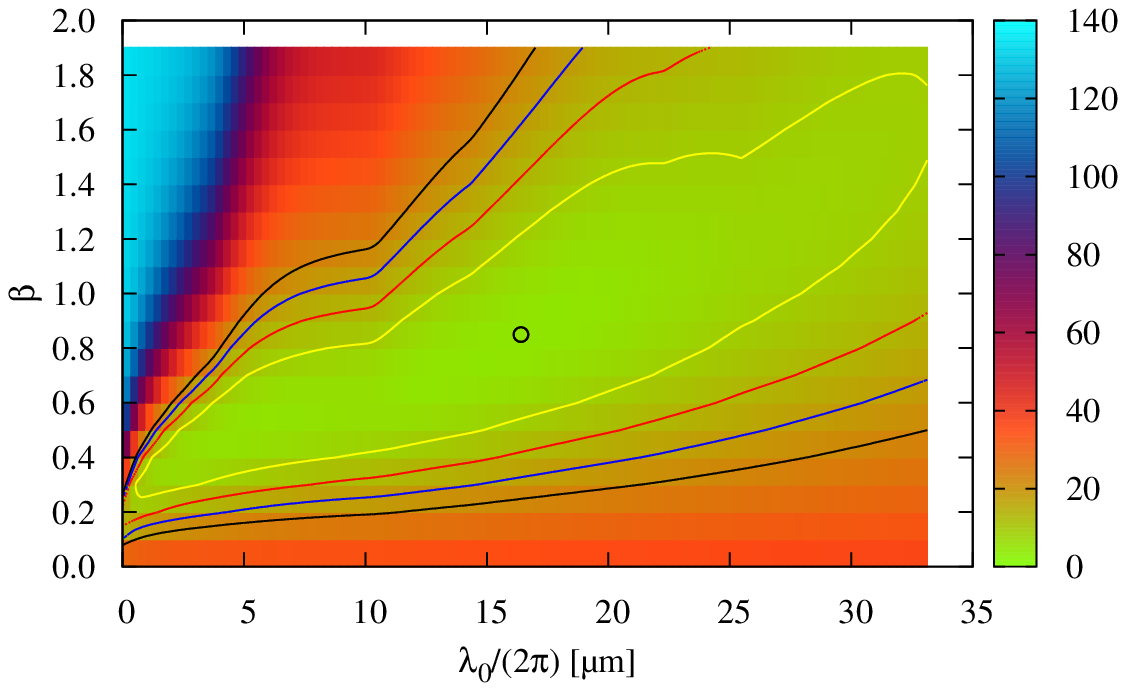}\\[-5mm]   
   
   \includegraphics[width=0.33\textwidth, angle=0]{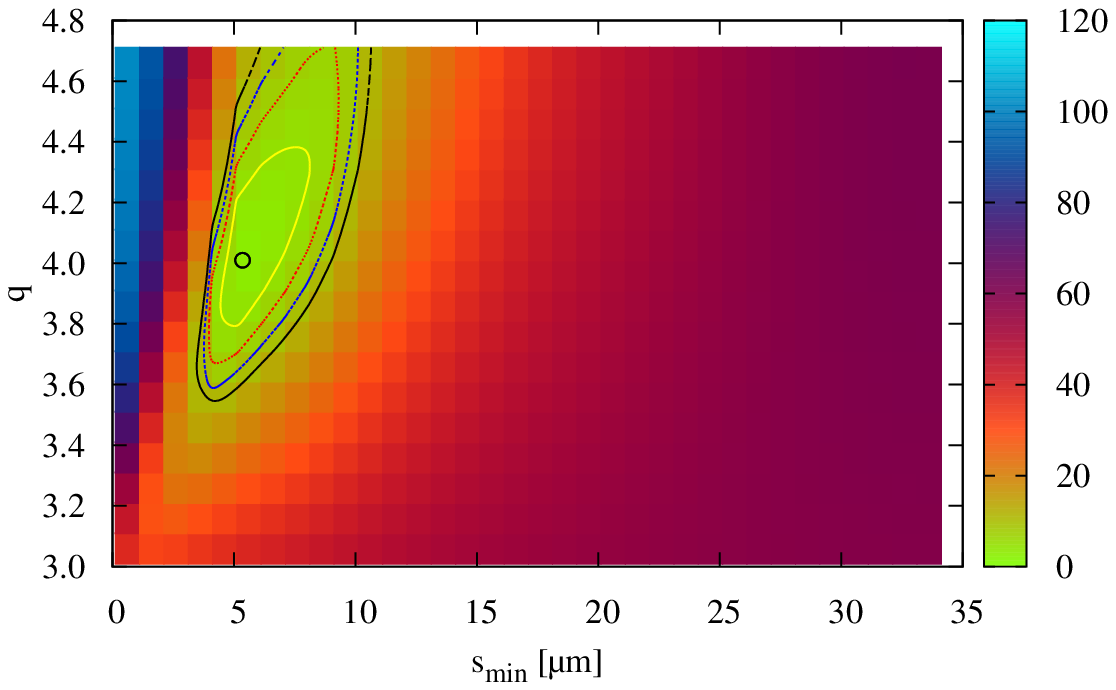}
   \hspace*{-5mm}
   \includegraphics[width=0.33\textwidth, angle=0]{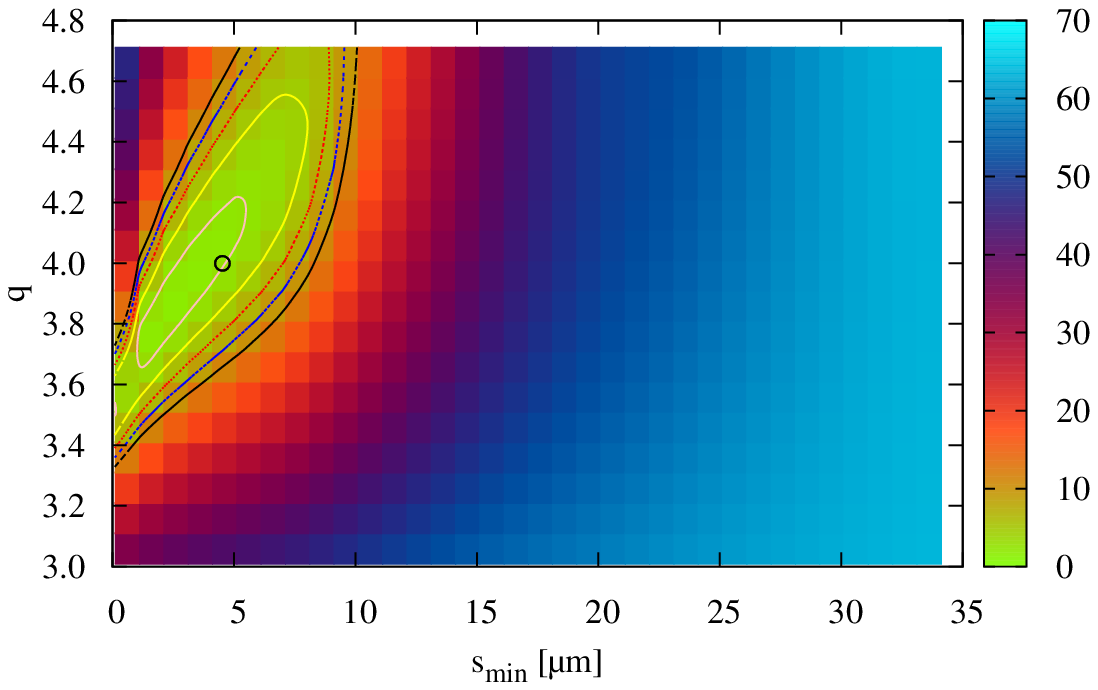}
   \hspace*{-5mm}
   \includegraphics[width=0.33\textwidth, angle=0]{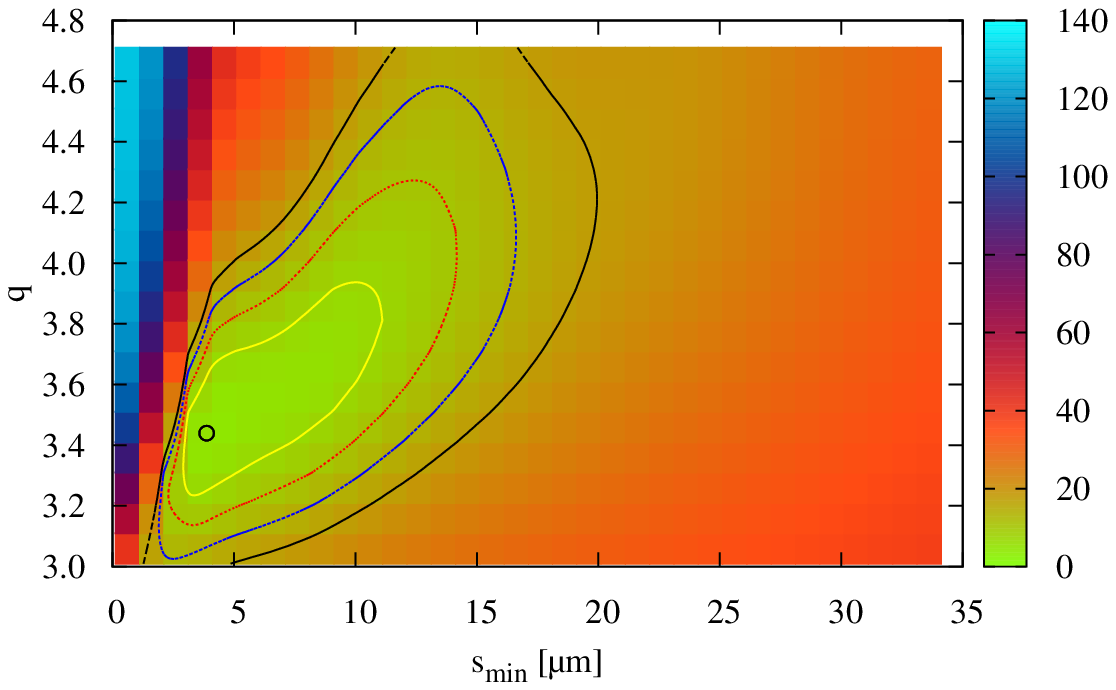}
   \caption{The $\chi^2$-maps for HD~50571 (left), HD~195627 (middle), and Fomalhaut 
   (right) in the MBB (top) and SD (bottom) method.
   The lines show the $n\cdot \chi^2_{\text{red}}$ isolines for $n = 1$ 
   (small circle), $n = 2$ (yellow line), $n = 3$ (red line), $n = 4$ (blue line) and $n = 5$ (black line),
   where $\chi^2_{\text{red}}$ is the best-fit value.
   \label{fig:chi2}
   }
   \end{figure*}

Figure~\ref{fig:chi2} presents typical $\chi^2$-maps
for three stars with different luminosities,
obtained with both fitting methods.
The contours of equal $\chi^2$ are elongated from the bottom left to
the top right, which is easy to understand.
Indeed, when $\lambda_0$ (or $\smin$)
is larger than the best-fit value, the fitting routine tries to compensate it by taking
a steeper size distribution, with a larger $\beta$ (or $q$).
Not only the bottom left -- top right orientation, but also the shape of the isolines can be 
better understood with the following arguments.
We take the SD method as an example.
Assuming that the emission of the different-sized grains is proportional to their cross section
(which is only a rough approximation to reality, of course), we can define 
the ``effective'' grain size, $s_0$.
The effective grain size is the one in which
we replace the grains with a size distribution (\ref{n(s)}) with the same number of
equal-sized grains of radius $s_0$, requiring that the latter have the same
total cross section. This gives
\be
  \int\limits_{\smin}^\infty s^{2-q} ds  = s_0^2   \int\limits_{\smin}^\infty s^{-q} ds .
\label{size eq 0}
\ee
For $q > 3$, this results in
\be
  s_0 \approx \smin \; \sqrt{(q-1) / (q-3)},
\label{s0 BB}
\ee
so that, for instance, $q=3.5$ corresponds to 
$s_0 \approx 2.2 \smin$.
We can expect that different pairs $(\smin, q)$ with the same $s_0$ should lead
to SEDs that reproduce the observed one equally well. In other words, the isolines
of $\chi^2$ should roughly follow the equation
\be
\smin \; \sqrt{(q-1)/ (q-3)} = \text{const}.
\label{curve}
\ee
The shape of the $q(\smin)$ curve determined by Eq.~(\ref{curve}) is exactly the one
that is seen in Figure~\ref{fig:chi2}.
In particular, $q$ rises more steeply at larger $\smin$
(a banana shape best
seen in yellow- and red-filled regions corresponding to moderate $\chi^2$ values).
We can also explain
why the tilt of the isolines to the x-axis gets
smaller for stars of higher luminosities (i.e. from the left to the right panels 
of Figure~\ref{fig:chi2}).
This is because $s_0$ is larger around those stars, so that at the same $\smin$
the derivative $d q / d \smin$ is smaller.

\end{document}